 \documentclass[twocolumn,showpacs,prb,eqsecnum,
 superscriptaddress,
 nofootinbib
 ]{revtex4}%

\newcommand{\Str}{\mbox{Str}}
\newcommand{\bphi}{\mbox{\boldmath $\phi$}}
\newcommand{\bvarphi}{\mbox{\boldmath $\varphi$}}
\newcommand{\bPsi}{\mbox{\boldmath $\Psi$}}
\newcommand{\bUpsilon}{\mbox{\boldmath $\Upsilon$}}
\newcommand{\bpsi}{\mbox{\boldmath $\psi$}}
\newcommand{\nnabla}{\mathbf \nabla}
\newcommand{\ptau}{\partial_\tau}
\newcommand{\bs}{\mbox{\boldmath $\sigma$}}
\newcommand{\req}[1]{Eq.~(\ref{#1})}
\newcommand{\reqs}[1]{Eqs.~(\ref{#1})}
\newcommand{\rref}[1]{(\ref{#1})}
\newcommand{\h}{\mathbf{h}}
\renewcommand{\S}{\mathbf{S}}
\newcommand{\w}{\omega}
\newcommand{\W}{\Omega}
\newcommand{\q}{\mathbf{q}}
\newcommand{\p}{\mathbf{p}}
\newcommand{\n}{\mathbf{n}}
\renewcommand{\r}{\mathbf{r}}
\renewcommand{\k}{\mathbf{k}}
\newcommand{\bchi}{\mbox{\boldmath $\chi$}}

\newcommand{\beq}{\begin{equation}}
\newcommand{\eeq}{\end{equation}}
\newcommand{\be}{\begin{equation}}
\newcommand{\ee}{\end{equation}}
\newcommand{\beqa}{\begin{eqnarray}}
\newcommand{\eeqa}{\end{eqnarray}}
\newcommand{\bea}{\begin{eqnarray}}
\newcommand{\eea}{\end{eqnarray}}
\usepackage{amssymb}
\usepackage{array}
\usepackage{amsmath}
\usepackage{graphicx}
\usepackage{dcolumn}
\usepackage{bm}
\usepackage[mathscr]{eucal}

\usepackage{amsfonts}%
\providecommand{\U}[1]{\protect\rule{.1in}{.1in}}
\begin{document}
\title{
Supersymmetric low-energy theory
 and renormalization group for a clean Fermi gas with a repulsion
in arbitrary dimensions.}
\author{I.~L.~Aleiner}
\affiliation{Physics Department, Columbia University, New York, NY 10027, USA}

\author{K.~B.~Efetov}

\affiliation{Theoretische Physik III, Ruhr-Universit\"at Bochum, 44780 Bochum,
Germany }
\affiliation{L. D. Landau Institute for Theoretical Physics, 117940 Moscow, Russia}
\date{\today}

\begin{abstract}
We suggest a new method of calculations for a clean Fermi gas with
a repulsion in any dimension. This method is based on writing
equations for quasiclassical Green functions and reducing them to
equations for collective spin and charge excitations. The spin
excitations interact with each other and this leads to non-trivial
physics. Writing the solution of the equations and the partition
function in terms of a functional integral over supervectors and
averaging over fluctuating fields we come to an effective field
theory describing the spin excitations. In some respects, the
theory is  similar to bosonization but also includes the ``ghost''
excitations which prevents overcounting of the degrees of freedom.
Expansion in the interaction reveals logarithmic in temperature
corrections. This enables us to suggest a renormalization group
scheme and derive renormalization group equations. Solving these
equations and using their solutions for calculating thermodynamic
quantities we obtain explicit expression for the specific heat
containing only an effective amplitude of the backward scattering.
This amplitude has a complicated dependence on the logarithm of
temperature, which leads to a non-trivial temperature dependence
of the specific heat.

\end{abstract}

\maketitle

\section{Introduction}

Landau theory of the Fermi liquid (FL) suggested 50 years ago \cite{landau} is
the basis for description of normal metals. Roughly speaking, the main
statement of the FL theory is that the low energy behavior of interacting
fermions is similar to that for the ideal Fermi gas. This theory explains very
successfully properties of a large number of metals and $He^{3}$. It is quite
common to discuss experimental systems just forgetting about the interaction
and using phenomenological parameters like, {\em e.g.} effective mass, density of
states at the Fermi surface, etc., instead.

Yet, a recent progress in study of unconventional metals like high temperature
superconductors, heavy-fermion materials has revealed considerable deviations
of their properties from those predicted by the FL theory (for a recent review
see, e.g. Refs.~\onlinecite{timusk,norman}. As a result, quite a few theoretical works
have appeared recently where the validity of the Landau FL theory was
discussed\cite{anderson,shankar,metzner,houghton,kim,abanov}.

At first glance, being phenomenological from the beginning, the Landau theory
has been later confirmed by analyzing diagrammatic expansions\cite{landau1}
(see, also Refs.~\onlinecite{agd,pines,fetter,baym}) and looks very well established.
However, a very strong assumption was used in this discussion, namely, that
one could single out a singular particle-hole channel and sum proper ladder
diagrams. Two particle irreducible
vertices entering the ladder diagrams should remain finite and analytic
in the limit of small momenta and frequencies. Of course, this is not always
true and under certain conditions the system may become superconductor,
antiferromagnet, etc. In many cases, the failure of the Fermi liquid
description can be checked by a more careful consideration of the perturbation
theory and, {\em e.g.}, the existence of the superconducting transition can be
established in this way \cite{agd,pines,fetter}.

Nevertheless, it is believed that the system of fermions with a
repulsion should behave like a Fermi liquid provided the
dimensionality $d>1$ and there are no van-Hove type singularities
on the Fermi surface. Naturally, the similarity between the Fermi
liquid and ideal Fermi gas cannot be exact and, clearly, there are
corrections at finite temperatures, finite frequencies or momenta.
These corrections  become especially interesting when they are
non-analytic  functions of the values of the temperatures,
frequencies or momenta.

For the ideal Fermi gas, such quantities as $C\left(  T\right)  /T$ and
$\chi\left(  T\right)$, where $C\left(  T\right)  $ is the specific heat and
$\chi\left(  T\right)  $ is the spin susceptibility, can be represented in a
form of asymptotic series in $T^{2}/\varepsilon_{F}^{2}$
($\varepsilon_{F}$ is the Fermi
energy). Fermion-fermion interactions lead to additional contributions to
these quantities that are not necessarily analytic in $T^{2}$.
It is well established that in $D=3$
next-to-leading term in $C\left(  T\right)  /T$ is $T^{2}\ln T$,
see Refs.~\onlinecite{eliash,doniach,brink,amit,chubukov4}. It was claimed in
Ref.~\onlinecite{belitz}
that the non-uniform spin
susceptibility, $\chi\left(  Q\right)$, depends on the momentum $Q$ as
$Q^{2}\ln Q$. In $2D$, non-analytical corrections to $C\left(
T\right)  /T$ and $\chi\left(  Q,T\right)  $ found so far scale as
$T$, see Refs.~
\onlinecite{coffey,chubukov1,sarma,Catelani}, and $\max\left\{  Q,T\right\}  $,
see
Refs.~\onlinecite{belitz,chubukov1,baranov,chitov,chubukov2,chubukov3},
respectively.

The existence of the non-analytical corrections to the physical
quantities is not accidental. 
In fact, all of the singular
corrections to the thermodynamic quantities can be understood in
terms of the dynamics of the low lying collective excitations, see
{\em e.g.} Ref.~\onlinecite{Catelani,chubukov3}.
A detailed analysis of non-analytic corrections to the specific
heat of a three-dimensional Fermi liquid is given recently in
Ref.~\onlinecite{chubukov4}. It was shown in 
Ref.~\onlinecite{chubukov3} 
that all the other contribtutions contain
integrations over the entire Fermi surface are regular in $T^{2}$,
unlike the contributions of the collective modes which
contain $2k_{F}$ scattering.  

Explicit calculations for systems such low lying modes are not
simple even in the lowest orders of the perturbation theory. This
situation is analogous to that in theory of disordered metals,
where the low energy behavior of the system is governed by the
multiple interference of the electron waves scattered by
impurities. In the diagrammatic language, this effect can be
expressed in terms of an interaction between electrons and
diffusion modes (so called cooperons and diffusons
\cite{glk,aar}). Calculations in high orders in the diffusion
modes (weak localization corrections) using the diagrammatic
expansions are also quite involved.

However, another approach has been developed in the theory of disordered
metals based on integrating out electron degrees of freedom and deriving an
effective Lagrangian describing the diffusion modes. This reduction simplifies
calculations because only low lying excitations are left in the theory. The
Lagrangian has the form of a so called $\sigma$-model, first introduced in the
theory of disordered metals in Ref.~\onlinecite{wegner} using the replica trick.
Another, supermatrix form of the $\sigma$-model, is based on a supervector
representation\cite{review} of Green functions, and this method has found
numerous applications (for a review, see, e.g., Ref.~\onlinecite{book}).

One can see a certain analogy between calculating the
non-analytical corrections for the Fermi gas with interaction and
the weak localization corrections in theory of disordered metals.
Following this analogy it seems quite natural to try to develop a
scheme that would allow us to reduce the initial model of the
interacting Fermi gas to a model describing only low lying
excitations. Then, we would have in the theory not the initial
fermions but the collective excitations like, the zero sound, may
be, weakly interacting with each other. Apparently, the latter are
bosons and we would have, as a result, a system of bosons instead
of the initial fermionic system. Reducing fermion models to boson
ones is usually referred to as \textit{bosonization} and we will
loosely use this word in the subsequent discussion, though the
final theory we derive will be necessarily supersymmetric, and
contain also the fermionic degrees of freedom -- ``ghosts''.

The notion of the supersymmetry appears quite naturally already in
the consideration of the leading non-analytic corrections to, say,
specific heat, $C$. Namely, for the spinless electrons, the
ring-diagrams correction to $C$ is of the form \be
\begin{split}
&\delta C = -T\frac{\partial^2\delta \W}{\partial T^2},
\\
&\delta\W=\frac{T}{2} \sum_{\w_n} \int\!\frac{d^d k d\n}{(2\pi)^d}
 \ln
\left[ 1 + {\hat{\mathbb F}} \hat{\Pi} (\w_n, \k)\right]_{\n\n},
\end{split}
\label{eq:1.1} \ee where the unit vector $\n$ characterizes the
position of the momentum on the Fermi surface, and $\w_n=2\pi n T$
is a bosonic Matsubara frequency. The quasiparticle polarization
operator acts on any function $b(\n,\q)$ as
\[
 \hat{\Pi} (\w_n, \k)b(\n,\k)=
\left[1-\frac{i\w_n}{i\w_n-v_F \k\cdot \n}\right]b(\n,\k)
\]
with $v_F$ being the Fermi velocity. The operator ${\hat{\mathbb
F}}$ is defined as
\[
\left[{\hat{\mathbb F}}b\right](\n_1,\k)=\int d\n_2
{\mathbb F}\left(\widehat{\n_1\n_2}\right) b(\n_2,\k)
\]
and ${\mathbb F}\left(\widehat{\n_1\n_2}\right)$ is the
Fermi-liquid function describing the interaction between the
quasiparticles moving in directions $\n_1$ and $\n_2$. We can
imply the proper normalization of the solid angle, {\em i.e.}
\begin{equation}
\int d\mathbf{n}=1, \label{a11a}%
\end{equation}
when integrating over the momentum directions

Equation \rref{eq:1.1} can be re-written identically as
\be
\begin{split}
&\delta\W= \W_\rho- \W_g;
\\
&\W_\rho=\frac{T}{2} \sum_{\w_n} \int\!\frac{d^d k d\n}{(2\pi)^d}
\left[\ln\left(i\w_n- v_F \left(1+\hat{\mathbb F}\right)\n\cdot \k\right)
\right];\\
&\W_g=\frac{T}{2} \sum_{\w_n} \int\!\frac{d^d k d\n}{(2\pi)^d}
\left[\ln\left(i\w_n- v_F\n\cdot \k\right)
\right].
\label{eq:1.2}
\end{split}
\ee [In all the consideration we will not write factors of volume,
whenever they are self-evident for an educated reader]. The form
of this expression is quite instructive. The first term, $
\W_\rho$ is nothing but the contribution of the non-interacting
bosons, whose spectrum is determined by the kinetic equation in
Landau theory. However, those bosons are made out of electrons
which are already included in the leading term of the specific
heat. That is why the second term, $\W_g$, simply subtracts the
contribution of the electron-hole pairs in the absence of the
interactions to avoid a double counting. Since the contribution of
the second term is opposite to the contribution of the physical
bosons, it will be natural to treat them as {\em pseudofermions},
or ``ghosts'' and include them into the field theory description
on equal footing with the bosonic field\footnote{The partition of
the low-energy excitations in the Fermi liquid in terms of
physical bosons and artificial ghosts (to avoid overcounting of
the degrees of freedom) was suggested in
Ref.~\onlinecite{Catelani}, however, the consideration there was
limited to the  theory of non-interacting bosons only.}.

However, one cannot proceed in a direct analogy with the supersymmetry
method of Ref.~\onlinecite{book} because the latter is essentially based on the use
of a sufficiently strong disorder leading to a diffusion motion at not very
long distances. Fortunately, the method can be generalized to the clean case
by writing equations for quasiclassical Green functions and representing their
solution in terms of functional integrals. Equations for the quasiclassical
Green functions for the disorder problems were introduced in Ref.~\onlinecite{mk}.
The authors of Ref.~\onlinecite{mk} suggested writing their solutions from the
condition for a minimum of functional having a form of a ballistic $\sigma
$-model. This could be written in a form of a functional integral provided
this integral could be calculated by the saddle point method. At the same
time, it was not clear why the saddle point approximation could work well for
the clean of weakly disordered case.

Later it was realized\cite{ek} that the solution of the quasiclassical
equations can be \textit{exactly} written in terms of a functional integral
with a Lagrangian having the form of the ballistic $\sigma$-model. Within such
an approach one reduces the initial electron model to a model describing
\textquotedblleft collective excitations\textquotedblright\ and therefore it
is relevant to call it loosely bosonization. A smooth potential could be
considered in this approach but no interaction was included.

In the present paper we develop a new method for studying clean
fermion systems with a fermion-fermion repulsive interaction. It
is based on decoupling the interaction by a proper
Hubbard-Stratonovich transformation. Both the forward and backward
scattering are taken into account and therefore the slow
decoupling field $\Phi\left(  \mathbf{r,}\tau\right)  $ has a spin
structure (where $\mathbf{r}$ is the coordinate and $\tau$ is the
imaginary time). After the decoupling we derive equations for the
quasiclassical Green functions. In order to solve the equations we
use a new trick based on a slow dependence of the field
$\Phi\left(  \mathbf{r,}\tau\right)$. As a result, we obtain
linear non-homogeneous equations for spin and charge excitations.
The spin excitations are most important and we represent the
solution of the corresponding equation in terms of an integral
over supervectors. Averaging over the field $\Phi$, we obtain an
effective theory with a $\psi^3+ \psi^{4}$ interaction. This
theory describes collective bose excitations. It is important to
emphasize that the $\psi^3$ and $\psi^{4}$ terms arise due to
spin-spin interactions. For fermion models containing only a
density-density interaction, those terms vanish and the theory
becomes free.

Making
expansions in the $\psi^3+\psi^{4}$ interaction we found that the theory is
logarithmic in any dimension, which allowed us to use a renormalization group
approach to sum up the parquet series.
Writing and solving the RG equations we are able to express the
non-analytical contribution to the specific heat with a logarithmic accuracy.

It is relevant to mention that the method we develop now is completely
different from the $\sigma$-model approach for disordered systems with
interaction\cite{finkel} (see also a recent supersymmetric
formulation\cite{schwiete}). In the present approach the collective
excitations are described by a Lagrangian with the $\psi^3+\psi^{4}$ interaction and
not by a $\sigma$-model.

Attempts to bosonize fermionic models in the dimensionality $d>1$ have been
undertaken in the past starting from the work\cite{luther}. In this first work
the one dimensional bosonization was directly extended to higher dimensions.
This idea was further developed more recently in a number of publications
\cite{haldane,houghton1,houghton,neto,kopietz,kopietzb,
khveshchenko,khveshchenko1,castellani,metzner}.
Our approach is completely different and more general. The high dimensional
bosonization developed previously can be applicable only for a long range
interaction when the backward scattering is absent. In this case the
fermion-fermion interaction is replaced by an interaction of the local
densities (local in space and in the position on the Fermi surface). This
means that effects related to electron spins are beyond the possibility of
that method. In contrast, the backward scattering is included in our approach
and plays a very important role.

The article is organized as follows:

In Section \ref{sec2}, we formulate the model and single out slow pairs in the
interaction term. We perform a Hubbard-Stratonovich transformation and reduce
the model with the interaction to a model with slowly fluctuating fields. Then
we derive quasiclassical equations for electron Green functions.

In Section \ref{sec3}, we introduce an eikonal type method
(a.k.a. Schwinger Ansatz) for solving the
equations. As a result, we obtain equations for effective charges and spins in
the presence of fluctuating fields. We express the partition function in terms
of the solutions of these equations and calculate it neglecting the
interaction between the excitations.

In Section \ref{sec4}, we write the solutions of the equations for the collective modes
and the partition function in terms of functional integrals over supervectors.
We average over the fluctuating fields and derive an effective field theory
containing interaction terms.

Section \ref{sec5} contains an explanation how one obtains logarithmic contributions.
Then, we develop a RG scheme integrating over fast variables and writing
renormalized coupling constants.

In Section \ref{sec6}, we derive renormalization group equations and find their solutions.

Section \ref{sec7} is devoted to calculation of the thermodynamic
potential and specific heat using the solutions of the RG
equations. Explicit formulae are obtained in two, three and,
separately, in one dimensions.

Our findings are discussed in Section \ref{sec8}.

\section{The model and basic equations.}
\label{sec2}
\subsection{Formulation of the model and singling out slow modes}

We start  with formulating the model we would like to investigate.
This is the most general model for fermions with a short range
interaction \footnote{Inclusion of the long-range Coulomb
interaction is straightforward and does not lead to any
consequences relevant for our study} in an arbitrary dimension
$d$. The Fermi surface is assumed to have no singularities. In
order to avoid unnecessary trivial generality we consider the
Fermi surface to be a just $d$ -dimensional sphere.

It will be  convenient for us to express the physical quantities
in terms of a functional integral over anticommuting variables
$\chi$ with an Euclidian action $\mathcal{S}$. In this
formulation, one can use imaginary time $\tau$ in the interval
$0<\tau<1/T$, where $T$ is the temperature and write the
temperature Green
function $G_{\sigma\sigma^{\prime}}\left(  x,x^{\prime}\right)$, as follows%
\begin{equation}
G_{\sigma,\sigma^{\prime}}\left(  x,x^{\prime}\right)  =Z^{-1}\int\chi
_{\sigma}\left(  x\right)  \chi_{\sigma^{\prime}}^{\ast}\left(  x^{\prime
}\right)  \exp\left(  -\mathcal{S}\right)  D\chi D\chi^{\ast}, \label{a1}%
\end{equation}
where
\begin{equation}
x=\left(  \mathbf{r},\tau\right),\quad
\int dx \cdot \equiv \int d^d\r \int_0^{1/T} d\tau \cdot  \label{a1ab}%
\end{equation}
$\mathbf{r}$ is the spatial coordinate, and $\sigma$ labels the spin.

The partition function $Z$ entering Eq. (\ref{a1}) has the form
\begin{equation}
Z=\int\exp\left(  -\mathcal{S}\right)  D\chi D\chi^{\ast}. \label{a1a}
\end{equation}

The action $\mathcal{S}$ in Eq. (\ref{a1}) can be written as%
\begin{equation}
\mathcal{S}=\int {\mathcal L}_{0}dx +\mathcal{S}_{int}, \label{a2}%
\end{equation}
where the term $\mathcal{L}_{0},$
\begin{equation}
\mathcal{L}_{0}=\sum_{\sigma}\chi_{\sigma}^{\ast}\left(  x\right)  \left(
-\frac{\partial}{\partial\tau}-\hat{H}_{0}\right)  \chi_{\sigma}\left(
x\right), \label{a3}%
\end{equation}%
\begin{equation}
\hat{H}_{0}=\frac{\mathbf{\hat{p}}^{2}}{2m}-\varepsilon_{F} \label{a3a}%
\end{equation}
stands for the Lagrangian density of free fermions ($\varepsilon_{F}$ is the Fermi
energy, $m$ is the mass and $\mathbf{\hat{p}}$ is the momentum operator) and
$\mathcal{S}_{int}$ describes the fermion-fermion interaction,
\be
\begin{split}
\mathcal{S}_{int}&=\frac{1}{2}\sum_{\sigma,\sigma^{\prime}}
\int  dxdx^{\prime} v\left(
x-x^{\prime}\right)
\\
& \times\left[\chi_{\sigma}^{\ast}\left(
x\right)  \chi_{\sigma^{\prime}}^{\ast}\left(  x^{\prime}\right)
 \chi_{\sigma^{\prime}}\left(  x^{\prime}\right)
\chi_{\sigma}\left(  x\right)
\right]
,
\end{split}
\label{a4}
\ee
where $v\left(  x-x^{\prime}\right)  =U\left(  \mathbf{r-r}^{\prime}\right)
\delta\left(  \tau-\tau^{\prime}\right)  $ and $U\left(  \mathbf{r-r}^{\prime
}\right)  $ is the potential of the interaction.

The field variable $\chi$ must be antiperiodic in $\tau$ with the period $1/T$%
\begin{equation}
\chi\left(  \mathbf{r,}\tau\right)  =-\chi\left(  \mathbf{r},\tau+1/T\right).
\label{a5}%
\end{equation}
The thermodynamic potential $\Omega$ can be written as
\begin{equation}
\Omega=-T\ln Z. \label{a6}%
\end{equation}

The functional integrals over $\chi$ with the Lagrangian $\mathcal{L}$,
\reqs{a2}--\rref{a4}, are too complicated to be calculated exactly and
making
controllable
approximations is inevitable.
For performing further formal manipulations we restrict ourselves with the case of a weak
interaction, and discuss the changes of the theory for stronger
interactions in the end of Sec.~\ref{sec3b}.
As we have mentioned in the Introduction, the most interesting
contributions come from the interaction of the fermions with low lying
collective excitations and we would like to concentrate on such
contributions.
Thus,  we will try to simplify the interaction term $\mathcal{S}_{int}$
to display these collective modes explicitly.

This can be achieved by singling out in the interaction term $\mathcal{S}
_{int}$ pairs of the variables $\chi$ slowly varying in space. Using the
Fourier representation we write the effective interaction $\mathcal{\tilde{S}%
}_{int}$ containing the slow pairs as

\begin{align}
&  \mathcal{S}_{int}\mathcal{\rightarrow}\mathcal{\tilde{S}}_{int}%
\mathcal{=}\frac{1}{2}\sum_{\sigma,\sigma^{\prime}}\int dP_{1}dP_{2}%
dK\label{a7}\\
& \times \Big\{{V}\left(\k\right)
\chi_{\sigma}^{\ast}\left(  P_{1}\right)  \chi_{\sigma}\left(
P_{1}+K\right)  \chi_{\sigma^{\prime}}^{\ast}\left(  P_{2}\right)
\chi_{\sigma^{\prime}}\left(  P_{2}-K\right) \nonumber\\
&  -V\left( \p_{12}\right)
\chi_{\sigma}^{\ast}\left(  P_{1}\right)  \chi_{\sigma^{\prime}}\left(  P_{1}+K\right)
\chi_{\sigma^{\prime}}^{\ast}\left(  P_{2}+K\right)  \chi_{\sigma}\left(
P_{2}\right)
\Big\},\nonumber
\end{align}
where
\[
V(\p )=\int d\r e^{-i\p\r} U(\r),
\]
and $\p_{12}\equiv \p_1-\p_2$.
In Eq. (\ref{a7}), $P_{i}=\left(  \mathbf{p}_{i}\mathbf{,}\varepsilon_{n_{i}%
}\right) ,$ where $\mathbf{p}_{i}$ is the momentum and $\varepsilon_{n_{i}%
}=\pi T\left(  2n_{i}+1\right)  $ are Matsubara fermionic frequencies
$(i=1,2)$. Short hand notation $K$ reads $K=\left(  \mathbf{k,}\ \omega
_{n}\right)  $, where $\omega_{n}=2\pi Tn$ are Matsubara bosonic frequencies.

The symbol of the integration $\int dP_{i}$ in Eq. (\ref{a7}) reads as

\begin{equation}
\int dP_{i}\left(  ...\right)  =T\sum_{\varepsilon_{n_{i}}}\int\frac
{d^{d}\mathbf{p}}{\left(  2\pi\right)  ^{d}}\left(  ...\right)  \label{a7a}%
\end{equation}
whereas the symbol $\int dK$ has the  meaning%
\begin{equation}
\int dK\left(  ...\right)  =T\sum_{\omega_{n}}\int f\left(
\mathbf{k}\right)  \frac{d^{d}\mathbf{k}}{\left(  2\pi\right)  ^{d}}\left(
...\right).  \label{a7b}%
\end{equation}

In \req{a7b} we define the cut-off function
\begin{equation}
f\left(  \mathbf{k}\right)  =f_{0}\left(  kr_{0}\right),
\quad
k=\left\vert \mathbf{k}\right\vert.
 \label{a7c}%
\end{equation}
The function $f_{0}\left(
t\right)$ has the following asymptotics: $f_{0}\left(  t\right)  =1$ at
$t=0$ and $f\left(  t\right)  \rightarrow0$ at $t\rightarrow\infty$.
This function is written in
order to cut large momenta $k.$ The parameter $r_{0}$ is the minimal length in
the theory and we assume that $r_{0}$ much larger than
the Fermi wavelength  $\lambda_{F}=1/p_F$.  In other words, the momenta $k$
are cut by the maximal momentum $k_{c}=r_{0}^{-1} \ll p_F$, and
the partition \rref{a7c} is not threatened by double counting,
see also the discussion in the end of this subsection.

\begin{subequations}
\label{a910}
Equation \rref{a7} permits the further simplifications for the short
range interaction potential. In the first term one can neglect the
dependence on the transmitted momentum $\mathbf{k}$,
\be
V(\k \ll p_F) = V_2.
\label{a9}
\ee
 In the second
term one notices that momenta $\p_{12}$ are close to the Fermi
surface, so one can write\footnote{We borrow the one-dimensional
  notation of Ref.~\onlinecite{dl} for the corresponding amplitudes.}
\be
V\left(\p_{12}\right)=V_1\left(\widehat{\p_1\p_2}\right);
\quad V_1\left(\theta\right)\equiv V\left(2p_F\sin\frac{\theta}{2}\right).
\label{a10}
\ee
\end{subequations}

In \req{a7} we recombine the terms with the help of the identity
\[
2\delta_{\sigma_1\sigma_4}\delta_{\sigma_2\sigma_3}=
\delta_{\sigma_1\sigma_2}\delta_{\sigma_3\sigma_4}
+ \bs_{\sigma_1\sigma_2}\cdot\bs_{\sigma_3\sigma_4},
\]
where $\bs=\left(\sigma_x.\sigma_y,\sigma_z\right)$, and
$\sigma_{i}$, $i=x,y,z$ are the Pauli matrices.
Utilizing definitions \rref{a910}, we re-write \req{a7}
as
\be
\begin{split}
 \mathcal{\tilde{S}}_{int}&=\frac{1}{2}
\int dP_{1}dP_{2}dK
\Big[
\rho\left(P_1,K\right)V_s\left(\theta_{12}\right)\rho\left(P_2,K\right)
\\
& \quad + \sum_{i=x,y,z}{S}_i\left(P_1,K\right)V_t\left
(\theta_{12}\right)
{S}_i\left(P_2,K\right)
\Big],
\end{split}
\label{a12}
\ee
where, as before, $\theta_{12}=\widehat{\p_1 \p_2}$.
and the definitions
\be
\begin{split}
\rho\left(P,K\right)&=\sum_\sigma
\chi_{\sigma}^{\ast}\left(  P+\frac{K}{2}\right)
\chi_\sigma\left(  P-\frac{K}{2}\right)\\
 S_i\left(P,K\right)&=\sum_{\sigma_1,\sigma_2}\sigma_i^{\sigma_1\sigma_2}
\chi_{\sigma_1}^{\ast}\left(  P+\frac{K}{2}\right)
\chi_{\sigma_2}\left(  P-\frac{K}{2}\right)
\end{split}
\label{a121} \ee are introduced. The functions $V_{s}(\theta)$ and
$V_{t}(\theta)$ are known  as amplitudes of the singlet and
triplet scattering, respectively. They are related to the
amplitudes $V_{1}$ and $V_{2}$ as
\begin{equation}
V_{s}(\theta) =V_{2}-\frac{1}{2}V_{1}(\theta), \quad
V_{t}(\theta)=-\frac{1}{2}V_{1}(\theta). \label{a13}%
\end{equation}
In what follows we assume the operator $\hat{V}_s(\theta)$
\be
\left[\hat{V}_{s,t}b\right](\n_1)=\int d\n_2 V_{s,t}\left(\widehat{\n_1\n_2}\right)b(\n_2).
\label{hatVst}
\ee
to be positive definite and the operator $\hat{V}_t(\theta)$
to be negative definite. According to \req{a13}, those assumptions
imply the repulsive interaction.

Equation \rref{a12} can be recast in a more transparent form.
Performing Fourier transform over $K$ in \req{a12}, we obtain \be
\begin{split}
 \mathcal{\tilde{S}}_{int}&=\frac{1}{2}
\int dP_{1}dP_{2}\int d\r\int_{0}^{1/T}d\tau
\\
&
\times
\Big[
\rho\left(P_1;\r,\tau\right)V_s\left(\theta_{12}\right)\rho\left(P_2;\r,\tau\right)
\\
& \quad + \sum_{i=x,y,z}{S}_i\left(P_1;\r,\tau\right)V_t\left
(\theta_{12}\right)
{S}_i\left(P_2;\r,\tau \right)
\Big].
\end{split}
\label{a122}
\ee
 The entries in \req{a122},
\be
\begin{split}
\rho\left(P;\r,\tau\right)&=T\sum_{\w_n}\int\frac{d^dk}{(2\pi)^d}e^{i\k\r-i\w_n\tau}
f^{1/2}(\k)
\rho\left(P,K\right);
\\
S_j\left(P;\r,\tau\right)&=T\sum_{\w_n}\int\frac{d^dk}{(2\pi)^d}e^{i\k\r-i\w_n\tau}
f^{1/2}(\k)
S_j\left(P,K\right),
\end{split}
\label{a123} \ee have the meaning of the smooth charge and spin
density accumulated in the phase space, and \req{a122} is
equivalent to the Landau description of the interacting
quasiparticles. Appearance of the cutoff function $f(\k)$,
\reqs{a7b}, \rref{a7c}, means that those densities may vary only
with the spatial scale much larger than the Fermi wavelength
$\lambda_F$.

Equations \rref{a12} -- \rref{a123} constitutes the reduction of
the original interaction to the interaction involving the soft
electron-hole pair only. These are the only terms that may produce
the non-analytic contributions to the observable quantities. In
what follows we will manipulate with interaction \rref{a122} to
obtain the low-energy theory in terms of the charge and the spin
densities in the phase space.




\begin{figure}[t]
\setlength{\unitlength}{2.3em}
\begin{picture}(12,10)

\put(5,7.5){\large a)}
\put(1.5,8){\includegraphics[width=10\unitlength]{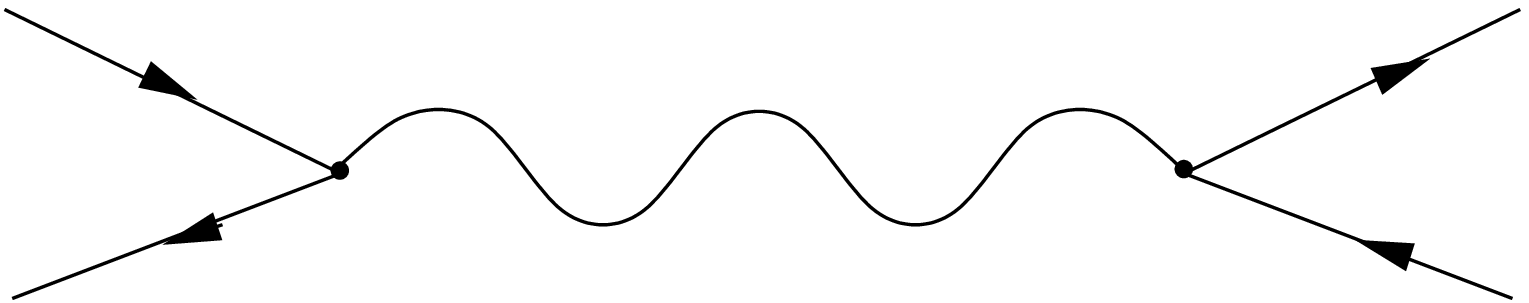}}
\put(0,9.5){$P_1,\sigma_1$} \put(7.5,9.5){$P_2-K,\sigma_2$}
\put(0,7.7){$P_1+K,\sigma_1$} \put(8,7.7){$P_2,\sigma_2$}

\put(1,3.5){\large b)}
\put(1.2,1.5){\includegraphics[width=2.3\unitlength]{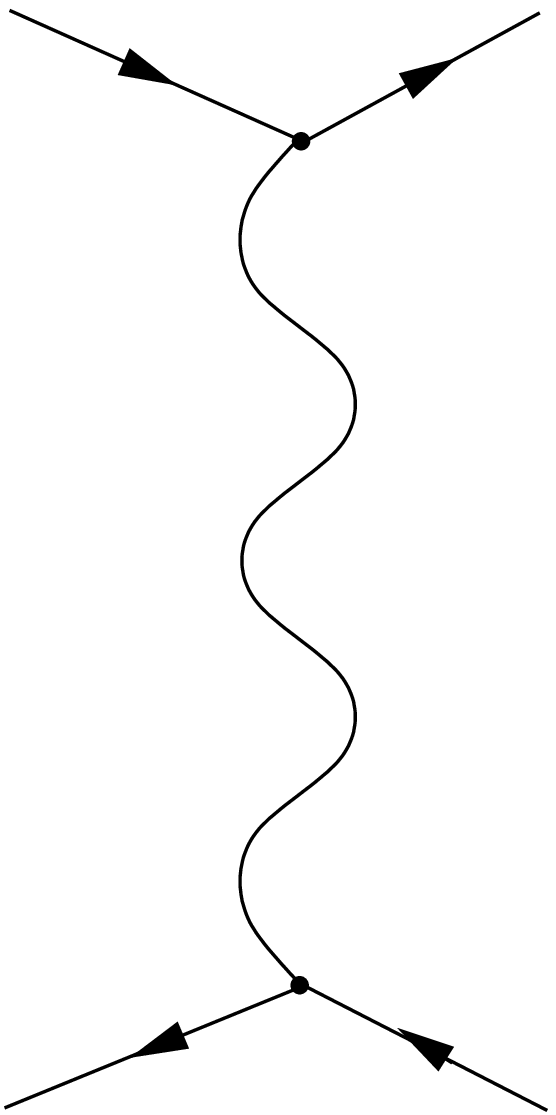}}
\put(0.7,5.3){$P_1,\sigma_1$} \put(3.0,5.3){$P_2-K,\sigma_1$}
\put(0,2.2){$P_1+K,\sigma_2$} \put(3.3,2.2){$P_2,\sigma_2$}

\put(6,3.5){\large c)}
\put(6.2,1.5){\includegraphics[width=2.3\unitlength]{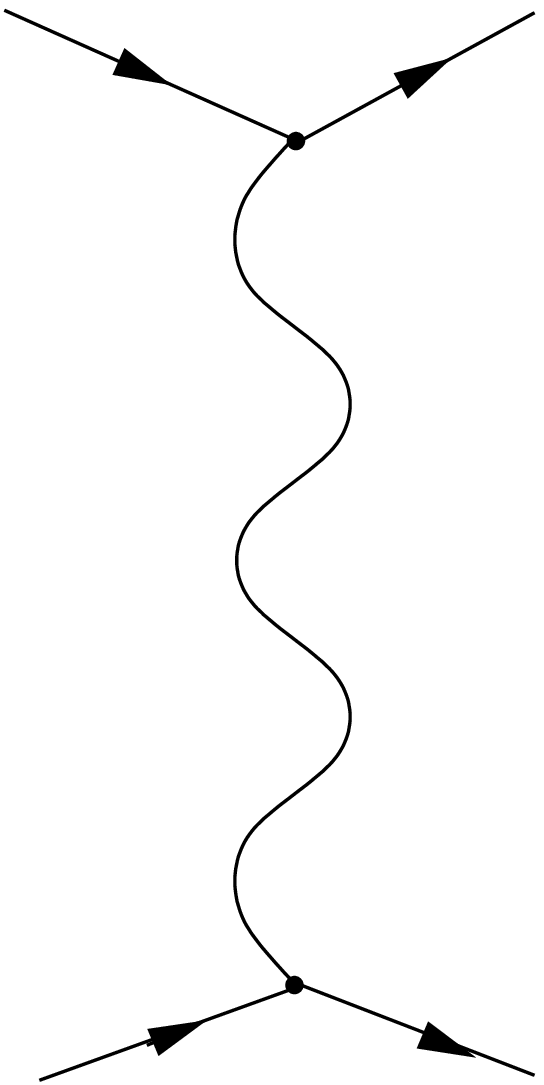}}
\put(5.3,5.3){$-P_1,\sigma_1$} \put(8.0,5.3){$-P_2,\sigma_1$}
\put(5,2.2){$P_1+K,\sigma_2$} \put(8.3,2.2){$P_2+K,\sigma_2$}

\end{picture}

\caption{Vertices describing separation into soft modes, see \reqs{a7}
  or \rref{a122}. Momenta $k$ are
small, $k<k_{c}\ll p_F$. Diagrams a,b) correspond to the
first and second terms in \req{a122}. The vertex c) was rightfully
omitted in the low-energy effective theory, see text.}

\label{fig1}
\end{figure}

Closing this subsection, we discuss a very crucial issue that
might start worrying an attentive reader at this point -- what is
the fate of the Cooper channel? Indeed, examination of the
scattering processes induced by the Hamiltonian \rref{a7}, see
Fig.~\ref{fig1} shows that the vertex $V_3$, describing the
particle-particle interaction between the pairs
$\chi(P_1)\chi(-P_1)$ with $\chi^{\ast}(P_2)\chi^{\ast} (-P_2)$ is
missing.
 These are  just terms that generate Cooperons for the system with the
time reversal symmetry\cite{book,finkel}.

\begin{figure}[t]
\setlength{\unitlength}{2.3em}
\begin{picture}(10,12)

\put(5,12){\large a)}
\put(1.8,8.2){\includegraphics[width=4.8\unitlength]{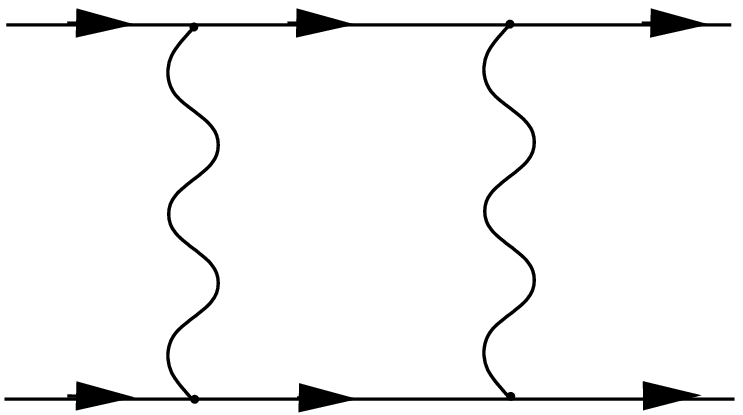}}
\put(0.9,10.9){$P,\sigma_1$} \put(6.7,10.9){$P^\prime,\sigma_1$}
  \put(3.7,11.1){$P_1$}

\put(0.9,7.9){$-P,\sigma_2$}
 \put(3.7,7.7){$-P_1$}
\put(6.7,7.9){$-P^\prime,\sigma_2$}

\put(5,6.6){\large b)}
\put(1.4,1.2){\includegraphics[width=6\unitlength]{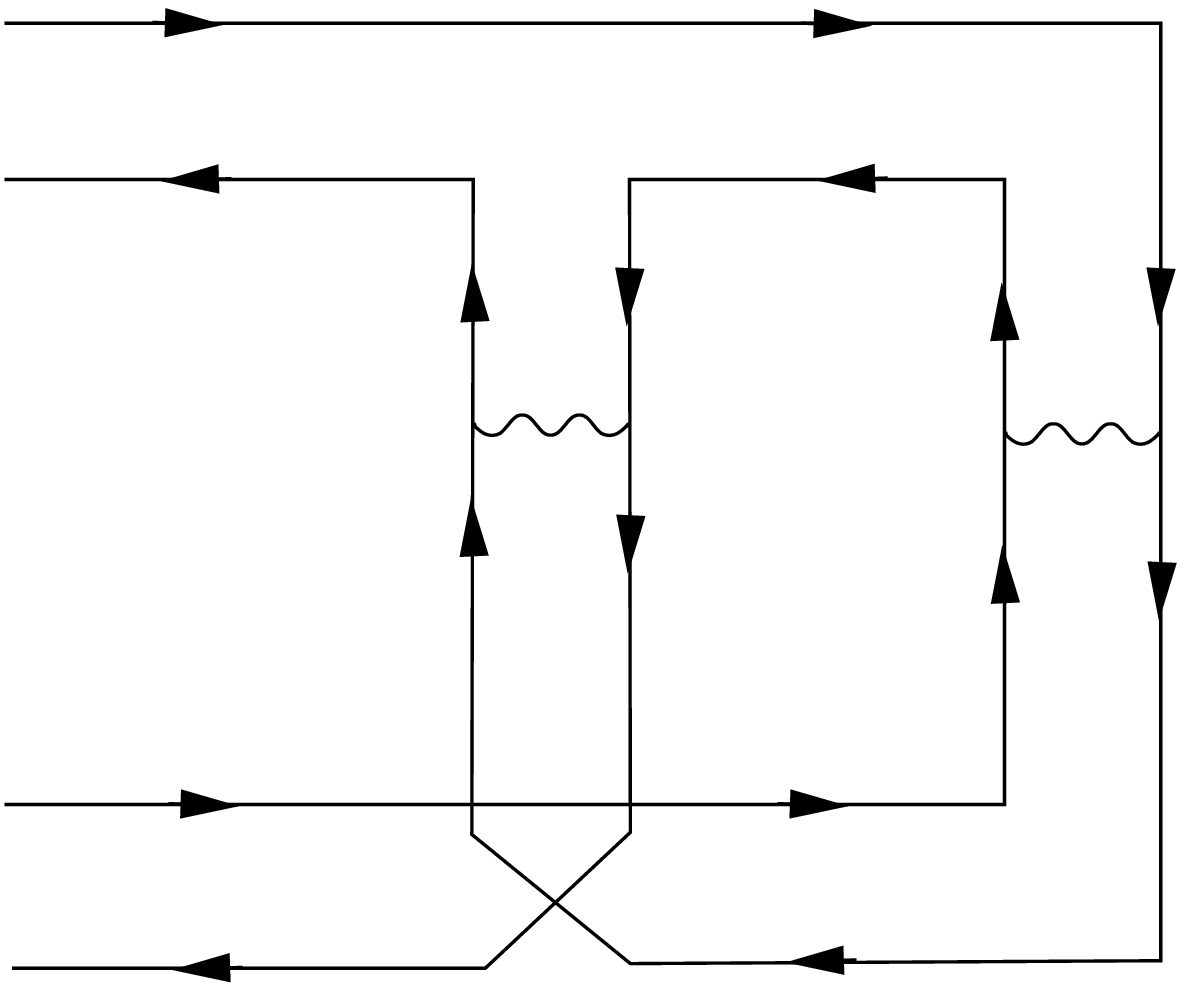}}

\put(0.5,4.9){$P+K,\sigma_1$}
\put(0.5,6.0){$P,\sigma_1$}

\put(4.9,4.7){$P+K_1$}
\put(4.9,0.7){$-P-K_1$}

\put(0.2,0.9){$-P-K,\sigma_2$}
\put(0.2,2.0){$-P,\sigma_2$}

\end{picture}

\caption{a) Lowest logarithmic
 contribution due to the
  Cooper channel, $\sigma_1\neq\sigma_2$.
b) The same contribution obtained as  an
  interaction of the spin modes generated by vertices
  Fig.~\ref{fig1}b, and $k \ll p_F$.
This diagram coincides with the renormalization
of the quadratic part of the spin wave Lagrangian, see
Fig.~\ref{fig6}b) re-written in terms of electronic lines. }

\label{fig2}
\end{figure}

\begin{figure}[t]
\setlength{\unitlength}{2.3em}
\begin{picture}(10,12)

\put(5,12){\large a)}
\put(1.6,8.7){\includegraphics[width=5.5\unitlength]{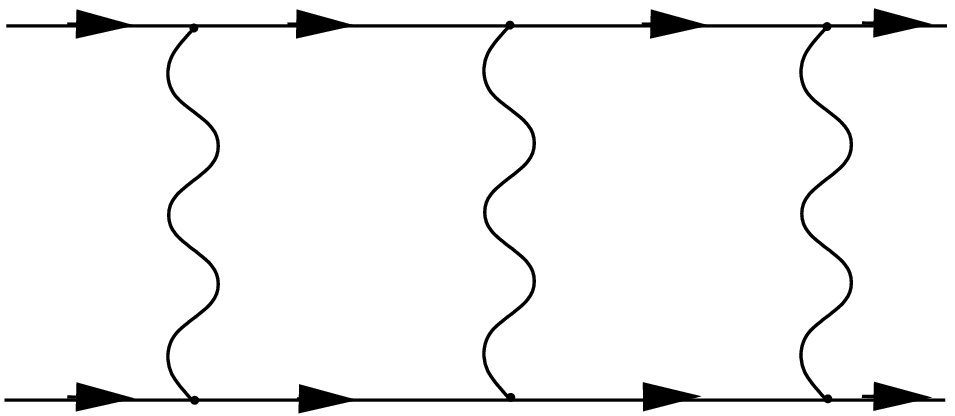}}

\put(0.4,11.2){$P,\sigma_1$} \put(3,11.2){$P_1$}
\put(5.5,11.2){$P_2$} \put(6.9,11.2){$P^\prime,\sigma_1$}

\put(0.4,8.2){$-P,\sigma_2$} \put(3,8.2){$-P_1$}
\put(5.5,8.2){$-P_2$} \put(6.9,8.2){$-P^\prime,\sigma_2$}

\put(5,7.0){\large b)}
\put(1.4,1.2){\includegraphics[width=7.8\unitlength]{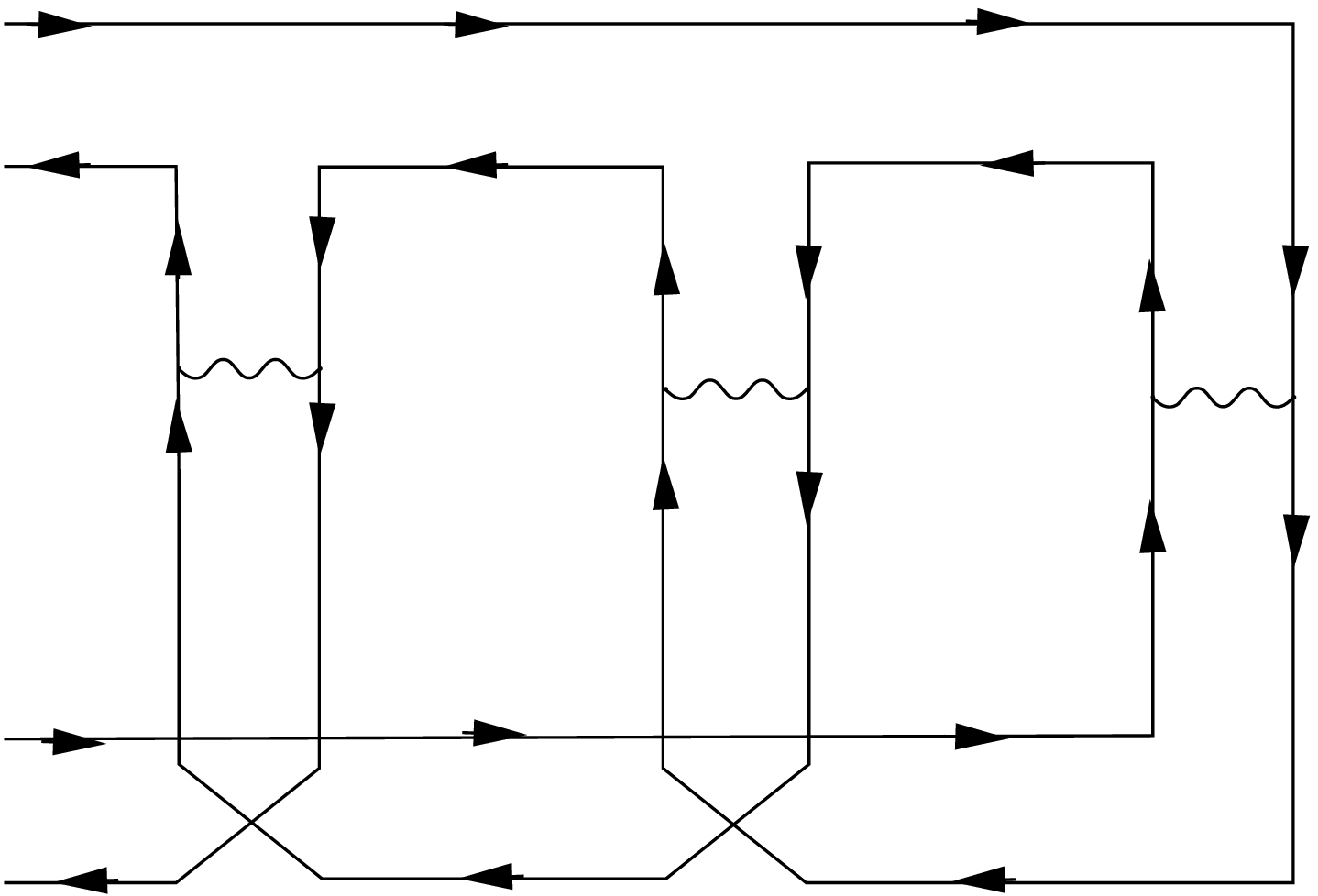}}

\put(0.5,5.){$P+K,\sigma_2$}
\put(0.5,6.4){$P,\sigma_1$}

\put(6.6,5.1){$P+K_1$}  \put(3.7,5.1){$P+K_2$}
\put(6.6,0.7){$-P-K_1$} \put(3.8,0.7){$-P-K_2$}

\put(0.2,0.7){$-P-K,\sigma_1$}
\put(0.2,2.){$-P,\sigma_2$}

\end{picture}

\caption{a) Second order logarithmic contribution
due to the
  Cooper channel, $\sigma_1\neq\sigma_2$.
b) The same contribution obtained as  an
  interaction of the spin modes generated by vertices
  Fig.~\ref{fig1}b, and $k \ll p_F$.
This diagram coincides with the renormalization
of the quadratic and quartic parts of the spin wave Lagrangian, see
Figs.~\ref{fig6} b), ~\ref{fig4}a) re-written in terms of electronic lines. }

\label{fig20}
\end{figure}

At first glance, we might miss important contributions because the Cooper
channel generates logarithms in any dimension\cite{agd}
and it looks as if we neglected
them. However, this is not so, although the reason for
necessity to neglect the third vertex $V_{3}$, Fig.~\ref{fig1}c is rather non-trivial.

As we will see, the most interesting contributions to interaction
vertices and, finally to the
thermodynamic quantities originate from the scattering on
angle either $0$ or $\pi$ and not from an integral over the entire Fermi surface,
so only such scattering amplitudes will be important.

Therefore, we must investigate the effect of the interactions in
the Cooper channel on these particular amplitude, see
Figs.~\ref{fig2}a, \ref{fig20}a. Direct comparison of those
contributions with the diagram generated by vertex $V_1$,  see
Figs.~\ref{fig2}b, \ref{fig20}b, shows that the perturbation
theory in $V_1$ generate in particular all the terms \footnote{In
fact, those terms will be properly accounted for in the
renormalization group treatment of Secs.~\ref{sec5} and
\ref{sec6}. } the Cooper channel is responsible for. (These
analogies in the two lowest orders of the logarithmic expansion
are easily followed to all the higher orders). The only difference
is the region of the integration over the intermediate momenta. In
the diagrams of Figs.~\ref{fig2}a, \ref{fig20}a, integrals over
the intermediate momenta $\p_1,\p_2$ span all over the Fermi
surface. On the other hand, (due to the condition $|\k|\simeq
1/r_0 \ll p_F$), the change in the momenta in   Figs.~\ref{fig2}b,
\ref{fig20}b is very close to the backscattering. To be concrete,
Fig.~\ref{fig20}b) describes the region of momenta
\[
|\pi-\widehat{\p\p_1}|,\ |\pi-\widehat{\p_2\p_1}|,\ |\pi-\widehat{\p_2\p^\prime}|
\leq \theta_*
\]
in Fig.~\ref{fig20}a), where the maximal angle appearing in our
theory is $\theta_*\simeq \left(p_Fr_0\right)^{-1}$.

Taking into account the third vertex $V_{3}$ in such region of the
momentum space would mean double counting the contribution of this
region that is most important for our consideration \footnote{This
double counting would not be important if all other regions of the
Fermi surface contributed as well. In this case an error due to
the double counting would be of order $\left(  r_{0}p_{F}\right)
^{1-d}\ll  1$. In dirty samples, the scattering on impurities
leads to such an isotropization and therefore one should include
in the effective Lagrangian the third term with the vertex
$V_{3}$. The possibility to neglect it is specific for the clean
systems considered here. The same approximation has been used in
Ref.~\onlinecite{chubukov1} for computation of diagrams of the
conventional perturbation theory.}. On the other hand, the region
with the scattering angle exceeding $\theta_*$
 does not appear very interesting. One could
integrate over this region from the beginning and this would simply
renormalize the coupling constant $V_{t}$. As we work in the limit of a
weak coupling, this renormalization cannot lead to any non-trivial effects and
we assume in the subsequent discussion that it has already been
performed.
We will return to the discussion of this renormalization in Sec.~\ref{sec7c0}.

One more uncertainty in the channel separation appears
when all four momenta in ${\mathcal S}_{int}$ of \req{a4}
are close to each other. In this limit the separation
into slowly varying pairs in \req{a7} is ambiguous.
We resolve this unecertainty by attributing this region of
 the momenta to the vertex
$V_1$. As a consequence, the vertex $V_2$ of \req{a9} should vanish for the
momenta $\p_1$ and $\p_2$, such that $|\p_1-\p_2| \simeq k \ll p_F $.
However, the vertex $V_2$ enters only the singlet channel, see \req{a13}. As we will
see, in integrals for physical quantities containing the singlet channel, the
main contribution comes from integration over the momenta  $p_1$ and $p_2$ well
separated from each other on the Fermi surface. The contribution from small
$|p_1-p_2|$ is not singular and is small.
So, for  dimensionalities $d>1$, we do not need taking special care about this
region when performing calculations for the singlet channel.
This argument cannot be used in $d=1$ but, in this case, the region of the momenta
all close to each other does not give logarithms and the contribution coming
from this region can be neglected anyway.

Finally, we notice that singling out the small and large angle
scattering amplitudes performed here is very similar to the one
carried out for one dimensional systems\cite{dl}. In $1d$ the
vertices $V_{1}$ and $V_{3}$ are indistinguishable at all.


\subsection{Hubbard-Stratonovich transformation}

Having reduced the interaction $\mathcal{L}_{int}$, Eq.
(\ref{a4}), to the interaction $\mathcal{\tilde{S}}_{int}$, Eq.
(\ref{a12}), we can decouple the quartic term by integration over
an additional field $\phi_{\sigma,\sigma^{\prime}}
\left(x;\n\right)$ {\em slowly varying in space}. We will still
use the notation \rref{a1ab}, and the unit vector $\n$ labels the
direction of the momentum on the Fermi surface. We introduce a
matrix operator $\hat{\Phi}$ by defining its action on any
function $\eta_\sigma\left(x\right)$ as
\begin{equation}
\begin{split}
\left[\hat{\Phi} \eta\right]_\sigma \left(x\right)
&=\sum_{\sigma^\prime}\int \frac{dx_{1}d^{d}\mathbf{p}}{\left(  2\pi\right)  ^{d}}
e^{i\p\left(\r-\r_{1}\right)  }
\\
&\times
\left\{
\phi_{\sigma
,\sigma^{\prime}}\left(\frac{\r_{1}+\r}{2},\tau;\,\frac{\p}{|\p|}\right)
\eta_{\sigma^\prime}\left(  \r_{1},\tau\right)\right\}. \label{a14a}%
\end{split}
\end{equation}

Then we represent the $2\times 2$ matrix in the spin space
$\hat{\phi}\left(x;\n\right)$ in the form
\begin{equation}
\hat\phi\left(  x,\n \right)  =i\varphi\left(x,\n\right)\openone_{\sigma}
+\bs\cdot  \mathbf{h}\left(  x,\n \right),  \label{a14b}%
\end{equation}
where $\openone_{\sigma}$ is the $2\times 2$ unit matrix in the
spin space, $\varphi\left(x,\n\right)  $ is a {\em real} function,
$\mathbf{h}\left(  x,\n\right)  $ is a three dimensional {\em
real} vector function ${\mathbf h}= \left(h_x,h_y,h_z\right)$, and
$\bs$ is the vector of the Pauli matrices, as it was defined after
\req{a10}. The auxiliary fields are defined as bosonic:
\begin{equation}
\hat{\phi}\left({\r},\tau;\, \n\right)  =
\hat{\phi}
\left({\r},\tau+\frac{1}{T};\, \n\right).
  \label{a21}%
\end{equation}

Then, the
partition function $Z$, Eq.~(\ref{a1a}), can be re-written as
a functional integral over the smooth fields:
\begin{equation}
\begin{split}
Z&=
\frac{1}{Z_{st}}
\int Z\left\{\hat{\phi}\right\}  W_s\left\{\phi\right\}
W_t\left\{{\mathbf h}\right\}  D\varphi
  D{\mathbf h}
\\
Z_{st}&=\int
 W_s\left\{\phi\right\}
W_t\left\{{\mathbf h}\right\}
 D\varphi
  D{\mathbf h}  .\label{a15}%
\end{split}
\end{equation}
Here, the functional $Z\left\{\hat{\phi}\right\}$ is the partition
function of the non-interacting fermions subjected to the
smooth field $\hat{\phi}$:
\begin{equation}
Z
\left\{\hat{\phi}\right\}
=\int\exp\left(  - \int dx \mathcal{L}_{eff}\left\{\hat{\phi}\right\}
\right)  D\chi D\chi^{\ast}. \label{a23}%
\end{equation}
The Lagrangian density $\mathcal{L}_{eff}$
 for a given configuration of the fields $\varphi$, ${\mathbf h}$ reads
\begin{equation}
\mathcal{L}_{eff} \left\{\hat{\phi}\right\} =\mathcal{L}_{0}
+\sum_{\sigma}\chi_{\sigma}^{\ast}\left(  x\right)
\left[\hat{\Phi}\chi\right]_{\sigma}(x), \label{a16}%
\end{equation}
where the Largangian density for the fermions ${\mathcal L}_0$ is
defined in \req{a3}.

The weights for the bosonic fields $W_{st}$ are of the form
\begin{subequations}
\label{a1819}
\begin{equation}
W_{s}  =\exp\left\{  -\frac{1}{2}\int
\varphi
\left(x,\n\right)
\left[ \hat{\mathbf V}_{s}^{-1}\varphi\right]\left(  x,\n\right)
dx d\n \right\}
 , \label{a18}%
\end{equation}%
\begin{equation}
W_{t}  =
\exp\left\{  -\frac{1}{2}\!\sum_{i=x,y,z}\int
h_i
\left(x,\n\right)
\left[ \hat{\mathbf V}_{t}^{-1}h_i\right]\!\left(  x,n\right)dx d\n
\right\}.
\label{a19}%
\end{equation}
Here we use the notation \rref{a1ab} and the convention
\rref{a11a} for the integration over the direction of the momentum
$\n$.

The operators $\hat{\mathbf V}_{s,t}$ in Eqs. (\ref{a18}, \ref{a19}) are defined
by its action on any function $a\left(x,\n\right)$
\end{subequations}
\be
\left[\hat{\mathbf V}_{\substack{\\s\\t}}\,a\right]\left(x,\n \right)
=\pm\int d\r_1
\bar{f}\left(  \mathbf{r}-\mathbf{r}_{1}\right)
\left[\hat{V}_{\substack{\\s\\t}}\, a\right]\left(\r_1,\tau;\n \right)
, \label{a20}%
\ee where $\hat{V}_{s,t}$ are defined in \req{hatVst}. Different
signs for the singlet ($+$) and triplet ($-$) channels correspond
to the fact that the operator $\hat{V}_s$ is the positive definite
and $\hat{V}_t$ is the negative definite, see \req{a13}. The
choice of the weights \rref{a1819} together with the fact that the
fields $\varphi,\ {\mathbf h}$ must be real is guided by the
requirement that the functional integrals to be absolutely
convergent. The function $\bar {f}\left( \mathbf{r}\right)$ is the
Fourier transform of the function $f\left(  \mathbf{k}\right)$
defined in Eq.~(\ref{a7c}):
\begin{equation}
\bar{f}\left(  \mathbf{r}\right)  =r_{0}^{-d}\bar{f}_{0}\left(  r/r_{0}%
\right)  \label{a20a}%
\end{equation}
and $\bar{f}_{0}(r)$ is the Fourier transform of $f_{0}(k)$. The function $\bar
{f}\left(  \mathbf{r}\right)  $ tends to a constant $r_{0}^{-d}\bar{f}%
_{0}\left(  0\right)  $ in the limit $\left\vert \mathbf{r}\right\vert
/r_{0}\rightarrow0$ and vanishes in the limit $\left\vert \mathbf{r}%
\right\vert \rightarrow\infty$. The role of this function is to
regularize the theory at small distances leaving all interesting
long distance physics intact.

Notice that \reqs{a1819}--\rref{a20} are local in time and, therefore,
the factors $Z_{s,t}$ of \req{a15} are not relevant for the
determining the properties of the system. Those factors will be
usually suppressed in the subsequent formulas.

Thus, we have decoupled the interaction term
$\tilde{\mathcal{S}}_{int},$ Eq. (\ref{a12}), in the action
$\mathcal{S}$, Eq. (\ref{a2}), with the Hubbard- Stratonovich
transformation, Eqs. (\ref{a15})--\rref{a19}. As the field
$\hat{\phi}\left(  x,\n\right)$ varies slowly in space, we can
apply quasiclassical description for the electron Green functions
in such fields.

\subsection{Quasiclassical Green functions}

Let us introduce Green functions $G_{\sigma,\sigma^{\prime}}\left(
x,x^{\prime}|\left\{\hat{\phi}\right\}\right) $
corresponding to the Lagrangian density $\mathcal{L}%
_{eff}\left\{  \phi\right\},$ Eq. (\ref{a16}), as
\begin{align}
&  G_{\sigma,\sigma^{\prime}}\left(
  x,x^{\prime}|\left\{\hat{\phi}\right\}\right)
= Z^{-1}
\left\{  \hat\phi\right\} \label{a22}\\
&  \times\int\chi_{\sigma}\left(  x\right)  \chi_{\sigma^{\prime}}^{\ast
}\left(  x^{\prime}\right)
\exp\left(  -\int dx\mathcal{L}_{eff}\left[  \Phi\right]
\right)  D\chi D\chi^{\ast}.\nonumber
\end{align}
In what follows we will suppress the argument
$|\left\{\hat{\phi}\right\}$ whenever its presence is
self-evident.

The Green function $\hat{G}\left(
  x,x^{\prime}|\left\{\hat{\phi}\right\}\right)$,
is the matrix $2\times 2$ in the free space, is the functional of
real fields $\varphi,{\mathbf h}$,  and it satisfies the equations
\begin{subequations}
\label{a2425}
\bea
\left(-\frac{\partial}{\partial\tau}-\hat{H}_{0\mathbf{r}}
\right)\hat{G}
+\hat{\Phi} \hat{G}
  &=&\delta\left(
x-x^{\prime}\right)\openone_{\sigma},  \label{a24}%
\\
\left(\frac{\partial}{\partial\tau^\prime}-\hat{H}_{0\mathbf{r^\prime}}
\right)\hat{G}
+\hat{G}\hat{\Phi}
  &=&\delta\left(
x-x^{\prime}\right)\openone_{\sigma},
 \label{a25}%
\eea
where $\hat{H}_{0\mathbf{r}}$, Eq.~(\ref{a3a}), acts on $\mathbf{r}$.
The action of the operator $\Phi$ on the Green function is
determined by \req{a14a} as
\be
\begin{split}
&\left[\hat{\Phi} \hat{G}\right]_{\sigma\sigma^\prime}(x,x^\prime)
=\sum_{\sigma^{\prime\prime}}\int \frac{dx_{1}d^{d}\mathbf{p}}{\left(  2\pi\right)  ^{d}}
e^{i\p\left(\r-\r_{1}\right)  }
\\
&
\quad\times
\left\{
\phi_{\sigma
,\sigma^{\prime\prime}}\left(\frac{\r_{1}+\r}{2},\tau;\,\frac{\p}{|\p|}\right)
{G}_{\sigma^{\prime\prime}\sigma^\prime}(\r_1,\tau; x^\prime)
\right\};
\\
&\left[\hat{G}\hat{\Phi}\right]_{\sigma\sigma^\prime}(x,x^\prime)
=\sum_{\sigma^{\prime\prime}}\int \frac{dx_{1}d^{d}\mathbf{p}}{\left(  2\pi\right)  ^{d}}
e^{i\p\left(\r_1-\r^\prime\right)  }
\\
&
\quad\times
\left\{
{G}_{\sigma\sigma^{\prime\prime}}(x; \r_1,\tau^\prime)
\phi_{\sigma^{\prime\prime}
,\sigma^\prime}\left(\frac{\r_{1}+\r^\prime}{2},\tau^\prime;\,\frac{\p}{|\p|}\right)
\right\}.
\end{split}
\label{a250}
\ee
\end{subequations}

 The derivation of the equations for the quasiclassical Green
functions can be carried out in the same way as in
Ref.~\onlinecite{lo}.
Subtracting
Eq.~(\ref{a24}) from Eq.~(\ref{a25}) and making a Wigner
transformation
\be
{G}\left(x; x^\prime
\right) =\int \frac{d^d\p}{(2\pi)^d}e^{i\p\cdot\r}
{G}\left(\tau,\tau^\prime;\frac{\r+\r^\prime}{2},\p\right)
\label{WT}
\ee
 of the result, we obtain
\begin{align}
&0=  \left(  \frac{\partial}{\partial\tau}+\frac{\partial}{\partial\tau^{\prime
}}-\frac{i\mathbf{p\nabla}_{\mathbf{r}}}{m}\right)
\hat{G}\left( \tau,\tau^{\prime};\r,\p\right) \label{a26}\\
&  +\left[
\hat{G}\left( \tau,\tau^{\prime};\r,\p\right)
\hat{\phi}\left(\r,\tau^{\prime};\n\right)
-\hat{\phi}\left(\r,\tau;\n\right)
\hat{G}\left( \tau,\tau^{\prime};\r,\p\right)
 \right].\nonumber
\end{align}

Equation (\ref{a26}) is justified provided the dependence of the
field $\hat{\phi}\left(x,\n\right)$ and, hence, of
$\hat{G}\left(\tau,\tau^{\prime};r;\p\right) $ on the coordinate
$\mathbf{r}$  is slow on the scale of the order of Fermi
wavelength. This is guarded by the cutoff scale $r_{0}$ in
Eq.~(\ref{a7c}). In principle, one could derive Eq.~(\ref{a26})
more accurately, which would produce additional terms containing
phase space derivatives of the functions $\phi$ and $\hat{G} $ in
the second line. However, the additional derivatives would
suppress the infrared singularities we are interested in, and that
is why we neglected them. At the same time, no higher derivatives
arise in the first bracket in Eq. (\ref{a26}) and this term is
exact for the quadratic spectrum of the fermions, \req{a3a}.

The next step is to reduce the Green function $G(\p,\r)$ to a
function involving the degrees of freedom describing the motion of
the system along the Fermi surface. To accomplish this task we
linearize the spectrum by putting \be \frac{\p}{m} \approx v_F\n
\label{a260} \ee in \req{a26}, where $\n$ is the unit vector.
 The justification of such approximation is that the
fluctuating fields can mix the electron states only in the
vicinity of the Fermi surface whereas the states deep in the Fermi sea
remain intact. After approximation \rref{a260} the operators
in \req{a26} do not depend on the variable
\begin{equation}
\xi=\mathbf{p}^{2}/2m-\varepsilon_{F}, \label{a26a}%
\end{equation}
that describes the evolution perpendicular to the Fermi level and
the latter can be integrated over. Then, one
obtains\cite{lo,eilenberger}
\begin{align}
&0=  \left(  \frac{\partial}{\partial\tau}+\frac{\partial}{\partial\tau^{\prime
}}-iv_F\n\cdot\mathbf{p\nabla}_{\mathbf{r}}\right)
\hat{g}\left( \tau,\tau^{\prime};\r,\n\right) \label{a28}\\
&  +\left[
\hat{g}\left( \tau,\tau^{\prime};\r,\n\right)
\hat{\phi}\left(\r,\tau^{\prime};\n\right)
-\hat{\phi}\left(\r,\tau;\n\right)
\hat{g}\left( \tau,\tau^{\prime};\r,\n\right)
 \right],\nonumber
\end{align}
where
\begin{equation}
\begin{split}
&{\hat g}\left(\tau,\tau^{\prime};\r,\n\Big|\left\{\hat{\phi}\right\}\right)
\\
&\quad=\frac{i}{\pi}\int_{-\infty}^{\infty}
{\hat G}\left[\tau,\tau^{\prime};\r;\left(p_F+\frac{\xi}{v_F}\right)
\n\Big|\left\{\hat{\phi}\right\}\right]
 d\xi. \label{a27}%
\end{split}
\end{equation}

The function $\hat{g}$ must obey the antiperiodicity conditions
\begin{align}
\hat{g}\left( \tau,\tau^{\prime};\r,\n\right)   &
=-\hat{g}\left( \tau+1/T,\tau^{\prime};\r,\n\right)
\nonumber\\
&  =-\hat{g}\left( \tau,\tau^{\prime}+1/T;\r,\n\right)
\label{a28a}%
\end{align}
that follow from Eqs.~(\ref{a5}), \rref{a22}, \rref{WT}, and
\rref{a27}.
Clearly, this condition is consistent with
Eq.~(\ref{a28}) for the periodic fluctuating fields $\hat{\phi}$, see Eq.~(\ref{a21}).

Equation~(\ref{a28}) is linear and therefore is not sufficient to
find $\hat{g}\left( \tau,\tau^{\prime};\r,\n\right)$
unambiguously. In order to define the problem completely, one has
to complement Eq.~\rref{a28a} with a certain constraint. To derive
this constraint, we introduce a new function \be
\begin{split}
&\hat{B}\left( \tau,\tau^{\prime};\r,\n \Big|\left\{\hat{\phi}\right\}\right)
\\
&=
\int_0^{1/T} d\tau^{\prime\prime}
\hat{g}\left( \tau,\tau^{\prime\prime};\r,\n \Big|\left\{\hat{\phi}\right\}\right)
\hat{g}\left( \tau^{\prime\prime},\tau^{\prime};
\r,\n\Big|\left\{\hat{\phi}\right\} \right).
\end{split}
\label{a280}
\ee

Using the definition \rref{a280}, we obtain from \req{a28}
\begin{align}
&0=  \left(  \frac{\partial}{\partial\tau}+\frac{\partial}{\partial\tau^{\prime
}}-iv_F\n\cdot\mathbf{p\nabla}_{\mathbf{r}}\right)
\hat{B}\left( \tau,\tau^{\prime};\r,\n\right) \label{a281}\\
&  +\left[
\hat{B}\left( \tau,\tau^{\prime};\r,\n\right)
\hat{\phi}\left(\r,\tau^{\prime};\n\right)
-\hat{\phi}\left(\r,\tau;\n\right)
\hat{B}\left( \tau,\tau^{\prime};\r,\n\right)
 \right].\nonumber
\end{align}

In the absence of the fluctuating field, $\hat{\phi}=0$, the Green
function can be easily found from \reqs{a24}, \rref{WT} and \rref{a27}
\be
\begin{split}
&\hat{g}\left( \tau,\tau^{\prime};\r,\n
  \Big|0\right)=\frac{i\openone_{\sigma}}{\pi}T\sum_{\epsilon_n}
e^{i\epsilon_n(\tau^\prime-\tau)}
\int \frac{d\xi}{i\epsilon_n-\xi}
 \\
&\quad =-i\openone_{\sigma}\operatorname{Re}\left[
\frac{T}
{\sin\pi T\left(  \tau-\tau^{\prime}+i 0\right)}  \right].
\end{split}
\label{a282}
\ee
Substituting \req{a282} into definition \req{a280} and performing
the integration, we find
\be
\hat{B}\left( \tau,\tau^{\prime};\r,\n
  \Big|0\right)=\openone_{\sigma}
\delta\left(\tau-\tau^\prime\right).
\label{a310}
\ee
Substitution of \req{a310} into \req{a281} shows that \req{a31}
remains valid for the arbitrary field $\phi$,
\be
\hat{B}\left( \tau,\tau^{\prime};\r,\n
  \Big|\left\{\hat{\phi}\right\}\right)=\openone_{\sigma}
\delta\left(\tau-\tau^\prime\right).
\label{a31}
\ee
i.e. no perturbation by the Hubbard-Stratonovich fields can violate
the condition \rref{a310}.

Equations~(\ref{a280}) and (\ref{a31}) complement Eq.~(\ref{a28})
and these equations are sufficient to find the function $\hat{g}$.
Eq. (\ref{a28}) is much simpler than the original Schr\"odinger
equation \rref{a24} as it operates with the smooth quantities and
involves only the first derivatives. The further program is to
solve \reqs{a28}, (\ref{a280}) and (\ref{a31}) for arbitrary
configurations of the fields $\phi$.
 After that, in order to calculate physical
quantities, one should perform a proper averaging over fields
$\phi$ with the weights defined in \reqs{a1819}. All this is still
not a simple task and in the next Sections we will express the
solution of these equations in terms of a functional integral over
supervectors, in order to obtain the {\em local} theory in terms
of only the bosonic variables describing the collective
excitations.

\section{Charge and spin collective variables. Partition function}
\label{sec3}
\subsection{Further simplification of the quasiclassical equations}

Solutions of Eqs.~(\ref{a28}), \rref{a280}, and \rref{a31}
describe collective excitations and
our task is to find them at least symbolically in order to facilitate
calculation of the partition function $Z\left\{\phi\right\} $,
see \req{a23} and the averaging over the auxiliary field
$\phi$.

 At first glance, we could simply follow
the scheme developed in Ref. \onlinecite{ek} writing the solution
of these equations in terms of a functional integral over
constrained supermatrices. However, in the present situation this
scheme is not convenient due to the dependence the
Hubbard-Stratonovich field $\phi$ on $\tau$. 

Instead, we look for the solution of Eqs.~(\ref{a28}),~\rref{a31},
and \rref{a280}
 in a form
\begin{equation}
\begin{split}
&\hat{g}\left( \tau,\tau^{\prime};\r,\n \Big|\left\{\hat{\phi}\right\}\right)
\\
&
=
\hat{\mathcal T}\left(\tau;\r,\n \Big|\left\{\hat{\phi}\right\}\right)
\hat{g}\left( \tau,\tau^{\prime}\Big|0\right)
\hat{\mathcal T}^{-1}\left(\tau^\prime;\r,\n \Big|\left\{\hat{\phi}\right\}\right)
\label{b3}%
\end{split}
\end{equation}
where the Green function for the free electrons, $\hat{g}\left(
\tau,\tau^{\prime}\Big|0\right)$ is defined in \req{a282}. The
$2\times 2$ matrix in the spin space, $\hat{\mathcal T}$ satisfies
the condition
\begin{equation}
\hat{\mathcal T}\left(\tau;\r,\n \Big|\left\{\hat{\phi}\right\}\right)
=
\hat{\mathcal T}\left(\tau+\frac{1}{T};\r,\n \Big|\left\{\hat{\phi}\right\}\right)
 \label{b4}%
\end{equation}
so that the antiperiodicity of the Green function \rref{a28a} is
preserved. In the remainder of this subsection, we will suppress
the argument $\Big|\left\{\hat{\phi}\right\}$ whenever it is
self-evident.

The representation of the Green function in the form of
Eq.~(\ref{b3}) is nothing but the matrix form of the eikonal
approximation, which can also be viewed as a generalization of the
Schwinger Ansatz \cite{schwinger}. It easy to check that the Green
function, Eq. (\ref{b3}) is consistent with Eqs.~(\ref{a31}), and
\rref{a280}, and what remains to be done is to find the proper
matrix $\hat{\mathcal T}$, such that Eq. (\ref{a28}) is satisfied.

Substituting Eq. (\ref{b3}) into Eq.~(\ref{a28}) we obtain%
\begin{equation}
\hat{g}\left( \tau,\tau^{\prime}\Big|0\right)
 \left[  \hat{K}\left(\tau,\r,\n\right)-
\hat{K}\left(\tau^{\prime},\r,\n\right)  \right]  =0,
\label{b4a}%
\end{equation}
where the $2\times 2$ matrix in the spin space $\hat{K}$ is given
by
\begin{align}
& \hat{K}\left( x,\n\right)  =
\hat{\mathcal T}^{-1}\left(x,\n\right)
\left(
\partial_\tau  -iv_{F}\mathbf{n\nabla}_{\mathbf{r}}\right)\hat{\mathcal T}\left(x,\n\right)
\label{b4c}\\
&  - \hat{\mathcal T}^{-1}\left(x,\n \right)
\hat{\phi}\left(x,\n \right) \hat{\mathcal T}\left(x,\n\right),
\nonumber
\end{align}
and we use the short-hand notation \rref{a1ab}.

\begin{subequations}
Equation (\ref{b4a}) must be fulfilled for any $\tau$ and $\tau^{\prime}$. This is
possible only for $\partial_\tau \hat{K}(x,\n)=0$. Using \req{b4c}, we obtain
\begin{align}
&  \left(  -{\partial}_{\tau}+iv_{F}\mathbf{n\nabla}_{\mathbf{r}%
}\right) \hat{\mathcal T}\left(x,\n \right)
 \label{b5}\\
&  = \hat{\mathcal T}\left(x,\n \right)
\hat{A}\left(  \mathbf{r},\n\right)
- \hat{\phi}(x,\n)
\hat{\mathcal T}\left(x,\n\right)\nonumber
\end{align}
where $\hat{A}\left(  \mathbf{r},\n\right)  $ is an arbitrary time
independent matrix.

We can transform Eq.~(\ref{b5}) to a more convenient form writing the
corresponding equation for $\hat{\mathcal T}^{-1}\left(  x,\n \right)  $
\begin{align}
&  \left( \partial_{\tau}-iv_{F}\mathbf{n\nabla}_{\mathbf{r}%
}\right) \hat{\mathcal T}^{-1}\left(x,\n \right)  \label{b6}\\
&  = \hat{A}\left(  \mathbf{r},\n\right) \hat{\mathcal T}^{-1}\left(x,\n \right)
- \hat{\mathcal T}^-1\left(x,\n\right)\hat{\phi}(x,\n)
 \nonumber
\end{align}
\end{subequations}

We differentiate Eq.~(\ref{b5}) with respect to $\tau$ and
post-multiply it by $\hat{\mathcal T}^{-1}\left(  x,\n \right)$.
Then we pre-multiply Eq.~(\ref{b6}) by
$\partial_\tau \hat{\mathcal T}\left(  x,\n \right)$
and subtract thus obtained equations from each other. As the result,
we find
\begin{align}
&  \left(  -{\partial}_{\tau}+iv_{F}\mathbf{n\nabla}_{\mathbf{r}%
}\right) \hat{M}\left(  x,\n\right) \label{b7}\\
&  +\left[  \hat{\phi}\left(  x,\n\right),\ M_{\mathbf{n}}\left(
x\right)  \right]  =-
\partial_\tau\hat{\phi}(x,\n)
\nonumber
\end{align}
where
\begin{equation}
\hat{M}\left(  x,\n \right)  =\frac{\partial {\hat{\mathcal T}}\left(
x,\n \right)  }{\partial\tau}{\hat {\mathcal T}}^{-1}\left(  x,\n\right)  \label{b8}%
\end{equation}
and the symbol $\left[\dots,\dots\right]  $ stands for the commutator.

Using the representation (\ref{a14b}) for the matrix $\hat{\phi}
\left(  x,\n\right)  $. we look for  the matrix $\hat{M}\left(
  x,\n\right)  $
in the form
\begin{equation}
\hat{M}\left(  x,\n\right)  =i\rho\left(  x,\n\right)\openone_{\sigma}
+{\mathbf S}\left(  x,\n\right) \cdot \bs \label{b9}%
\end{equation}
where $\rho\left(x,\n\right)  $ is a scalar real field and
$\mathbf{S}_{\mathbf{n}}\left(x,\n\right)  $ is a real three dimensional
vector field. As follows from \reqs{b9} and \rref{b4}, those fields
are periodic
\be
\begin{split}
\rho\left(\tau,\r,\n\right)=\rho\left(\tau+\frac{1}{T},\r,\n\right);
\\
\mathbf{S}\left(\tau,\r,\n\right)=
\mathbf{S}
\left(\tau+\frac{1}{T},\r,\n\right).
\end{split}
\label{b90}
\ee

Substituting \req{b9} into \req{b7}
we obtain two independent equations for $\rho(x,\n)$ and
$\mathbf{S}
\left(  x,\n\right)$:
\begin{subequations}
\begin{equation}
\left(  -\frac{\partial}{\partial\tau}+iv_{F}\mathbf{n\nabla}_{\mathbf{r}%
}\right)  \rho_{\mathbf{n}}\left(  x\right)
=-\frac{\partial\varphi\left(  x,\n\right)  }{\partial\tau} \label{b10}%
\end{equation}%
\begin{align}
&  \left(  -\frac{\partial}{\partial\tau}+iv_{F}\mathbf{n\nabla}_{\mathbf{R}%
}\right)  \mathbf{S}_{\mathbf{n}}\left(  x\right) \nonumber\\
&  +2i\left[  \mathbf{h}_{\mathbf{n}}\left(  x\right)  \mathbf{\times
S}_{\mathbf{n}}\left(  x\right)  \right]  =-\frac{\partial\mathbf{h}%
_{\mathbf{n}}\left(  x\right)  }{\partial\tau} \label{b11}%
\end{align}
It is easy to see that \reqs{b1011} are consistent with the
periodicity requirements \rref{b90} and \rref{a14b}.

Equations \rref{b1011} are the final quasiclassical equations that
will be used for further calculations. We emphasize that
\reqs{b1011} are obtained from Eqs. (\ref{a28}). \rref{a280}.
\rref{a31} without making any further approximation. The field
$\rho\left(  x,\n\right)  $ corresponds to the density fluctuation
in the phase space, whereas the field $\mathbf{S}\left(
  x,\n\right) $ describes the spin fluctuations.
\label{b1011}
\end{subequations}

Equations (\ref{b10}) and (\ref{b11}) determining these
fluctuations due to the Hubbard-Stratonovich fields are remarkably
different from each other. Equation (\ref{b10}) for the density is
rather simple, and can be solved immediately by the Fourier
transform. This is what one obtains using the high dimensional
bosonization of Refs. \onlinecite{haldane,houghton1,houghton,neto,
kopietz,kopietzb,khveshchenko,khveshchenko1,castellani,metzner}
from an eikonal equation. Of course, we could take into account
gradients of the field $\varphi\left(  x,\n\right) $ and this
would lead to additional terms in the L.H.S. of Eq.~(\ref{b10}).
However, this does not lead to new physical effects.

In contrast, Eq.~(\ref{b11}) is not readily solvable due to the
presence of $\mathbf{h}\left(  x,\n\right)  $ in the
left-hand-side (L.H.S) of this equation. Actually, the L.H.S of
Eq.~(\ref{b11}) is just the equation of motion of a classical
spin-density in the external magnetic field ${\mathbf h}$. We will
see that the presence of this form will result in  non-trivial
effects that will be considered later. To the best of our
knowledge, this difference between the charge and spin excitations
in $d>1$ has not been emphasized in literature.

\subsection{Partition function}
\label{sec3b}

Having found the semiclassical representation for the Green
functions, we are prepared to  express the partition function
$Z\left\{ \hat{\phi}\right\}  $ from  Eq. (\ref{a23}) in terms of
the collective variables $\rho\left(x,\n\right)$ and
$\mathbf{S}\left(  x,\n\right)  $. Integrating over $\chi$,
$\chi^{\ast}$ in Eq. (\ref{a23}) and using Eqs. (\ref{a16}),
(\ref{a3}) for the Lagrangian density
$\mathcal{L}_{eff}\left\{\hat{\phi}\right\}$, we write $Z\left\{
\hat{\phi}\right\} $ in the form%
\begin{equation}
\ln Z\left\{\hat{\phi}\right\}=Tr\int\ln\left(-\partial_\tau \openone_{\sigma}
-H_{0} \openone_{\sigma} +\hat{\Phi}\right),
\label{b12}%
\end{equation}
where the Hamiltonian $H_{0}$ is introduced in Eq. (\ref{a3a}),
operator $\hat{\Phi}$ is defined by \req{a14a}, and $Tr$ includes
the trace in the spin space as well as the integration over
$\r,\tau $.

Equation (\ref{b12}) can be rewritten using the standard trick of
integration over coupling the constant as \be
\begin{split}
\ln Z\left\{
\hat{\phi}\right\}&-\ln Z\left\{0\right\}
\\
&
=  \int_0^1 du \partial_u Tr\int\ln\left(-\partial_\tau \openone_{\sigma}
-H_{0} \openone_{\sigma} +u \hat{\Phi}\right)
\\
&=\int_0^1 du
\sum_\sigma\int dx \left[\hat{\Phi}\hat{G}\left\{
u \hat{\phi}\right\}\right]_{\sigma\sigma}(x,x),
\label{b13}
\end{split}
\ee where the Green function $\hat{\Phi}\hat{G}\left\{u
\hat{\phi}\right\}$ is obtained from that of \req{a24} by the
rescaling of the Hubbard-Stratonovich fields: $\hat{\Phi}\to
u\hat{\Phi}$, and the action of the operator $\hat{\Phi}$ is
defined by \req{a250}. The term $\ln Z\left\{0\right\}$ describes
the thermodynamics of the non-interacting fermions, and we will
suppress this term in all the subsequent formulae.

Using \reqs{a250} and \rref{WT}, we obtain from \req{b13}
\be
\begin{split}
&\ln Z\left\{\hat{\phi}\right\}
\\
&=\int_0^1 du
\int dx \int\frac{d^d\p}{(2\pi)^d}
 Tr_\sigma\hat{\phi}\left(x;\n\right)
\hat{G}\left(\tau,\tau; \r,\p|\left\{
u \hat{\phi}\right\}\right),
\label{b130}
\end{split}
\ee
where  $Tr_\sigma$ denotes the trace in the spin space,
and the short hand notation \rref{a1ab} is used.
We represent the integration over the momentum as
\[
\int\frac{d^d\p}{(2\pi)^d}\dots =\int\nu(\xi) d\xi\int d\n  \dots,
\]
where $\nu(\xi)$ is the density of states (DoS) per one spin orientation,
$\xi=0$ corresponds to the Fermi level
and we use the convention \rref{a11a} for the integration over the
direction over the momentum on Fermi surface, $\n$.
Neglecting the energy dependence of DoS, we obtain from \req{b130}
\be
\begin{split}
&\ln Z\left\{\hat{\phi}\right\}
\\
&=-i\pi\nu\int\limits_0^1 du
\int \!dx\! \int\! \!d\n
 Tr_\sigma\hat{\phi}\left(x;\n\right)
\hat{g}\left(\tau,\tau; \r,\n|\left\{
u \hat{\phi}\right\}\right),
\label{b131}
\end{split}
\ee
where $\nu\equiv \nu(\xi=0)$ is the DoS on the Fermi level
per one spin orientation,
Green function
$\hat{g}\left(\tau,\tau^\prime; \r,\p|\left\{u
    \hat{\phi}\right\}\right)$ satisfies the constraints
\rref{a280} and \rref{a31} and satisfies \req{a28} with the
rescaling $\hat{\phi} \to u\hat{\phi}$. According to \req{a282}
$\hat{g}$ is a singular function at coinciding time, so that the
equal time value should be understood as \be
\hat{g}(\tau,\tau)\equiv \frac{1}{2}\lim_{\delta \to 0}
\left[\hat{g}(\tau,\tau+\delta) +\hat{g}(\tau,\tau-\delta)
\right]. \label{b132} \ee

Next, we substitute Eq.~(\ref{b3}) into \req{b131}.
Using \req{a282} and the rule \rref{b132}, we find
\[
\begin{split}
&\hat{g}\left( \tau,\tau;\r,\n \Big|\left\{u\hat{\phi}\right\}\right)
\\
&
=\frac{i}{2}\lim_{\delta \to 0}
\frac{1}{\delta}
\Bigg[
\hat{\mathcal T}\left(\tau;\r,\n \Big|\left\{u\hat{\phi}\right\}\right)
\hat{\mathcal T}^{-1}\left(\tau+\delta;\r,\n \Big|\left\{u\hat{\phi}\right\}\right)
\\
& \quad
- \hat{\mathcal T}\left(\tau;\r,\n \Big|\left\{u\hat{\phi}\right\}\right)
\hat{\mathcal T}^{-1}\left(\tau-\delta;\r,\n \Big|\left\{u\hat{\phi}\right\}\right)
\Bigg]
\\
&= - i\frac{\partial\hat{\mathcal T}\left(\tau;\r,\n
  \Big|\left\{\hat{u\phi}\right\}\right)}{\partial \tau}
\hat{\mathcal T}^{-1}\left(\tau;\r,\n
  \Big|\left\{\hat{u\phi}\right\}\right).
\end{split}
\]
Using the definition, Eq. \rref{b8}, the representation \rref{b9},
and equations of motion \rref{b1011}, we obtain finally from
\req{b131}
\begin{subequations}
\begin{equation}
Z\left\{\hat{\phi}\right\}  =Z_{0}Z_{\rho}\left\{\varphi\right\}Z_s
\left\{{\mathbf h}\right\}.
\label{b17}%
\end{equation}
Here
\begin{equation}
Z_{\rho}=\exp\left[2\nu\int_{0}^{1}
du \int \rho\left(
x,\n;u\right)  \varphi\left(  x,\n\right)  dxd\mathbf{n}\right]  , \label{b18}%
\end{equation}%
\begin{equation}
Z_{s}=\exp\left[  -2\nu\int_{0}^{1}du\int
\mathbf{S}\left(
x,\n;u\right)  \mathbf{h}\left(  x,\n\right)  dxd\mathbf{n}\right].
\label{b19}%
\end{equation}
\label{b171819}
\end{subequations}
The functions $\rho\left(  x,\n; u\right)  $ and
$\mathbf{S}\left(  x,\n; u\right)  $ should be found from the equations
[cf. \reqs{b1011}]

\begin{equation}
\left(-\ptau + iv_F\n\nnabla_\r\right)\rho \left(  x,\n; u \right)  =-u
\ptau
\varphi
\left( x,\n\right),
\label{b20}
\end{equation}
\begin{equation}
\hat{L}_{u}\mathbf{S}\left(  x,\n; u \right)  =-u
\ptau \mathbf{h}\left(  x,\n\right). \label{b21}%
\end{equation}
In Eq.~(\ref{b21}), the operator $\hat{L}_{u}$ equals
\begin{equation}
\hat{L}_{u}=\left(-\ptau+
iv_F\n\nnabla_\r\right)\openone_s
 +2iu\hat{h}(x,\n) \label{b23}%
\end{equation}
where the matrix $\hat{h}$ has the following form
\begin{equation}
\hat{h}(x,\n)=\begin{pmatrix}
0 & -h_{z}(x,\n) & h_{y}(x,\n)\\
h_{z}(x,\n) & 0 & -h_{x}(x,\n)\\
-h_{y}(x,\n) & h_{x} (x,\n)& 0
\end{pmatrix}_s, \label{b24}%
\end{equation}
and $h_{x},h_{y},$ and $h_{z}$ are the components of the real vector $\mathbf{h}$
($\hat{h}\mathbf{a=}\left[  \mathbf{h\times a}\right]  $ for any vector
$\mathbf{a}$). We will call this space of three dimensional vectors
the ``spin space'' as the $2\times 2$ spin space for the original
electron will be no longer needed in further considerations.
The functions $\mathbf{S}(x,\n;u)$ and $\rho(x,\n;u)$
satisfy the periodicity condition \rref{b90}.

The operator $\hat{L}_{u}$, Eq.~(\ref{b23}), is antisymmetric%
\begin{equation}
\hat{L}_{u}^{T}=-\hat{L}_{u} \label{b24b}%
\end{equation}
where the transposition $"T"$ includes both the changing of the sign of the
derivatives and the transposition of the spin indices. However, this
operator is neither Hermitian nor anti-Hermitian. The importance of
this subtlety  will be underlined in the next section.

Thus, in order to calculate the partition  function,
Eq.~(\ref{a1a}), for the system of interacting fermions in the
quasiclassical approximation, one should solve Eqs.~(\ref{b20}),
(\ref{b21}) and substitute their solutions into Eqs.
(\ref{b17})--(\ref{b19}). Then, one should use Eq.~(\ref{a15}) and
average over the fields $\varphi$ and $\mathbf{h}$ with the weight
given by \req{a1819}.

Before proceeding, we notice that there is a well-known flaw in
the quasiclassical approximation \rref{b131} to the exact
\req{b13} (this flaw is usually referred to as an ultraviolet
anomaly). In \req{b13}, the two times in the Green function are
put equal to each other {\em before} the integration over the
momentum is performed, whereas \reqs{b131}--\rref{b132} imply the
opposite order of the limits. Those operations do not commute as
they treat contributions from the region far from the Fermi
surface differently: in the quasiclassical approximation the
information that the electron states are limited from below at
$\xi > - \epsilon_F$ is lost. Lost contributions, however, are
coming from the transitions with the large energy and therefore
are perfectly analytic functions of fields $\varphi$, $\mathbf h$,
and their gradients. As the result, \reqs{b17} is modified as \be
\begin{split}
Z\left\{\hat{\phi}\right\}  & =Z_{0}Z_{\rho}\left\{\varphi\right\}
Z_{\rho}^{uv} \left\{\varphi\right\}
Z_s \left\{{\mathbf h}\right\} Z_s^{uv}\left\{{\mathbf h}\right\}\\
\ln Z_{\rho}^{uv} & =
- \nu \int d x d\n d\n^\prime \phi(x,\n)d_\rho
\left(\widehat{\n\n^\prime}\right)
\phi(x,\n^\prime )
+\dots\\
\ln Z_{s}^{uv} & =
\nu \int d x d \n d\n^\prime {\mathbf h}(x,\n)d_s
\left(\widehat{\n\n^\prime}\right)
{\mathbf h}(x,\n^\prime )
+\dots,
\end{split}
\label{b240}
\ee
where $\dots$ stand for the terms containing higher gradients of the
field or the higher powers of the field. All such terms however will
be small as $1/\epsilon_F$ and that is why keeping them would be
the overstepping of the accuracy of the quasiclassical equations
\rref{a28}.

Functions $d_{\rho,s}\left(\theta\right)$ depend on the details of the
ultraviolet cut-off [for the weakly interacting gas
$d_\rho=d_\sigma$]. One property, however, remains intact -- the
response of the system on the fields independent on the coordinate
but arbitrary periodic function of time $\varphi(\tau), {\mathbf
  h}(\tau);
\int_0^{1/T}\varphi(\tau)d\tau = \int_0^{1/T}{\mathbf
  h}(\tau)=0$
should vanish, because the total charge and the total spin commute
with the Hamiltonian: \be
\begin{split}
Z_{\rho}\left\{\varphi(\tau)\right\}Z_{\rho}^{uv}
\left\{\varphi(\tau)\right\}
= Z_{s}\left\{{\mathbf h}(\tau)\right\}Z_{s}^{uv}
\left\{{\mathbf h}(\tau)\right\}=1.
\label{b241}
\end{split}
\ee For such fields, \reqs{b20} -- \rref{b21} are trivially solved
$\rho(\tau,u)=-u\phi(\tau)$, ${\mathbf S}(\tau,u)=-u{\mathbf
h}(\tau)$ and we obtain from \reqs{b241}, \rref{b18}, \rref{b19}
and \rref{b240} \be \int d_\rho
\left(\widehat{\n\n^\prime}\right)d\n =\int d_s
\left(\widehat{\n\n^\prime}\right)d\n=1. \label{b242} \ee All the
other properties of $d_{\rho,s}$ are model dependent and can be
established by direct perturbative calculation for stationary
fields for which the semiclassical contributions \rref{b18}--
\rref{b19} vanish.

However, it would be a redundant exercise, as Eq.~\ref{b240} have
the same form as the weights \rref{a1819}, and the role of terms
\rref{b240} is just a renormalization of the constants in those
weights. The contribution of the interaction terms with the
high-momentum transfer not included into the Hubbard-Stratonovich
transformation leads to the similar effects. It means, that the
form of the weights for the fields $h,\varphi$ should be
established not from the first principles but from the requirement
that the quadratic part of the theory should reproduce the bosonic
modes obtained from the kinetic equation in the Landau theory of
Fermi liquid. It leads to the replacement of \reqs{a1819} with
\begin{subequations}
\label{b243}
\be
{\mathcal W}_{s}  =\exp\left\{  -\frac{\nu}{2}\int
\varphi
\left(x,\n\right)
\left[ \hat{\Gamma}_{s}^{-1}\varphi\right]\left(  x,\n\right)
dx d\n \right\}
\label{b243a}
\ee
\be
{\mathcal W}_{t}  =
\exp\left\{  -\frac{\nu}{2}\sum_{i=x,y,z}\int
h_i
\left(x,\n\right)
\left[ \hat{\Gamma}_{t}^{-1}h_i\right]\left(  x,n\right)dx d\n
\right\},
\label{b243b}
\ee
where the operators $\hat{\Gamma}_{s,t}$  are defined
by its action on any function $a\left(x,\n\right)$ as
\bea
&&2\hat{\Gamma}_{s}
={\hat{f}}\frac{\hat{\mathbb F}^\rho}{1+\hat{\mathbb F}^\rho}
\label{b243c}\\
&&2\hat{\Gamma}_{t}
=-
{\hat{f}}\frac{\hat{\mathbb F}^\sigma}{1+\hat{\mathbb F}^\sigma} .
\label{b243d}
\eea
Here operators $\hat{f}$ and
 ${\hat{\mathbb F}}$ are defined by their action on an arbitrary function
 $b(\tau, \r;\n)$ as
\be
\begin{split}
&\left[\hat{f}b\right](\r,\tau;\n)
=\int d\r_1 \bar{f}\left(\r-\r_1\right) b(\r_1,\tau; \n)
\\
&\left[{\hat{\mathbb F}}^{\rho.\sigma}b\right](\r,\tau; \n)=\int d\n_2
{\mathbb F}^{\rho,\sigma}\left(\widehat{\n\n_2}\right) b(\r,\tau; \n_2),
\end{split}
\label{b244} \ee and the convention \rref{a11a} is used. The
cutoff function $\bar{f}$ is defined by \req{a20a}, while the
functions ${\mathbb F}^{\rho,\sigma}$ are the Fermi liquid
functions describing the interaction between two quasiparticles in
the singlet or triplet states. We will see in the next subsection
that the choice \rref{b243} indeed reproduces the correct
propagators for the collective modes in the Fermi liquid theory.
In what follows we assume the operators $\hat{\Gamma}_{s,t}$ to be
positive definite, and the system far from Pomeranchuk
instabilities.
\end{subequations}

\begin{subequations}
\label{b245} With the help of the quasiclassical consideration we
can recast \req{a15} to the form \be \Omega= -T\ln Z
=\Omega_0+\Omega_\rho+\Omega_s \label{b245a} \ee where
$\Omega_0=-T\ln Z_0$ describes the leading contribution of the
quasiparticles. The leading singular corrections are associated
with the collective modes and they are given by \bea
\exp\left(-\frac{\Omega_\rho}{T}\right) &=&\int D\varphi {\cal
  W}_s\left\{\varphi\right\}Z_\rho\left\{\varphi\right\}
\label{b245b}
\\
\exp\left(-\frac{\Omega_s}{T}\right)
&=&\int D{{\mathbf h}} {\cal
  W}_t\left\{{\mathbf h}\right\}Z_s\left\{{\mathbf h}\right\},
\label{b245c}
\eea
\end{subequations}
where the functionals $Z_{\rho,s}$ are given by
\reqs{b18}--\rref{b19}.
In writing the expression for the partition function we ignored the
terms, [e.g. $Z_st$ in \req{a15}], which do not lead to change of any
observable quantities.

The results of the present Section show that study of the system of the
interacting fermions can be reduced to investigation of a system of bosonic
charge and spin excitations. Therefore the word \textquotedblleft
bosonization\textquotedblright\ is most suitable for our approach. We see that
the method should work in any dimension. At the same time, it is more general
than the scheme of the high dimensional bosonization of Refs.~
\onlinecite{haldane,houghton1,houghton,neto,kopietz,kopietzb,khveshchenko,khveshchenko1,castellani,metzner}
because we can consider the spin excitations that are much
less trivial than the charge ones.

\subsection{Thermodynamics of free modes.}
\label{sec3c}

Before we start formulating the proper field theory description
for calculating the partition functions \rref{b245b} --
\rref{b245c}, it is instructive to try to determine it by the
brute force analysis of \reqs{b18}--\rref{b19}.

For the charge mode we immediately solve \req{b20} by the Fourier
transform: \be \rho\left(\w_n,k;\n\right)=u
\frac{i\w_n}{i\w_n-v_F\k\n} \varphi\left(\w_n,k;\n\right).
\label{b280} \ee where $\w_n=2\pi T n$ is the bosonic Matsubara
frequency, \req{b90}, and \be
\varphi(\omega_n,\k;\n)=\varphi^*(-\omega_n,-\k;\n) \label{b281}
\ee because the field $\varphi(x;\n)$ is real. Substituting
\req{b280} into \req{b18}, we find \be Z_{\rho}=\exp\left[\nu T
\sum_{\w_n} \int \frac{d^{d}\mathbf{k}d\mathbf{n}}{\left(
2\pi\right)  ^{d}}
\frac{i\omega_{n}\left|\varphi(\omega_n.\k;\n)\right|^2
}{i\omega_{n}-v_{F}\k\n} \right]. \label{b282} \ee Both the
partition function, Eq. \rref{b282}, and the weight, Eq.
(\rref{b243a}) are the Gaussian functionals and, therefore, the
functional integration in \req{b245b} can be readily performed
with the result
\begin{equation}
\Omega_{\rho}=\frac{T}{2}\sum_{\omega_{n}}\int
\frac{d^{d}\mathbf{k}d\mathbf{n}}{\left(  2\pi\right)  ^{d}}
 \ln\left[  1+f\left(
\mathbf{k}\right)\hat{\mathbb F}^\rho \frac{v_{F}\mathbf{kn}%
}{-i\omega_{n}+v_{F}\mathbf{kn}}\right].  \label{b30}%
\end{equation}
Here, factor of $1/2$ originates due to the constraint
\rref{b281}, and the action of the interaction function
$\hat{\mathbb F}^\rho$ is defined by \req{b244}. The function
$f\left( \mathbf{k}\right),$ Eq. (\ref{a7c}), cuts momenta $k$
exceeding $r_{0}^{-1}$. However, its presence in Eq. (\ref{b30})
is important only for calculation of corrections to the
coefficient in the linear term in the specific heat. Non-trivial
contributions to the specific heat $C$ come from the momenta
$k\sim T/v_{F}\ll r_{0}^{-1}$ and do not depend on the function
$f$. The explicit formulas for the specific heat given by
\req{b30} will be derived in Sec.~\ref{sec7}, however, here we
will give the equivalent representation of \req{b30} more
convenient for the comparison with future material:
\begin{align}
\Omega_{\rho}&=\frac{T}{2}\sum_{\omega_{n}}\int
\frac{d^{d}\mathbf{k}d\mathbf{n}}{\left(  2\pi\right)  ^{d}}
 \ln\left[  1+f\left(
\mathbf{k}\right)\hat{\gamma }^\rho \frac{i\omega_{n}+v_{F}\mathbf{kn}%
}{-i\omega_{n}+v_{F}\mathbf{kn}}\right];
\nonumber\\
\hat\gamma^\rho&=
\frac{\frac{1}{2}\hat{\mathbb F}^\rho}{\frac{1}{2}+\hat{\mathbb F}^\rho}
\tag{\ref{b30}$^\prime$}
\label{b30prime}
\end{align}

Equation \rref{b30} has a very simple form and describes the
thermodynamic potential of the collective non-interacting charge
mode. In the conventional diagrammatic language, \req{b30}
corresponds to the contribution of ring diagrams. The advantage of
the derivation here is the explicit demonstration that \req{b30}
completely solves the problem of the singular corrections in the
charge channel [which is the only one present for the spinless
electrons]. No further terms are present for the linearized
spectrum and all of the other corrections have additional
smallness of the order of $T/\epsilon_F$ in comparison with
\req{b30}. This means that no further consideration of the singlet
channel is necessary.

Let us turn to the triplet channel. One can see that due to the
presence of the Hubbard-Stratonovich field $\h$ in the operator
\rref{b23}, one can solve \req{b21} only approximately. In
particular for $|\h|\to 0$, one finds [cf. \req{b280}]: \be
\S\left(\w_n,k;\n\right)=u \frac{i\w_n}{i\w_n-v_F\k\n}
\h\left(\w_n,k;\n\right)+\dots. \label{b300} \ee where $\dots$
stand for the functionals of the second and higher orders in field
$\h$, and \be \h(\omega_n,\k;\n)=\h^*(-\omega_n,-\k;\n).
\label{b301} \ee Substitution of \req{b300} into \req{b19} yields
\be
\begin{split}
Z_{s}=\exp\Bigg[&-\nu T \sum_{\w_n}
\int \frac{d^{d}\mathbf{k}d\mathbf{n}}{\left(  2\pi\right)  ^{d}}
\\
&
\times\frac{i\omega_{n}
}{i\omega_{n}-v_{F}\k\n}\sum_{j=x,y,z}\left|h_j(\omega_n.\k;\n)\right|^2
 + \dots
\Bigg],
\end{split}
\label{b302} \ee where $\dots$ denote the functional of the third
and higher orders in $\h$. It is important to emphasize that such
non-linear terms do not have any additional smallness in
$T/\epsilon_F$ in contrast to the singlet channel formula
\rref{b282}.

If we ignore those non-linear terms we end up with the Gaussian
functional integral in \req{b245c} and we obtain analogously to
\req{b30}:
\begin{equation}
\Omega_{s}^{\left(  0\right)  }=\frac{3 T}{2}\sum_{\omega_{n}}
\frac{d^{d}\mathbf{k}d\mathbf{n}}{\left(  2\pi\right)  ^{d}}
 \ln\left[  1+f\left(
\mathbf{k}\right)\hat{\mathbb F}^\sigma \frac{v_{F}\mathbf{kn}%
}{-i\omega_{n}+v_{F}\mathbf{kn}}\right].
  \label{b31}%
\end{equation}
and the action of the interaction function $\hat{\mathbb F}^\sigma$ is defined
by \req{b244}. The additional factor of $3$ in comparison with
\req{b30} stands for the three independent components of the spin
density.
The equivalent representation for \req{b31} is [cf. \req{b30prime}]
\begin{align}
\Omega_{\rho}&=\frac{3T}{2}\sum_{\omega_{n}}\int
\frac{d^{d}\mathbf{k}d\mathbf{n}}{\left(  2\pi\right)  ^{d}}
 \ln\left[  1+f\left(
\mathbf{k}\right)\hat{\gamma }\frac{i\omega_{n}+v_{F}\mathbf{kn}%
}{i\omega_{n}-v_{F}\mathbf{kn}}\right];
\nonumber\\
\hat\gamma&=-
\frac{\frac{1}{2}\hat{\mathbb F}^\sigma}{\frac{1}{2}+\hat{\mathbb F}^\sigma}.
\tag{\ref{b31}$^\prime$}
\label{b31prime}
\end{align}
Let us mention for the future comparison with previous works, that
for the weakly interacting systems, the kernels $\gamma$ are the
linear function of the scattering amplitudes \rref{a13} \be
\gamma_\rho=\nu V_s; \quad \gamma=-\nu V_t. \label{b310} \ee

However, due to the presence of the ignored non-small terms in
\req{b302}, formulas \rref{b31},\rref{b31prime} are by no means exact or correct
low temperature asymptotic expression. The next section present an efficient
calculational scheme to deal with this non-linearity.

\section{Supersymmetry approach for the bosonic excitations.}
\label{sec4}

As we have already explained, the exact derivation of the
functional $Z_s\left\{{\mathbf h}\right\}$ determining the
thermodynamics of the triplet mode \req{b245c} is not possible.
Moreover, obtaining the non-linear terms by the further expansion
of the solution in powers of $\h$ is not a correct  way to proceed
because the resulting theory in terms of $\h$ will be not only
non-linear but also {\em non-local} which would obscure such
important features of the theory as its renormalizability.

An analogous problem exists in the theory of disordered metals but
in many cases it can be overcome using the supersymmetry method
\cite{review,book}. The main idea of the method is to express the
solution of a linear equation with a disorder in terms of a
functional integral over auxiliary supervectors containing both
conventional complex numbers and anticommuting Grassmann
variables. Then, one is able to average over the disorder and
reduce the disordered system to a regular model without any
disorder but with a {\em local} effective interaction.

We will borrow these ideas. The role of the disorder here will be
played by the field $\hat{h}$ itself, so its imaginary time dependence
will lead to certain modifications.

We will discuss the number and the properties of the necessary
fields in Subsection \ref{sec4a} in somewhat simpler form and will
write down the full-fledged effective Lagrangian in Subsection
\ref{sec4b}. The final form of the field theory obtained after the
averaging over field $\h$ is given in Subsection \ref{sec4c} and
it is generalized further in Subsection \ref{sec4d}.

\subsection{The number of auxiliary fields and Hermitization.}
\label{sec4a}

Solution of \req{b21} can still be written in a symbolic form as
\be \mathbf{S}\left(  x,\n; u \right)= -u \hat{L}_{u}^{-1}\ptau
\mathbf{h}\left(  x,\n\right) \label{c100} \ee where the {\em
local} operator $\hat{L}_{u}$ is given by \req{b23}. Substitution
of \req{c100} into \req{b19} yields \be Z_{s}=\exp\left[
2\nu\int_0^1 udu\int dxd\n\, \h(x,\n)\hat{L}_{u}^{-1} \ptau
\mathbf{h}\left(x,\n\right)\right], \label{c101} \ee and we use
the notation \rref{a1ab} throughout this section.

The argument of the exponent is non-local and our goal is to get
rid of such a non-locality. The standard route to proceed would be
to re-write \req{c101} as a functional integral \be
\begin{split}
Z_{s}\left\{\h\right\}& \stackrel{?}{=}
\int D\S^\dagger D\S D\bchi D\bchi^\dagger
\exp\Bigg[-
2\nu\int_0^1du\int dxd\n
\\
&
\times
\left\{
\S^*\hat{L}_{u}\S + \bchi^* \hat{L}_{u}\bchi
 +u \hat{S} \ptau \mathbf{h} +  \hat{S}^*\mathbf{h}
\right\}
\Bigg].
\end{split}
\label{c102}
\ee
Here
\be
\begin{split}
&\S=\begin{pmatrix}S_{x}\\S_{y}\\S_{z}\end{pmatrix}_s;
\quad \bchi=\begin{pmatrix}\chi_{x}\\\chi_{y}\\
\chi_{z}\end{pmatrix}_s;
\\
&
\S^\dagger=\left(S_{x}^*,S_{y}^*,S_{z}^*\right)_s;
\quad \bchi^\dagger=\left(\chi_{x}^*,\chi_{y}^*,\chi_{z}^*\right)_s.
\end{split}
\label{c1020}
\ee
 The spin space $s$ was introduced after \req{b24}.
The fields $\S^*,\S$ are the usual complex vector fields and
$\bchi, \bchi^*$ are anticommuting Grassmann fields needed to
cancel out the operator determinant. All the fields are functions
of $x,\n, u$ and satisfy the periodicity conditions \footnote{Note
that although the theory is still Gaussian we had to introduce one
more coordinate $u$, such that the fields depend on $u$ for the
representation of the needed determinants. The other possible way
would be to use representation \rref{b8}--\rref{b9} for ${\mathbf
S}$ and write WZNW action for matrices ${\hat{\mathcal T}}$, see
{\em e.g.} Ref.~\onlinecite{gntbook}. We chose not to pursue this
line because of the difficulty of the identification of the
manifold of the integration over the matrices ${\hat{\mathcal T}}$
such that the ``saddle point'' matrix ${\hat{\mathcal T}}$ solving
\req{b6} would belong to the integration manifold.}.
\be
\begin{split}
&\S\left(x,\n,u\right)=\S\left(\r,\tau+\frac{1}{T},\n,u\right);
\\
&\S^*\left(x,\n,u\right)=\S^*\left(\r,\tau+\frac{1}{T},\n,u\right);
\\
&\bchi\left(x,\n,u\right)=\bchi\left(\r,\tau+\frac{1}{T},\n,u\right);
\\
&\bchi^*\left(x,\n,u\right)=\bchi^*\left(\r,\tau+\frac{1}{T},\n,u\right).
\end{split}
\label{c103} \ee Because the Grassmann fields are periodic rather
than anti-periodic we will call them pseudofermions. The argument
of the exponent \rref{c102} would be, then, the local functional
linear in $\h$, so  the integration in \req{b245c} could be easily
performed yielding the {\em local} expression in terms of powers
of $\S$ and $\bchi$ only.

However, \req{c102} is rather deceptive. Indeed, the possibility
to write such a functional integral is based on the assumption
that the integration over the bosonic fields is defined for an
arbitrary configuration of the field $\h$, in particular the
directions of such integrations cannot depend on the field $\h$ at
all. Therefore, \req{c102} would be correct only if the operator
$\hat{L}_u$ of \req{b23} were positive definite which is not the
case. Moreover, as we have mentioned after \req{b24b}, $\hat{L}_u$
is not even Hermitian and we do not have {\em a priori} the
knowledge about the signs of real and imaginary parts of
eigenvalues.

Those complications make expression \rref{c102} mathematically
meaningless. Fortunately, the proper procedure for non-hermitian operators containing
first-order derivatives has also been worked out previously\cite{hermit} and
we will use this method for further calculations.

Let us double the number of the bosonic and pseudofermionic fields as
\be
\begin{split}
&\S=\begin{pmatrix}
\S^1\\
\S^2
\end{pmatrix}_H;
\quad \bchi=\begin{pmatrix}
\bchi^1\\
\bchi^2
\end{pmatrix}_H;
\\
& {\S}^\dagger =\left([\S^1]^*,[\S^2]^* \right)_H; \quad
 {\bchi}^\dagger =\left([\bchi^1]^*,[\bchi^2]^* \right)_H,
\end{split}
\label{c104} \ee where each element has the structure of
\req{c1020}. We will call the additional space of the two-
component vectors ``hermitized'' space and use the subscript
$(\dots)_H$ for the writing explicitly the structure in this
space. We define the new operator $\hat{M}_u$ acting in this
doubled space \be \hat{M}_u=\frac{1}{2}
\begin{pmatrix}
\hat{L}_u+ \hat{L}_u^\dagger; & \hat{L}_u- \hat{L}_u^\dagger \\
 \hat{L}_u^\dagger- \hat{L}_u; & -\hat{L}_u -  \hat{L}_u^\dagger
\end{pmatrix}_H,
\label{c105} \ee where each element of this matrix is a matrix in
the spin space, see \req{b23}. By construction, the operator, Eq.
(\rref{c105}), is Hermitian, $M=M^\dagger$, and thus the
functional
\[
\int dx  \left({\S}^\dagger\hat{M}_u\S\right)
\]
is real. Therefore, the identity
\be
\begin{split}
&1=
\int D{\S}^\dagger D\S D{\bchi}D\bchi^\dagger
\exp\left[- {\cal S}_{\h}^{\rm eff}\right]
\\
&{\cal S}_{\h}^{\rm eff}= -
2 i\nu\int_0^1du\int dxd\n
\\
&
\quad \times
\left\{
{\S}^\dagger\left(\hat{M}_{u}
+i\delta
\right)\S +
{\bchi}^\dagger\left(\hat{M}_{u}
+i\delta
\right)\bchi
\right\}.
\end{split}
\label{c106} \ee holds for an arbitrary configuration of the field
$\h$ as the integral over the bosonic fields is always convergent.
Here $\delta$ is a positive real number and the limit $\delta\to
+0$ is to be taken at the end of the calculation. The fields
satisfy the periodicity conditions, Eq. (\rref{c103}). Finally,
using \req{c106} and obvious formula
\[
\frac{1}{\hat{M}_u}=\frac{1}{2}
\begin{pmatrix}
\displaystyle{\frac{1}{\hat{L}_u}+
\frac{1}{\hat{L}_u^\dagger}}; \quad &
\displaystyle{\frac{1}{\hat{L}_u^\dagger}  - \frac{1}{ \hat{L}_u}} \\
\\
\displaystyle{\frac{1}{\hat{L}_u}
-\frac{1}{\hat{L}_u^\dagger}}
; \quad & \displaystyle{-\frac{1}{\hat{L}_u}-
\frac{1}{\hat{L}_u^\dagger}}
\end{pmatrix}_H,
\]
we obtain instead of \req{c102}
\be
\begin{split}
Z_{s}\left\{\h\right\}& =
\int D\bar{\S} D\S D\bar{\bchi} D\bchi
\exp\left[- {\cal S}_0^{\h}\right]
\\
&
\exp\Bigg[
\nu\sqrt{2i}\int_0^1du\int dxd\n
\\
&
\times
\left\{
 u \left(\hat{\S}^1+\hat{\S}^2 \right) \ptau \mathbf{h}  +
 \left(\left[\hat{\S}^1\right]^*-
\left[\hat{\S}^2\right]^* \right)\mathbf{h}
\right\}
\Bigg].
\end{split}
\label{c107}
\ee

Equations \rref{c106}--\rref{c107} is the final result of this
subsection. We have succeeded in rewriting the original {\em
non-local} expression written in terms of the Hubbard-Stratonovich
field $\h$  into the theory local in terms of the new fields $\S,
\bchi$. As this action is a linear functional in $\h$, we will be
able to integrate it out and obtain the action in terms of those
new fields $\S, \bchi$ only. Before doing so, we will recast
\reqs{c106}--\rref{c107} in a more compact form.



\subsection{Supervectors and the effective Lagrangian.}
\label{sec4b}

We introduce the superspace  (graded space) as the space of
vectors having the same number of complex and Grassmann
components\cite{book}. In particular, the field introduced in
\req{c104} can be compactified as one supervector\footnote{Using
the same notation for the
  supervector here and for the Hubbard-Stratonovich field in
  Sec.~\ref{sec2}
should not lead to a confusion as the latter will
not appear in any further consideration.}
\be
\bvarphi = \begin{pmatrix}\bchi \\ \S \end{pmatrix}_g;
\quad \bvarphi^\dagger = \left( {\bchi}^\dagger, \S^\dagger\right)_g
\label{c1070}
\ee
where subscript $g$ stands for the graded space, and each element
has the structure of \req{c104}, so the fields are
defined in a linear space obtained as a direct product of
 Hermitized (H)  spin  (s) and superspace (g).
In other words, notation $\bvarphi$ means the $12$ -component
supervector defined in a linear space $s \otimes  H \otimes g $.

Using notation \rref{c1070} the action from \req{c106} can be
re-written in a short form
\be
{\cal S}_{\h}^{\rm eff}= -
2 i\nu\int_0^1du\int dxd\n
\left[{\bvarphi}^\dagger\left(\hat{M}_{u}\otimes \openone_g
+i\delta
\right)\bvarphi\right].
\label{c108}
\ee
where $\openone_g$ is the $2\times 2$ unit matrix acting in the superspace.

We will see shortly that the most interesting contribution will come
from the scattering terms where the direction of the spin changes
its direction to opposite. Anticipating this fact, we will
join ${\bvarphi}(\n)$ and ${\bvarphi}(-\n)$ in one
vector of the larger dimensionality
\be
\bphi(\n)=\begin{pmatrix}\bvarphi(\n) \\ \bvarphi(-\n) \end{pmatrix}_n;
\quad \bphi^\dagger(\n)=\Big(\bvarphi^\dagger(\n); \bvarphi\dagger(-\n) \Big)_n,
\label{c109}
\ee
where each element has the structure of \req{c1070}.
and keep integration over $\n$ in all of the subsequent formulas
over the $d$-dimensional hemisphere,  say $n_x>0$. The final answers definitely
will not depend on  particular choice of the hemisphere.
From now on the integration over the momentum direction will mean
\be
\int d\n \dots \equiv \int_{n_x>0} d\n \dots;
\quad \int 1 d\n\, = \frac{1}{2}.
\label{c110}
\ee
We will call two-dimensional space defined in \req{c108} the
``left-right'' space, using the analogy with the one-dimensional
systems and will denote it by subscript $n$.

The final step in the definition of the supervector is once again
performed for the calculational convenience, and it is equivalent
to the introduction of the Gorkov-Nambu spinors in the theory of
the superconductivity\cite{GN}. We increase the size of the
supervector as \be \bpsi = \frac{1}{\sqrt{2}} \begin{pmatrix}
\bphi^* \\ \bphi \end{pmatrix}_{eh}, \quad \bpsi^\dagger =
\frac{1}{\sqrt{2}} \left( \left[\bphi^*\right]^*;\, \bphi^*
\right)_{eh}, \label{c111} \ee where each component has the
structure of \req{c109}. We will call the corresponding space
``electron-hole space'' and denote it by the subscript $eh$ when
written explicitly. The benefit of this doubling  of the size of
the supervector will become apparent when the perturbation theory
for the resulting model is developed in Section \ref{sec5}.
To avoid misunderstanding, we emphasize that the
``electron- hole'' introduced here has nothing to do with
the electrons and holes in the original system and
introduced here to label the fields of the spin excitations only.

Using the standard convention for the complex conjugate of the
Grassmann variables, \be \left[\chi^*\right]^*=-\chi, \label{c112}
\ee and definitions \rref{c109} -- \rref{c111} we  rewrite
\req{c108} as \be {\cal S}_{\h}^{\rm eff}= - 2 i\nu\int_0^1du\int
dxd\n \left[{\bpsi}^\dagger\left(\hat{{\mathcal M}}_{u}\otimes
\openone_g +i\delta \right)\bpsi\right]. \label{c113} \ee Here
$\hat{{\mathcal M}}_{u}$ is the matrix in the $H\otimes n \otimes
eh$ space and it has the structure \be
\begin{split}
\hat{{\mathcal M}}_{u} = \begin{pmatrix}
\hat{\mathbf M}_u & 0
\\ 0 & \left[\hat{\mathbf M}_u\right]^T\end{pmatrix}_{eh};
\\
\\
\hat{\mathbf M}_u(\n) = \begin{pmatrix}
\hat{ M}_u(\n) & 0
\\0 & \hat{ M}_u(-\n)
\end{pmatrix}_{n},
\end{split}
\label{c114}
\ee
with matrix $\hat{M}_u$  given by \req{c105}.

To make the notation  consistent with the previous work\cite{review,book,hermit}, we
introduce the  conjugated supervector as
\be
\overline{\bpsi}={\bpsi}^\dagger \hat{\Lambda};
\quad \hat\Lambda =
\openone_g\otimes\openone_n \otimes \openone_s\otimes \openone_{eh}
\otimes \begin{pmatrix}
1 & 0\\
0 & -1
\end{pmatrix}_H.
\label{c115} \ee Using the explicit structure of the supervectors,
Eq. (\rref{c111}), and the convention $\rref{c112}$ one can verify
that the conjugated supervector $\overline{\bpsi}$ is related to
$\bpsi$ as
\begin{equation}
\begin{split}
&\overline{\bpsi}=\left(  \hat{C}\bpsi\right)  ^{T},
\\
&\hat{C}= \openone_s \otimes \openone_n \otimes
\begin{pmatrix}
\hat{C}_{0} & 0\\
0 & -\hat{C}_{0}
\end{pmatrix}_H  ,\text{ }C_{0}=\begin{pmatrix}
\hat{c}_{1} & 0\\
0 & \hat{c}_{2}
\end{pmatrix}_g
\\
&
\hat{c}_{1}=\left(
\begin{array}
[c]{cc}%
0 & -1\\
1 & 0
\end{array}
\right)_{eh}  ,\text{ }\hat{c}_{2}=\left(
\begin{array}
[c]{cc}%
0 & 1\\
1 & 0
\end{array}
\right)_{eh}.
\end{split}
\label{c23}
\end{equation}
Accordingly, the conjugation of supermatrices is introduced as
\begin{equation}
\overline{A}=CA^{T}C^{T} \label{c24a}%
\end{equation}
for an arbitrary supermatrix $A$ acting in $s \otimes g \otimes H
\otimes n \otimes eh $ space. For the two supervectors
$\bpsi_{1,2}$ of the structure \rref{c111} one finds \be
\left(\overline{\bpsi}_1A {\bpsi}_2\right)=
\left(\overline{\bpsi}_2\overline{A} \bpsi_1\right). \label{c24b}
\ee

Substituting definition \rref{c115} into  \req{c113}
and finding  explicit form of $\hat{{\mathcal M}}_{u}$
from  \reqs{c114}, \rref{c105}
and \rref{b23}-- \rref{b24b}, we obtain
\be
{\cal S}_{\h}^{\rm eff}= -
2 i\nu\int\overline{\bpsi}
\left(
X\right)\hat{\cal L}_\h\bpsi\left(  X\right)  dX, \label{c25}%
\ee where we use the short hand notation \be
\begin{split}
X&=\left( \r,\tau,\n,u\right);
\\
 \int dX\dots &=\int d\r\int_0^{1/T}d\tau\int d\n\int_0^1 du\dots
\end{split}
\label{c250}
\ee
and the convention \rref{c110} for the angular integration.

The Lagrangian in \req{c25} is given by \be \hat{{\cal L}}_\h =
\hat{{\cal L}}_ 0 -2i u \hat{\tau}_{3} \delta\hat{{\cal L}}_\h
 -i \delta \hat{\Lambda},\label{c26}%
\ee where the matrix in the spin space $\hat{h}$ is defined in
\req{b24} and the free propagation Lagrangian has the form \be
\begin{split}
{\cal L}_{0}=-iv_{F}\left(  \mathbf{n\, \nabla}\right)  \hat{\tau}_{3}
\hat{\Sigma}_{3}-\ptau\hat{\Lambda}_{1}, \\
{\cal L}_{0}^\dagger=-iv_{F}\left(  \mathbf{n\, \nabla}\right)  \hat{\tau}_{3}
\hat{\Sigma}_{3}+\ptau\hat{\Lambda}_{1},
\end{split}
\label{c261} \ee
The rotation of the spin excitation by the
Hubbard-Stratonovich field [cf. \req{b24}] is described by \be
\begin{split}
&
\delta \hat{\Lambda}_{\h}(x,\n)=\begin{pmatrix}
0 & -\hat{\mathbb H}_{z}(x,\n) & \hat{\mathbb H}_{y}(x,\n)\\
\hat{\mathbb H}_{z}(x,\n) & 0 & -\hat{\mathbb H}_{x}(x,\n)\\
-\hat{\mathbb H}_y(x,\n) & \hat{\mathbb H}_{x} (x,\n)& 0
\end{pmatrix}_s,
\\
&\hat{\mathbb H}_\gamma(x,\n)
 =
\openone_g \otimes \openone_H
\otimes  \openone_{eh} \otimes
\begin{pmatrix}
h_\gamma(x,\n) & 0\\
0& h_\gamma(x,-\n)
 \end{pmatrix}_n.
\end{split}
\label{c262}
\ee
The supermatrices in \reqs{c261} and \rref{c262} are introduced as
\be
\begin{split}
&\hat{\tau}_{3}=
\openone_s \otimes \openone_g \otimes
 \openone_H \otimes \openone_n \otimes
\left(
\begin{array}
[c]{cc}%
1 & 0\\
0 & -1
\end{array}
\right)_{eh}
  ,\text{ }
\\
&
\hat{\Sigma}_{3}=
\openone_s \otimes \openone_g \otimes
 \openone_H
\otimes \left(
\begin{array}
[c]{cc}%
1 & 0\\
0 & -1
\end{array}
\right)_{n}\otimes \openone_{eh},\\
&
\hat{\Lambda}_{1}=
\openone_s \otimes \openone_g \otimes
  \left(
\begin{array}
[c]{cc}%
0 & 1\\
1 & 0
\end{array}
\right)_{H}\otimes \openone_{n} \otimes \openone_{eh}
,\\
\end{split}
\label{c28}%
\ee The action ${\cal S}_{\h}^{\rm eff}$ is supersymmetric, i.e.
invariant with respect to all  possible homogeneous rotations in
$g$-space.

To complete the derivation, we have to express the exponent in
\req{c107} in terms of the supervector $\bpsi$. Those are only
terms that break the supersymmetry and thus lead to finite
contributions to physical quantities. Using the definitions
\rref{c111}, \rref{c109}, \rref{c104}, \rref{c1020}, we find
\begin{subequations}
\label{c280}
\be
\label{c280a}
\begin{split}
&\int_0^1du\int dxd\n
\left(
\left[\hat{\S}^1\right]^*-\left[\hat{\S}^2\right]^* \right)\mathbf{h}
\\
& \quad=
2\int dX
\Bigg(\overline{\bpsi}_\gamma(X)
{\mathbf F}^1(X) \Bigg)
\\
&
F^1_\gamma =
\frac{1}{\sqrt{2}}\begin{pmatrix} 0 \\ 1 \end{pmatrix}_g
\otimes
\begin{pmatrix} 1 \\ 1 \end{pmatrix}_H
\otimes
\begin{pmatrix} 0 \\ 1 \end{pmatrix}_{eh}
\otimes
\begin{pmatrix}
h_\gamma(\n)
\\
h_\gamma(-\n)
\end{pmatrix}_n
,
\end{split}
\ee
and, analogously,
\be
\label{c280b}
\begin{split}
&\int_0^1du\int dxd\n
\left(
\hat{\S}^1+\hat{\S}^2 \right)\mathbf{h}
\\
& \quad=
2\int dX
\Bigg(\overline{\bpsi}_\gamma(X)
{\mathbf F}^2(X) \Bigg)
\\
&
F^2_\gamma =
\frac{1}{\sqrt{2}}\begin{pmatrix} 0 \\ 1 \end{pmatrix}_g
\otimes
\begin{pmatrix} 1 \\ - 1 \end{pmatrix}_H
\otimes
\begin{pmatrix} 1 \\ 0 \end{pmatrix}_{eh}
\otimes
\begin{pmatrix}
u\ptau h_\gamma(\n)
\\
u\ptau h_\gamma(-\n)
\end{pmatrix}_n.
\end{split}
\ee where Eq. \rref{c250} is used, and $\gamma=x,y,z$ labels the
components in the spin space.
\end{subequations}

Equations \rref{c280} and \rref{c25} enable us to obtain a
representation of formulas \rref{c106}--\rref{c107} in
supersymmetric notations \be
\begin{split}
Z_s\left\{\mathbf{h}\right\}
&  =\int
D\bpsi\exp\left[- {\cal S}_{\h}^{\rm eff}\right]
\\
&  \times
\exp\left[
2\nu\sqrt{2i}\int \overline{\bpsi}\left(
X\right)
{\mathbf F}\left(  X\right)  dX
\right],
\end{split}
\label{c33}
\ee
where ${\cal S}_{\h}^{\rm eff}$ is given by \req{c25} and
\be
{\mathbf F}(X)={\mathbf F}^1(X)+ {\mathbf F}^2(X).
\label{c330}
\ee
The superfields in \req{c33} satisfy the periodic boundary conditions
\be
\bpsi\left(\tau,\r,\n;u\right)=\bpsi\left(\tau+1/T,\r,\n;u\right).
\label{c331}
\ee
Equation \rref{c33} is a main result of this subsection and will be
used for the further manipulations.

It is worthwhile to notice that the functional \rref{c33} has an
interesting symmetry. Let us make a shift of the variables \be
\overline{\bpsi} \to \overline{\bpsi} - (1-\alpha) \frac{
u}{\sqrt{2i}}{\mathbf F}_1^T \label{c332} \ee in the functional
integral \rref{c33}, where $\alpha$ is an arbitrary constant.
Using \reqs{c280}, we find for the transformation \be
\begin{split}
&2\nu\sqrt{2i}\int \overline{\bpsi}\left(
X\right)
{\mathbf F}\left(  X\right)  dX
\to 2\nu\sqrt{2i} \int \overline{\bpsi}\left(
X\right)
{\mathbf F}\left(  X\right)  dX
\\
&\quad\quad
+(1-\alpha) \nu \int dx d\n
\left[\h^2(\n)+\h^2(-\n) \right],
\end{split}
\label{c333} \ee where the notation \rref{a1ab} and the convention
\rref{c110} for the angular integration are used. Analogously
using
\[
\hat{h}\mathbf{h}=0
\]
that can easily be checked using the definition of $\hat{h}$,
Eq.~(\ref{b24}),
one obtains from \req{c25}
\be
\begin{split}
&\int\overline{\bpsi}
\left(
X\right)\hat{\cal L}_\h\bpsi\left(  X\right)  dX
\to \int\overline{\bpsi}
\left(
X\right)\hat{\cal L}_\h\bpsi\left(  X\right)  dX
\\
&\quad - (1-\alpha) \frac{2 u}{\sqrt{2i}}
\int\overline{\bpsi}\left(X\right)\hat{\cal L}_0\bpsi
\hat{C}\,{\mathbf F}^1\left(X\right)  dX.
\label{c334}
\end{split}
\ee The extra term appearing in \req{c333} has the same functional
form as the interaction \rref{b243b} and can be incorporated into
renormalization of the interaction constant, whereas the extra
term in \req{c334} can be accommodated into redefinition of the
operator $F^2(X)\to F^2(X;\alpha)$ in \req{c280b} as \be
\begin{split}
F^2_\gamma\left(X;\alpha\right)& =
\frac{1}{\sqrt{2}}\begin{pmatrix} 0 \\ 1 \end{pmatrix}_g
\otimes
\begin{pmatrix} 1 \\ - 1 \end{pmatrix}_H
\otimes
\begin{pmatrix} 1 \\ 0 \end{pmatrix}_{eh}
\\
&
\otimes
\begin{pmatrix}
\displaystyle{u
\left[\alpha\ptau+i(1-\alpha)v_F\n\nnabla\right] h_\gamma(x,\n)}
\\
\\
\displaystyle{u\left[\alpha\ptau -i(1-\alpha)v_F\n\nnabla\right]  h_\gamma(x,-\n)}
\end{pmatrix}_n.
\label{c335}
\end{split}
\ee

Accordingly, the low-energy representation of \req{b245c} can be
written for an arbitrary parameter $\alpha$ as \be
\exp\left(-\frac{\Omega_s}{T}\right)= \int D{{\mathbf h}} {\cal
  W}_t\left(\left\{{\mathbf h}\right\};\alpha\right)
Z_s\left(\left\{{\mathbf h}\right\};\alpha\right)
\label{c336}
\ee
where [cf. \reqs{b243b}, \rref{b243c}]
\be
\begin{split}
&{\mathcal W}_{t}(\alpha)  =
\exp\left\{  -\frac{\nu}{2}\int
{\h}
\left(x,\n\right)
\left[ \hat{\Gamma}_{t}^{-1}(\alpha){ \h}\right]\left(  x,n\right)dx d\n
\right\},\\
&
2\hat{\Gamma}_{t}(\alpha)
=- {\hat{f}}
\frac{\hat{\mathbb F}^\sigma}{1+\alpha\hat{\mathbb F}^\sigma},
\end{split}
\label{c337} \ee and the angular integration over the whole
$d$-dimensional sphere is meant, and convention \rref{a11a} is
implied.

The partition function $Z_s(\alpha)$ is a generalization of
\req{c33}: \be
\begin{split}
Z_s\left\{\mathbf{h},\alpha\right\}
&  =\int
D\bpsi\exp\left[- {\cal S}_{\h}^{\rm eff}\right]
\\
&  \times
\exp\left[
2\nu\sqrt{2i}\int \overline{\bpsi}\left(
X\right)
{\mathbf F}\left(  X;\alpha\right)  dX
\right],\\
&{\mathbf F}(X;\alpha)={\mathbf F}^1(X)+ {\mathbf F}^2(X;\alpha),
\end{split}
\label{c338} \ee where vectors ${\mathbf F}^1;{\mathbf
F}^2(\alpha)$ are given by \reqs{c280a} and \rref{c335},
respectively. A particular choice of the parameter $\alpha$ is
merely a matter of a convenience.

Closing this subsection, we recast the supersymmetry breaking
terms in \req{c338} into a form more convenient for further
application, We introduce a $16$-component  supervector ${\cal
F}_0$
\begin{equation}
\begin{split}
&{\cal F}_{0}=\frac{1}{\sqrt{2}}
\begin{pmatrix}
0\\ 1
\end{pmatrix}_g
\otimes
\begin{pmatrix}
1 \\ 1
\end{pmatrix}_n
\otimes
\begin{pmatrix}
\begin{pmatrix}
1\\ 1
\end{pmatrix}_{eh}
\\
\begin{pmatrix}
-1\\1
\end{pmatrix}_{eh}
\end{pmatrix}_H;
\\
& \overline{\cal F}_{0}=\frac{1}{\sqrt{2}}
 \left(
 0\ 1
 \right)_g
 \otimes
\left(
 1 \ 1
 \right)_n
 \otimes
\Big(
 \left(
 1\ 1
 \right)_{eh}
 \left(
 -1\ 1
 \right)_{eh}
 \Big)_H
 ;
\end{split}
\label{c41}%
\end{equation}
and the operator $\hat{l}_{u,\alpha}$ is given by
\be
\begin{split}
&\hat{l}_{u,\alpha}
 = \frac{u}{2}
\left[\left(2\alpha-1\right)\hat{\cal L}_{0}-  \hat{\cal L}_{0}^\dagger\right]\tau_+
+ \tau_-;\\
&\hat{\overline{l}}_{u,\alpha}
 =  \frac{u}{2}
\left[\left(2\alpha-1\right)\hat{\cal L}_{0}- \hat{\cal L}_{0}^\dagger\right]
\tau_-
+ \tau_+,
\\
&\hat{\tau}_{\pm}=\frac{1\pm \hat{\tau}_3}{2},
\end{split}
 \label{c410}
 \ee
and $\hat{\cal L}_{0},\,\hat{\cal L}_{0}^\dagger$ are defined in
\req{c261}. The conjugation for the matrix operator is given by
\req{c24a}, and the supermatrix $\hat\tau_3$ is given by
\req{c28}. Then it is easy to check by explicit calculation that
\reqs{c338}, \rref{c280a} and \rref{c335} can be re-written as \be
\begin{split}
\int \overline{\bpsi}\left(
X\right)
{\mathbf F}\left(  X;\alpha\right)  dX & =
\int \overline{\bpsi}_\gamma\left(
X\right)\hat{l}_{u,\alpha} \hat{\mathbb H}_\gamma(X) {\cal F}_{0}\  dX
\\
& = \int  \overline{\cal F}_{0}
\hat{\mathbb H}_\gamma(X) \hat{\overline{l}}_{u,\alpha} {\bpsi}_\gamma\left(
X\right)  dX
\end{split}
\label{c411} \ee where operator $\hat{\mathbb H}_\gamma$ is given
by \req{c262}, and summation over the repeated index
$\gamma=x,y,z$ is implied \footnote{ When we write  the index in
the spin space explicitly, we imply that matrix $\openone_s$
should be dropped from definitions \rref{c28} and the relevant
supermatrices have the dimensionality $16\times 16$}. The latter
formula is the most convenient for the integration over $\h$ which
will be performed in the next subsection.

\subsection{Averaging over the Hubbard-Stratonovich field.}
\label{sec4c}

The argument of  the exponential in \req{c338} is a linear
functional of the Hubbard-Stratonovich field $\h$, and thus the
integral over $\h$ in \req{c336} is purely Gaussian. The field
$\mathbf{h}$ enters both the function ${\mathbf F}$ and the
Lagrangian \rref{c26}.  This means that the new effective field
theory will contain quadratic, cubic and quartic in $\psi$ terms.
The quartic term originates from the averaging of the
supersymmetric part of the action, and therefore, it preserves the
supersymmetry, whereas the quadratic and cubic terms lift it.

Performing Gaussian integration over $\bf h$ in \req{c336} with the
help of \reqs{c411} and \rref{c262}, we find
the contribution of the spin modes to the thermodynamic potential
\begin{equation}
\Omega_s=-T \ln \left[\int\exp\left(  -{\mathcal S}\left[  \psi\right]  \right)
D\psi\right]\label{c35}%
\end{equation}
with%
\begin{equation}
{\mathcal S}\left[  \psi\right]  ={\mathcal S}_{0}\left\{  \psi\right\}
+{\mathcal S}_{2}\left[ \left\{ \psi\right\};\alpha\right]  +
{\mathcal S}_{3}\left[  \left\{\psi\right\},\alpha\right]
+{\mathcal S}_{4}\left[ \left\{\psi\right\};\alpha\right].  \label{c36}%
\end{equation}
In Eq.~(\ref{c36}), the free supersymmetric
part  of the
action can be written as
\begin{equation}
{\mathcal S}_{0}\left[  \psi\right]  =
-2i\nu\int\overline{\psi}_{\gamma}\left(
X\right)
\left[
\hat{\mathcal L}_{0}
-i\delta\hat{\Lambda}
\right]
\psi_{\gamma}\left(  X\right)  dX \label{c37}%
\end{equation}
where $\hat{\mathcal L}_{0}$ is given by Eq.~(\ref{c261}), the
summation is implied over the repeated spin subscripts $\gamma$,
[see also footnote after \req{c411}],  variables $X$ are defined
in \req{c250}, and the convention  \rref{c110} is used for the
angular integration.

The term ${\mathcal S}_{4}$ describes the quartic
interaction and it takes the form
\be
\begin{split}
&{\mathcal S}_{4}\left[ \left\{\psi\right\};\alpha\right]   =
-4\nu\varepsilon_{\delta\beta\gamma
}\varepsilon_{\delta\beta_{1}\gamma_{1}}\sum_{i,j=1}^2
\lambda_{ij} \int dX
\\
&\times
\left(  \overline{\psi}_{\beta
}\left(  X\right)  \hat{\tau}_{3}\hat{\Pi}_j\psi_{\gamma}\left(  X\right)  u\right)
{\hat{\Gamma}}_{i}\left(  u\overline{\psi}_{\beta_{1}}\left(
X\right)\hat{\tau}_{3}\hat{\Pi}_j\psi_{\gamma_{1}}\left(  X\right)  \right)
\label{c38}
\end{split}
\ee where \be \hat{\lambda}=\begin{pmatrix}1&1 \\ 1
&-1\end{pmatrix}, \label{c3800} \ee and we introduced the
$16\times 16$ self-conjugated supermatrices [see \reqs{c115},
\rref{c28}, and footnote after \req{c411}]
\begin{equation}
\hat{\Pi}_{1}=1\text{,  }
\hat{\Pi}_{2}=\hat{\Sigma}_{3}\text{,  }
\hat{\Pi}_{3}=\hat{\Lambda}_{1}\hat{\tau}_{3}\text{,  }
\hat{\Pi}_{4}=\hat{\Lambda}_{1}\hat{\tau}_{3}\hat{\Sigma}_{3}. \label{d7a}%
\end{equation}
The significance of matrices $\hat{\Pi}_{3,4}$ will become clear in
the next subsection.

\begin{subequations}
\label{c38ab} The operators ${\hat{\Gamma}}_i$ here are slight
modification of $\hat{\Gamma}_{t}(\alpha)$ in \reqs{c337},
\rref{b243c}: \be {\hat{\Gamma}}_i(\alpha) =
{\hat{f}}{\hat{\gamma}}_i(\alpha); \label{c38a} \ee where action
of the cut-off operator $\bar{f}$ is defined in \req{b244} and the
operators $\hat{\gamma}_{i}(\alpha)$ are defined by \be
\left[\hat{\gamma}_{i}b\right](X)= \int d\n_1\int\limits_0^1 du_1
{\gamma}_{i}\left(\widehat{\n\n_1};u,u_1\right)
b(\r,\tau,\n_1,u_1). \label{c38b} \ee Hereinafter, the convention
\rref{c110} is used for the angular integration. The kernels in
\req{c38b} are given by
\end{subequations}
\be
\begin{split}
{\gamma}_{1}\left(\widehat{\n\n_1};u,u_1;\alpha\right)
=
- \frac{1}{2}
\left\langle\n\Big|\frac{\mathbb F^\sigma}{\alpha + \mathbb F^\sigma}
\Big|\n_1\right\rangle\equiv {\gamma}_{f}^0;
\\
{\gamma}_{2}\left(\widehat{\n\n_1};u,u_1;\alpha\right)
=
- \frac{1}{2}
\left\langle -\n\Big|\frac{\hat{\mathbb F}^\sigma}{\alpha + \hat{\mathbb F}^\sigma}
\Big|\n_1\right\rangle \equiv  {\gamma}_{b}^0;
\end{split}
\label{c380} \ee and they are independent on the parameters
$u,u_1$. We will see that this
 will change when we consider the fluctuation corrections to the bare
 action \rref{c38}.
As we will see later, the most interesting effects will come from
$\n\simeq\n_1$ so that the notation $\gamma_{f(b)}$ for the
forward (backward) scattering will be self-evident.

The tensor $\varepsilon_{\alpha\beta\gamma}$ is the antisymmetric
tensor of the third rank ($\varepsilon_{123}=1$) and the summation
over repeated indices is implied in Eq. (\ref{c38}). Also, the
relation
\begin{equation}
\varepsilon_{\alpha\beta\gamma}\varepsilon_{\alpha\beta_{1}\gamma_{1}}%
=\delta_{\beta\beta_{1}}\delta_{\gamma\gamma_{1}}-\delta_{\beta\gamma_{1}%
}\delta_{\beta_{1}\gamma} \label{c39}%
\end{equation}
holds.

The term ${\mathcal S}_{3}\left[  \left\{\psi\right\},\alpha\right]  $
describes the cubic interaction
and we write it as
\be
\begin{split}
&{\mathcal S}_{3}\left[  \left\{\psi\right\},\alpha\right]
    =-4\nu\sqrt{2i}\varepsilon_{\delta \beta\gamma}\sum_{i,j=1}^2
 \lambda_{ij} \int dX
\\
&\times
\left(  \overline{\psi}_{\beta}\left(  X\right)
 \hat{\tau}_{3}\hat{\Pi}_j
 \psi_{\gamma}\left(  X\right)  u
\right)  {\hat{\Gamma}}_{i}
\left(
 \overline{\cal F}_{0} \hat{\overline{l}}_{u,\alpha}
 \tau_{3}\hat{\Pi}_j\psi_{\delta}\left(  X\right)
\right) ,
\end{split}
\label{c40}
\ee
where operator $\hat{\overline{l}}_{u,\alpha}$ and supervector ${\cal
  F}_0$ are given by \reqs{c410} and \rref{c41} respectively.

At last, the quadratic term
${\mathcal S}_{2}\left[ \left\{ \psi\right\};\alpha\right]  $ reads
\be
\begin{split}
{\mathcal S}_{2}&\left[ \left\{ \psi\right\};\alpha\right]     =
-2i\nu
\sum_{i,j=1}^2
\lambda_{ij}
\int dX
\\
&\times
\left(
 \overline{\cal F}_{0} \hat{\overline{l}}_{u,\alpha}
 \tau_{3}\hat{\Pi}_j\psi_{\delta}\left(  X\right)
\right)
 {\hat{\Gamma}}_{i}
\left(
 \overline{\cal F}_{0} \hat{\overline{l}}_{u,\alpha}
 \tau_{3}\hat{\Pi}_j\psi_{\delta}\left(  X\right)
\right).
\end{split}
\label{c42}
\ee

Equations \rref{c35}-- \rref{c42}) completely specify the field
theory that describes the collective spin excitations. We will see
that the interaction between the modes given by the terms
${\mathcal S}_{3,4}$, leads in the limit $T\rightarrow0$ to
logarithmically divergent terms of the perturbation theory in
these interactions. These divergencies make the theory non-trivial
and interesting. The logarithmic contributions can be summed up
using a renormalization group (RG) theory. This will be done in
Sec.~\ref{sec5}. Before doing so, however, we will slightly
generalize the action to a form reproducing itself under the
renormalization group procedure.

\subsection{Further generalization of the theory.}

\label{sec4d}

We start with generalization of the quartic interaction by including
all the matrices $\hat{\Pi}_k$, see \req{d7a}, to the interaction part
of the action.
We increase the dimensionality of  matrix $\hat\lambda$ of \req{c3800} as
\be
\hat{\lambda}=\left(
\begin{array}
[c]{cccc}%
1 & 1 & -1 & -1\\
1 & 1 & 1 & 1\\
1 & -1 & 1 & -1\\
1 & -1 & -1 & 1
\end{array}
\right)
\label{d70e}
\ee
with the properties
\begin{equation}
\sum_{i=1}^{4}\lambda_{ik}=4\delta_{k1};
\quad
\sum_{k=1}^{4}\lambda_{ik}=4\delta_{i2}.
 \label{d7e}%
\end{equation}
Then, \req{c38} can be re-written as
\be
\begin{split}
&{\mathcal S}_{4}\left[ \left\{\psi\right\};\alpha\right]   =
-2\nu\varepsilon_{\delta\beta\gamma
}\varepsilon_{\delta\beta_{1}\gamma_{1}}\sum_{i,j=1}^4
\lambda_{ij} \int dX
\\
&\times
\left(  \overline{\psi}_{\beta
}\left(  X\right)  \hat{\tau}_{3}\hat{\Pi}_j\psi_{\gamma}\left(  X\right)  u\right)
{\hat{\Gamma}}_{i}\left(  u\overline{\psi}_{\beta_{1}}\left(
X\right)\hat{\tau}_{3}\hat{\Pi}_j\psi_{\gamma_{1}}\left(  X\right)  \right),
\end{split}
\label{d8} \ee where the operators ${\hat{\Gamma}}_{i}$ are
defined by their action on any function $b(X)$ as [cf.
\reqs{c38ab}] \be
\begin{split}
\left[{\hat{\Gamma}}_{i} b\right](X)&=
\int\!\! d\n_1\int\limits_0^1 \!\!du_1\int\!\! d\r_1\bar{f}(\r_1)
\\
&\times
{\Gamma}_{i}\left(\widehat{\n\n_1};u,u_1;\r_1^\perp\right)
b(\r+\r_1,\tau,\n_1,u_1),
\end{split}
\label{d801} \ee where \be \r_\perp(\n) \equiv \r -\n\left(
\r\cdot\n\right) \label{d802} \ee denotes the coordinate
transverse to the direction of the momentum, the cut-off function
$\bar{f}(\r)$ is defined in \req{a20a},
 and [cf. \req{c380}]
\be
\begin{split}
&{\Gamma}_{1}
\left(\theta;u,u_1;\r_\perp\right)
={\Gamma}_{2}
\left(\theta;u,u_1;\r_\perp\right)
=\gamma_f^0(\theta);
\\
&{\Gamma}_{3}
\left(\theta;u,u_1;\r_\perp\right)
={\Gamma}_{4}
\left(\theta;u,u_1;\r_\perp\right)
=\gamma_b^0(\theta).
\end{split}
\label{d81}
\ee

Analogously, we rewrite \req{c40} as
\be
\begin{split}
&{\mathcal S}_{3}\left[  \left\{\psi\right\},\alpha\right]
    =-2\nu\sqrt{2i}\varepsilon_{ \beta\gamma\delta}\sum_{i,j=1}^4
 \lambda_{ij}\sum_{\sigma=\pm} \int dX
\\
&\times
\left(  \overline{\psi}_{\beta}\left(  X\right)
 \hat{\tau}_{3}\hat{\Pi}_j
 \psi_{\gamma}\left(  X\right)  u
\right)
\hat{{\mathcal B}}_{i}^\sigma
\left(\overline{\mathbb D}_{\sigma}
\tau_{3}\hat{\Pi}_j\psi_{\delta}\left(  X\right)
\right),
\\
&\overline{\mathbb D}_{\pm}\equiv \overline{\cal F}_{0}
\hat{\overline{l}}_{u,\alpha}\hat\tau_{\pm}
; \quad  {\mathbb D}_{\pm}\equiv
\hat\tau_{\mp}
\hat{{l}}_{u,\alpha}{\cal F}_{0}
\end{split}
\label{d13}
\ee
where, similarly to \req{d801}
\be
\begin{split}
\left[{\hat{\mathcal B}}_{i}^\sigma b\right](X)
&=
\int\!\! d\n_1\int\limits_0^1 \!\!du_1\int\!\! d\r_1\bar{f}(\r_1)
\\
&\times
{\mathcal B}_{i}^\sigma\left(\widehat{\n\n_1};u,u_1;\r_1^\perp\right)
b(\r+\r_1,\tau,\n_1,u_1),
\end{split}
\label{d12}
\ee
operator $\hat{\overline{l}}_{u,\alpha}$ and
matrices $\hat{\tau}_{\pm}$ are defined in \req{c410},
and the bare value of the coupling functions are
\be
{\mathcal B}_{i}^\sigma\left(\theta;u,u_1;\r_\perp\right) =
{\Gamma}_{i}\left(\theta;u,u_1;\r_\perp\right).
\label{d14}
\ee

Finally, the quadratic term, Eq. \rref{c42}, is recast as \be
\begin{split}
{\mathcal S}_{2}&\left[ \left\{ \psi\right\};\alpha\right]     =
- i\nu
\sum_{i,j=1}^4
\lambda_{ij}\sum_{\sigma_{1,2}=\pm}
\int dX
\\
&\times
\left(
\overline{\psi_{\delta}}\left(  X\right)\hat{\Pi}_j
 \tau_{3}
{\mathbb D}_{\sigma_1}
\right) \hat{\mathbf \Delta}_i^{\sigma_1\sigma_2}
\left(
 \overline{\mathbb D}_{\sigma_2}\tau_{3}\hat{\Pi}_j\psi_{\delta}\left(  X\right)
\right),
\end{split}
\label{d15} \ee where ${\mathbb D}_{\sigma}$ are defined in
\req{d13}. (The different overall sign in comparison with
\req{c42} appears because the matrix $\hat{\tau}_3$ is
anticonjugate, $\hat{\overline{\tau}}_3=-\hat{\tau}_3$).

Though it appears that four coupling matrices $\hat{\mathbf
  \Delta}_i^{\sigma_1\sigma_2},
\ i=1,2,3,4$ may be present in \req{d15}, only two of them
actually give a non vanishing contribution. Indeed, using
\reqs{d13}, \rref{c41} and  \rref{d7a}, we find
$\hat{\Pi}_3\hat{\mathbb D}\otimes\hat{\overline{\mathbb
    D}}\hat{\Pi}_3=
\hat{\Pi}_1\hat{\mathbb D}\otimes\hat{\overline{\mathbb
    D}}\hat{\Pi}_1
$,
$\hat{\Pi}_2\hat{\mathbb D}\otimes\hat{\overline{\mathbb
    D}}\hat{\Pi}_2=
\hat{\Pi}_4\hat{\mathbb D}\otimes\hat{\overline{\mathbb
    D}}\hat{\Pi}_4
$, which leads using \req{d70e}, to \be \sum_{k=1}^4 \lambda_{ik}
\hat{\Pi}_k\hat{\mathbb D}\otimes\hat{\overline{\mathbb
    D}}\hat{\Pi}_k=
0, \quad  i=1,4.
\label{d150a}
\ee


The non-vanishing operators $\hat{\mathbf
\Delta}_i^{\sigma_1\sigma_2}$, $i=2,3$, in \req{d15} are defined
as [cf. \reqs{d801}, \rref{d12}] \be
\begin{split}
&\left[{\hat{\mathbf \Delta}}_{i}^{\sigma_1\sigma_2} b\right](X)=
\int\!\! d\n_1\int\limits_0^1 \!\!du_1\int\!\! d\r_1\bar{f}(\r_1)
\\
&\quad\times
{\mathbf
  \Delta}_{i}^{\sigma_1\sigma_2}\left(\widehat{\n\n_1};u,u_1;\r_1^\perp\right)
b(\r+\r_1,\tau,\n_1,u_1)
.
\end{split}
\label{d15a}
\ee
Equations \rref{d15}--\rref{d15a} reproduce \req{c42} for
\be
{\mathbf \Delta}_{i}^{\sigma_1\sigma_2}
\left(\theta;u,u_1;\r_\perp\right)
=\Gamma_i\left(\theta;u,u_1;\r_\perp\right) ,
\quad i=2,3;
\label{d16}
\ee

We will see that the form of the action \rref{d8}, \rref{d13}, \rref{d15}
 for $\alpha=1/2$
will be reproduced by the renormalization group but the relation
between constants \rref{d81}, \rref{d14}, \rref{d16} will be
violated, so that \reqs{d81}, \rref{d14}, \rref{d16} will serve as
initial conditions for the renormalization group flow of
Sec.~\ref{sec5}.

The reason for the introduction of the $\Pi_k$ matrices \rref{d7a}
into the definition of the interaction actions is that they
separate the combination of the supervectors which may transform
the free Lagrangian $\hat{\mathcal L}_{0}$, Eq.~(\ref{c261}), into
$\hat{\mathcal L}_{0}^\dagger$. We will in the next section that
it will be a necessary condition to give rise to the logarithmic
divergence.

To understand such partition, note that
any supermatrix $\hat{P}$ can be represented as
\begin{equation}
\hat{P}=\sum_{i=1}^{4}\hat{P}^{\left(  i\right)  } \label{d7b}%
\end{equation}
where $ \hat P^{\left(  i\right)  }$ are supermatrices such as
\begin{align}
\left[  \hat P^{\left(  1\right)  }, \hat \Sigma_{3}\right]   &  =0,\text{ }\left\{
\hat P^{\left(  1\right)  }, \hat\tau_{3} \hat\Lambda_{1}\right\}  =0\label{d7c}\\
\left[ \hat P^{\left(  2\right)  }, \hat\Sigma_{3}\right]   &  =0,\text{ }\left[
\hat P^{\left(  2\right)  },\tau_{3} \hat\Lambda_{1}\right]  =0\nonumber\\
\left\{  \hat P^{\left(  3\right)  }, \hat\Sigma_{3}\right\}   &  =0,\text{ }\left[
 \hat P^{\left(  3\right)  },\tau_{3} \hat\Lambda_{1}\right]  =0\nonumber\\
\left\{  \hat P^{\left(  4\right)  }, \hat\Sigma_{3}\right\}   &  =0,\text{ }\left\{
 \hat P^{\left(  4\right)  },\tau_{3} \hat\Lambda_{1}\right\}  =0\nonumber
\end{align}
where $\left[\dots  ,\dots\right]  $ stands for the commutator,
$\left\{ \dots , \dots \right\} $ for the anticommutator, and the
relevant supermatrices are defined in \req{c28}.

It is not difficult to invert Eq.~(\ref{d7b}):
\begin{equation}
\hat P^{\left(  i\right)
}=\frac{1}{4}\sum_{k=1}^{4}\lambda_{ik}\hat\Pi_{k}\hat P\hat\Pi_{k},
\label{d7d}%
\end{equation}
where the $4\times4$ matrix $\hat \lambda$ is given by \req{d70e}.
Eq. (\ref{d7d}) can easily be checked by using the property
\rref{d7e} and the commutation relations of the matrices $\Pi_k$.

One can see that
\begin{align}
{\Str}\left(  \hat P^{\left(  i\right)  }\hat P^{\left(  j\right)  }\right)   &
=\delta_{ij}{\Str}\left( \hat P^{\left(  i\right)  }\right)  ^{2}\text{, }\nonumber\\
{\Str}\left(  \hat P\hat Q\right)   &  =\sum_{k}
{\Str}\left(  \hat P^{\left(
      k\right)  }
\hat Q^{\left(
k\right)  }\right)  \label{d7f}%
\end{align}
for arbitrary supermatrices $\hat P$ and $\hat Q$,
where the supertrace operation is defined\cite{book} as
\be
\Str \begin{pmatrix}
\hat{a} & \hat{\rho}\\
\hat{\sigma} & \hat{b}
\end{pmatrix}_g= \mathrm{Tr}\,\hat{a} - \mathrm{Tr} \,\hat{b}.
\label{d7f0}
\ee
We note that the following very useful relation
\begin{align}
&  \sum_{i_{1},i_{2},k_{1},k_{2}=1}^{4}a_{i_{1}}b_{i_{2}}\lambda_{i_{1}k_{1}%
}\lambda_{i_{2}k_{2}}
\Str\left(  \hat A \hat \Pi_{k_{1}}
\hat{\Pi}_{k_{2}}\hat{B}
\hat\Pi_{k_{1}}\hat{\Pi}_{k_{2}}
\right)
\nonumber\\
&  =4\sum_{i,k=1}^{4}a_{i}b_{i}\lambda_{ik}
\Str
\left( \hat A\hat\Pi_{k}\hat B\hat\Pi_{k}\right)
\label{d7h}%
\end{align}
is valid for arbitrary coefficients $a_{i}$ and $b_{i}$.
One can prove \req{d7h} by a direct calculation
using the fact that a product of two matrices \rref{d7a}
is once again one of the matrices \rref{d7a}.
Finally, combining relations \rref{d7d} and \rref{d7h},
one finds
\be
\Str \left(\hat{P}^{\left(  i\right)  }\hat{Q}^{\left(  j\right)  }\right)
=\frac{\delta_{ij}}{4}
\sum_{k=1}^{4}\lambda_{ik}\Str\left(  \hat P \hat \Pi_{k}
\hat Q
\hat\Pi_{k}
\right)
\label{d7g}
\ee

\section{Perturbation theory and renormalization group.}
\label{sec5}

This section contains the perturbative analysis of the field
theory derived in the previous section. We will start in
Sec.~\ref{sec5a} with the brief formulation of the rules of the
diagrammatic technique\cite{agd} emphasizing the aspects different
from the conventional models. We will show the origin of the
logarithmic divergence in Sec.~\ref{sec5b} and formulate the
renormalization group (RG) procedure of summation the leading
logarithmic series in Sec.~\ref{sec5c}. We will be able to derive
the RG equation for the coupling constant \reqs{d8}, \rref{d13},
\rref{d15}. The solution of the RG equation will be done in
Sec.~\ref{sec6}.

\subsection{Rules of perturbation theory.}
\label{sec5a}

As usual, we would like to construct the expansion of the
observable quantities in terms of the interaction vertices
produced by \reqs{d8}, \rref{d13}, \rref{d15} and the Green
functions (in our case supermatrices) of the free motion \be
\hat{\mathcal G}_{0}\left( X_1,X_2\right) = - 4i\nu\left\langle
\bpsi\left(X_1\right) \otimes
\overline{\bpsi}(X_2)\right\rangle_{0} \label{d100} \ee where
variables $X$ are defined in \req{c250}, and the averaging means
\be \langle \dots \rangle_0=\int \dots\ \exp\left(  -{\mathcal
S}_0\left[  \psi\right]  \right) D\psi, \label{d101} \ee the free
action is defined in \req{c37}, and the normalization is trivial
due to the supersymmetry:
\[
\int
\exp\left(  -{\mathcal S}_0\left[  \psi\right]  \right) D\psi=1.
\]
The factor $- 4i\nu$ is introduced in \req{d101} for the sake of
convenience. All the higher order averages are, then, to be found
using the Wick theorem and \req{d100}.

Using \reqs{c37}, \rref{c261}, one easily finds
\be
\begin{split}
&\hat{\mathcal G}_{0}\left( X_1,X_2\right)
=\delta(u_1-u_2)\delta\left(\n_1;\n_2\right)
\\
&\quad\times
T\sum_{\w_n}
\int\frac{d^dk}{\left(2\pi\right)^{d}}
e^{-i\w_n(\tau_1-\tau_2)+i\k(\r_1-\r_2)}
 \hat{G}_{0}\left(\w_n,\k;\n_1\right);
\\
&\hat{G}_{0}\left(\w,\k;\n\right)= \frac{1}{i\omega\hat{\Lambda}_{1}+v_{F}\k\n
\hat{\tau}_{3}\hat{\Sigma}_{3}-i\delta \hat{\Lambda}}
\\
&\quad\quad\quad\quad =\frac{
-i\omega\hat{\Lambda}_{1}+
v_{F}\k\n
\hat{\tau}_{3}\hat{\Sigma}_{3}+i\delta\hat{\Lambda}
}
{\omega^2+v_F^2(\k\n)^2+\delta^2}.
\end{split}
\label{d1} \ee From \req{d1} we see that terms involving $\delta
\to 0$ are dangerous only for zero Matsubara frequency
contribution. These contributions are associated with the real
scattering event with the energy transfer much smaller than
temperature. Such processes, though determining the kinetic of the
system, are not interesting for equilibrium thermodynamics, and
will be considered elsewhere\cite{elsewhere}. For the logarithmic
contributions considered further in this paper, the real processes
are not important and we will put $\delta=0$ from now on.

Let us discuss peculiar features of the perturbation theory.
First of all, due to the supersymmetry of ${\mathcal S}_0$, the averages of the
operators $\hat{A}_1$ not perturbing the supersymmetry vanish
 \be
 \left\langle \left(\overline{\bpsi}\hat{A} {\bpsi}\right)\right\rangle_{0}
 =0;
\quad\left\langle
\left(\overline{\bpsi}\hat{A}_1 {\bpsi}\right)
 \left(\overline{\bpsi}\hat{A}_2 {\bpsi}\right)
 \right\rangle_{0}=0;\ \dots.
 \label{d102}
 \ee
This leads to the cancellation of the closed loop contributions,
see {\em e.g.} Fig.~\ref{fig4}b).

Second feature originates from the dependence of the supervectors
$\bpsi,\ \overline{\bpsi}$ on each other, see \req{c23}. This
mutual dependence makes the rules of the Wick contractions of the
supervectors very similar to ones for the real fields (so that the
arrow in the Green function looses its meaning). To illustrate
this point, consider the connected average ($\hat{Q}_{1,2}$ are
the arbitrary self-conjugated supermatrices,
$\hat{Q}_{1,2}=\hat{\overline{Q}}_{1,2}$ ):
\begin{subequations}
\label{d103}
\bea
\setlength{\unitlength}{2.3em}
\begin{picture}(5,0.5)(2,0)
\put(0,0){\mbox{$
\left\langle
\left(\overline{\bpsi}\hat{Q}_1\bpsi\right)
\left(\overline{\bpsi}\hat{Q}_2\bpsi\right)
\right\rangle_0
=
\big(\overline{\bpsi}\,\hat{Q}_1\,\bpsi\big)
\big(\overline{\bpsi}\,\hat{Q}_2\,\bpsi\big)$}}
\put(5.3,-0.31){\line(1,0){2.8}}
\put(5.3,-0.31){\line(0,1){0.13}}
\put(8.1,-0.31){\line(0,1){0.13}}
\put(6.3,0.57){\line(1,0){0.8}}
\put(6.3,0.57){\line(0,-1){0.13}}
\put(7.1,0.57){\line(0,-1){0.13}}
\end{picture}
\label{d103a}
\\
\setlength{\unitlength}{2.3em}
\begin{picture}(5,1)(-2.4,0)
\put(0,0){\mbox{$+\big(\overline{\bpsi}\,\hat{Q}_1\,\bpsi\big)
\big(\overline{\bpsi}\,\hat{Q}_2\,\bpsi\big),$}}
\put(0.6,-0.31){\line(1,0){1.8}}
\put(0.6,-0.31){\line(0,1){0.13}}
\put(2.4,-0.31){\line(0,1){0.13}}
\put(1.7,0.57){\line(1,0){1.8}}
\put(1.7,0.57){\line(0,-1){0.13}}
\put(3.5,0.57){\line(0,-1){0.13}}
\end{picture}
\label{d103b}
\\
\nonumber \eea where over- and underbrackets stand for the Wick
contractions. The line \rref{d103b} can be transformed with the
help of \req{c24b} as
\[
\rref{d103b}=
\setlength{\unitlength}{2.3em}
\begin{picture}(5,0.3)(-0.1,0)
\put(0,0){\mbox{$\big(\overline{\bpsi}\,{\hat{\overline{Q}}}_1\,\bpsi\big)
\big(\overline{\bpsi}\,\hat{Q}_2\,\bpsi\big),$}}
\put(1.4,-0.31){\line(1,0){0.7}}
\put(1.4,-0.31){\line(0,1){0.13}}
\put(2.1,-0.31){\line(0,1){0.13}}
\put(0.3,0.57){\line(1,0){2.8}}
\put(0.3,0.57){\line(0,-1){0.13}}
\put(3.1,0.57){\line(0,-1){0.13}}
\end{picture}
\]
which coincides with the first term in the right-hand-side of
\req{d103},
because $\hat{Q}_1$ is self-conjugate.
\end{subequations}
As the result, we obtain \be \left\langle
\left(\overline{\bpsi}\hat{Q}_1\bpsi\right)
\left(\overline{\bpsi}\hat{Q}_2\bpsi\right) \right\rangle_0 =-2\
\Str \left[\hat{Q}_1{\cal \hat{G}}_0\hat{Q}_2{\cal
\hat{G}}_0\right], \label{d104} \ee where we omitted the trivial
factors of proportionality between the averages and the Green
function, \req{d100}. The appearance of the factor of $2$ in such
an average is the feature of the real fields. The minus sign in
\req{d104} originates from the definition of the supertrace
operation, Eq. \rref{d7f0}, where the commuting sector is taken
with the negative sign.

To further utilize the analogy with the real fields, let us
consider a connected average involving eight fields but with only
four fields contracted (such an averaging appears {\em e.g.} as a
correction to the interaction constant for the quartic scattering
term) \bea \setlength{\unitlength}{2.3em}
\begin{picture}(5,1)(2.1,0)
\put(0,0){\mbox{${\cal
      J}_1=
\left[\big(\overline{\bpsi}\,{\hat{{Q}}}_1\,\bpsi\big)
\big(\overline{\bpsi}\,{\hat{{Q}}}_1\,\bpsi\big)\right]
\left[\big(\overline{\bpsi}\,\hat{Q}_2\,\bpsi\big)
\big(\overline{\bpsi}\,\hat{Q}_2\,\bpsi\big)\right];
$}}
\put(3.4,-0.31){\line(1,0){3.2}}
\put(3.4,-0.31){\line(0,1){0.13}}
\put(6.6,-0.31){\line(0,1){0.13}}
\put(2.7,0.57){\line(1,0){4.7}}
\put(2.7,0.57){\line(0,-1){0.13}}
\put(7.4,0.57){\line(0,-1){0.13}}.
\end{picture}
\nonumber\\
\setlength{\unitlength}{2.3em}
\begin{picture}(5,1)(2.1,0)
\put(0,0){\mbox{${\cal
      J}_2=
\left[\big(\overline{\bpsi}\,{\hat{{Q}}}_1\,\bpsi\big)
\big(\overline{\bpsi}\,{\hat{{Q}}}_1\,\bpsi\big)\right]
\left[\big(\overline{\bpsi}\,\hat{Q}_2\,\bpsi\big)
\big(\overline{\bpsi}\,\hat{Q}_2\,\bpsi\big)\right];
$}}
\put(4.4,-0.31){\line(1,0){1.2}}
\put(4.4,-0.31){\line(0,1){0.13}}
\put(5.6,-0.31){\line(0,1){0.13}}
\put(2.7,0.57){\line(1,0){4.7}}
\put(2.7,0.57){\line(0,-1){0.13}}
\put(7.4,0.57){\line(0,-1){0.13}},
\end{picture}
\nonumber\\
\label{d105} \eea If  $\bpsi$ were usual fermionic field,
contributions ${\cal J}_1$ and ${\cal J}_2$ would be responsible
for  different processes (particle-hole and particle-particle,
respectively). For the superfields, however, we can transform
${\cal J}_2$ [cf.  derivation of \req{d104}] and thus obtain
\[
\setlength{\unitlength}{2.3em}
\begin{picture}(5,1)(2.5,0)
\put(0,0){\mbox{${\cal
      J}_2=
\left[\big(\overline{\bpsi}\,{\hat{{Q}}}_1\,\bpsi\big)
\big(\overline{\bpsi}\,{\hat{\overline{Q}}}_1\,\bpsi\big)\right]
\left[\big(\overline{\bpsi}\,\hat{\overline{Q}}_2\,\bpsi\big)
\big(\overline{\bpsi}\,\hat{Q}_2\,\bpsi\big)\right]
={\cal J}_1;
$}}
\put(3.4,-0.31){\line(1,0){3.2}}
\put(3.4,-0.31){\line(0,1){0.13}}
\put(6.6,-0.31){\line(0,1){0.13}}
\put(2.7,0.57){\line(1,0){4.7}}
\put(2.7,0.57){\line(0,-1){0.13}}
\put(7.4,0.57){\line(0,-1){0.13}},
\end{picture}
\]
i.e. the construction ${\cal J}_2$ describes precisely the same
contributions. Analogously one finds \bea
\setlength{\unitlength}{2.3em}
\begin{picture}(5,1)(2,0)
\put(0,0){\mbox{$
\left[\big(\overline{\bpsi}\,{\hat{{Q}}}_1\,\bpsi\big)
\big(\overline{\bpsi}\,{\hat{{Q}}}_1\,\bpsi\big)\right]
\left[\big(\overline{\bpsi}\,\hat{{Q}}_2\,\bpsi\big)
\big(\overline{\bpsi}\,\hat{Q}_2\,\bpsi\big)\right]
={\cal J}_1;
$}}
\put(3.4,-0.31){\line(1,0){2.2}}
\put(3.4,-0.31){\line(0,1){0.13}}
\put(5.6,-0.31){\line(0,1){0.13}}
\put(1.5,0.57){\line(1,0){5.9}}
\put(1.5,0.57){\line(0,-1){0.13}}
\put(7.4,0.57){\line(0,-1){0.13}}
\end{picture}
\nonumber\\
\setlength{\unitlength}{2.3em}
\begin{picture}(5,1)(2,0)
\put(0,0){\mbox{$
\left[\big(\overline{\bpsi}\,{\hat{{Q}}}_1\,\bpsi\big)
\big(\overline{\bpsi}\,{\hat{{Q}}}_1\,\bpsi\big)\right]
\left[\big(\overline{\bpsi}\,\hat{{Q}}_2\,\bpsi\big)
\big(\overline{\bpsi}\,\hat{Q}_2\,\bpsi\big)\right]
={\cal J}_1;
$}}
\put(0.6,0.57){\line(1,0){5.8}}
\put(0.6,0.57){\line(0,-1){0.13}}
\put(6.4,0.57){\line(0,-1){0.13}}
\put(2.3,-0.31){\line(1,0){2.2}}
\put(2.3,-0.31){\line(0,1){0.13}}
\put(4.5,-0.31){\line(0,1){0.13}}
\end{picture}
\nonumber \eea so that the permutations of the fields in the
vertices do not give rise to any new effects but simply lead to
the multiplication by a factor of $4$ -- number of trivial
symmetries of the interaction vertex. That decrease of the number
of different contractions is a great simplification in a further
derivation.

\begin{figure}[t]
\setlength{\unitlength}{2.3em}
\begin{picture}(10,16)
  \put(0,15) 
   {
    \begin{picture}(10,1)
      \put(0.2,0)
      {\includegraphics[width=3\unitlength
        ]{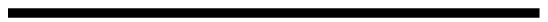}$\ =\ \frac{i}{4\nu}\hat{G}_0\left(\w,\k,\n\right)$}
      \put(1.1, 0.4){$\w,k,n,u$}
    \end{picture}
  }

\put(0.5,12.2) 
{
  \begin{picture}(10,3)
    \put(0.5,0){\includegraphics[width=2\unitlength]{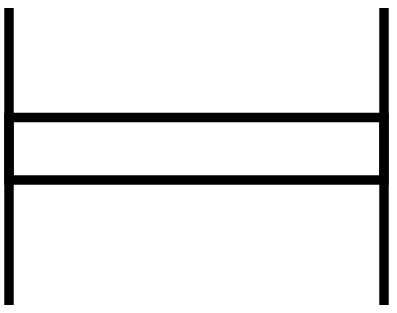}}
    \put(0,-0.3){$\beta$}\put(2.7,-0.3){$\beta_1,\k+\q$}
    \put(0,1.7){$\gamma$}\put(2.7,1.7){$\gamma_1,\k$}
    \put(0,0.9){$u_1$}\put(2.7,0,9){$u_2$}

  \end{picture}
}

\put(0,10.5)
{\begin{picture}(10,3)
    \put(5.6,0.4)
    {
      \begin{picture}(1,1)
        \put(1,1){\line(0,-1){0.5}}
        \put(1,0,5){\line(-1,-2){0.3}}
      \end{picture}
    }

\put(5.6,0){$\hat{\tau}_{3}\hat{\Pi}_j$}
    \put(5.1,-1.1)
    {
      \begin{picture}(1,1)
        \put(0,0){\line(0,1){0.5}}
        \put(0,0,5){\line(1,2){0.3}}
      \end{picture}
    }


 \put(8.6,0.4)
    {
      \begin{picture}(1,1)
        \put(1,1){\line(0,-1){0.5}}
        \put(1,0,5){\line(-1,-2){0.3}}
      \end{picture}
    }

\put(8.6,0){$\hat{\tau}_{3}\hat{\Pi}_j$}
    \put(8.1,-1.1)
    {
      \begin{picture}(1,1)
        \put(0,0){\line(0,1){0.5}}
        \put(0,0,5){\line(1,2){0.3}}
      \end{picture}
    }


\put(-0.7,0)
 {$
 \displaystyle{=[8]\times 2\nu f(q)\varepsilon_{\delta\beta\gamma
 }\varepsilon_{\delta\beta_{1}\gamma_{1}}\sum_{i,j=1}^4
 \lambda_{ij}
 }
 $}

\put(6.6,0){$\cdot\left(u_1{\hat{\Gamma}}_{i}u_2\right)\cdot$}

\end{picture}
}

\put(0.5,5.8) 
{
  \begin{picture}(10,3)
    \put(0.5,0){\includegraphics[width=2.4\unitlength]{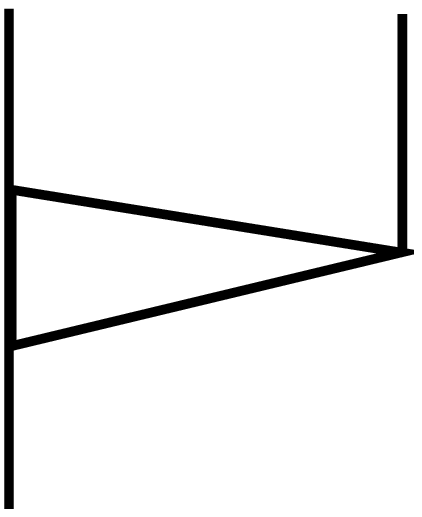}}
    \put(0,0.){$\beta$}
    \put(0,2.7){$\gamma$}\put(3.2,2.7){$\delta$}
    \put(0,1.3){$u_1$}\put(3.2,2,3){$u_2$}
    \put(0,0.9){$\k$}\put(3.2,1,9){$\q$}
  \end{picture}
}

\put(0,4.9)
{\begin{picture}(10,3)
    \put(5.6,0.4)
    {
      \begin{picture}(1,1)
        \put(1,1){\line(0,-1){0.5}}
        \put(1,0,5){\line(-1,-2){0.3}}
      \end{picture}
    }

\put(5.5,0){$\hat{\tau}_{3}\hat{\Pi}_j$}
    \put(5.1,-1.1)
    {
      \begin{picture}(1,1)
        \put(0,0){\line(0,1){0.5}}
        \put(0,0,5){\line(1,2){0.3}}
      \end{picture}
    }


 \put(8.9,0.2)
    {
      \begin{picture}(1,1)
        \put(1,1){\line(0,-1){0.5}}
        \put(1,0,5){\line(-1,-2){0.3}}
      \end{picture}
    }

\put(8.3,0){$
\overline{\mathbb D}_{\sigma}
\hat{\tau}_{3}\hat{\Pi}_j
$}


\put(-1.2,0)
 {$
 \displaystyle{=[2]\times
2\nu f(q)\sqrt{2i}\varepsilon_{ \beta\gamma\delta}\!\!\sum_{i,j=1}^4
\sum_{\sigma=\pm}
 \lambda_{ij}
 }
 $}

\put(6.6,0){$\!\cdot\!\left(u_1{\hat{\mathcal B}}_{i}^\sigma\right)\!\!\cdot$}

\end{picture}
}

\put(1.5,2.5) 
{
  \begin{picture}(10,3)
    \put(1.9,0){\includegraphics[width=2.4\unitlength]{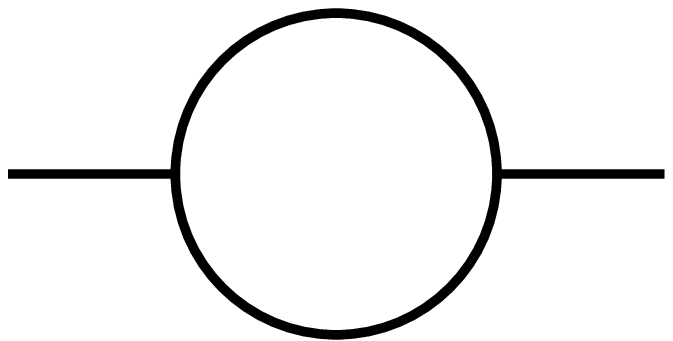}}
    \put(0,0.5){$u,\k, \n, \gamma$}\put(4.6,0.5){$u_1,\k,\n_1, \gamma$}
  \end{picture}
}

\put(0.7,0.9)
{\begin{picture}(10,3)



\put(-1.2,0)
 {$
 \displaystyle{=[2]\times
i\nu f(k)
\sum_{\sigma_{1,2}=\pm}
 \lambda_{ij}
 }
 $}

\put(3.5,0.2){\line(1,0){0.5}}
\put(8.6,0.2){\line(1,0){0.5}}
\put(4.1,0){$\hat{\Pi}_j
 \tau_{3}
{\mathbb D}_{\sigma_1} \hat{\mathbf \Delta}_i^{\sigma_1\sigma_2}
 \overline{\mathbb D}_{\sigma_2}\tau_{3}\hat{\Pi}_j
$}

\end{picture}
}
\end{picture}

\caption{Basic element for the diagrammatic technique,
External legs of the vertex are amputated.
The expression for the interaction vertices are
found from \reqs{d8}, \rref{d13}, \rref{d15}, and the
additional symmetry factors (in brackets) are discussed in text.
The lines in the analytic expressions for the diagrams indicate the
directions of the insertion of the Green function $G_0$.
Expressions for ${\mathbb D}_{\pm}$ are  obtained from \reqs{d13}
and \rref{c410} with
${\cal L}_0=i\w\hat{\Lambda}_1+v_F(\k\n)\hat{\Sigma}_3\hat{\tau}_3$,
${\cal L}_0^\dagger=-i\w\hat{\Lambda}_1+v_F(\k\n)\hat{\Sigma}_3\hat{\tau}_3$.
Cut-off function $f$ is defined in \req{a7c}.
}

\label{fig201}
\end{figure}

This observation enable us to formulate the simple diagrammatic
rules for the generation of the perturbation expansion,
see Fig.~\ref{fig201}.

As usual the summation of the non-fixed
by the external legs or conservation laws
coordinates $\w_i,\k_i,\n_i,u_i$,  must be performed, and the
supermatrix product to be calculated.
 The closed
loop of the supersymmetric Green functions bring $-\Str$ of the
product of all the terms in the loop, [cf. comment after \req{d104}].

\subsection{Identification of logarithmic divergence.}
\label{sec5b}

Having established the basic rules of the diagrammatic technique
we are ready to demonstrate the logarithmic singularity in any
dimensions.

\begin{figure}[ht]
\setlength{\unitlength}{2.3em}
\begin{picture}(10,7,5)
\put(1.8,1)
{
\includegraphics[width=6\unitlength]{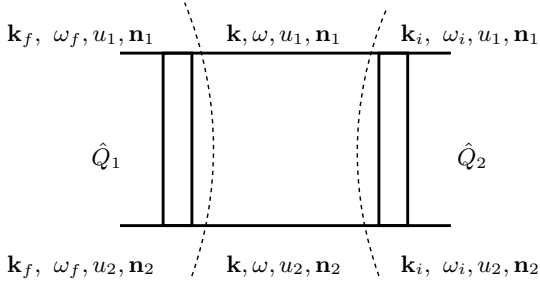}
}
\put(0,1.1){$\k_f,\ \w_f, u_2,\n_2$} \put(7,1.1){$\k_i,\ \w_i,
  u_2,\n_2$}
\put(3.9,1.1){$\k,\w,u_2,\n_2$}

\put(1.5,3){$\hat{Q}_1$} \put(8,3){$\hat{Q}_2$}

\put(0,5.2){$\k_f,\ \w_f, u_1,\n_1$}
\put(7,5.2){$\k_i,\ \w_i, u_1,\n_1$}
\put(3.9,5.2){$\k,\w,u_1,\n_1$}
\end{picture}

\caption{Lowest order correction
to the quartic interaction vertex (external legs are amputated).
This contribution logarithmically
diverge at $\ T\to 0, \n_1\to \n_2$, independently on dimensionality.
Objects separated by dotted arcs are named $Q_{1,2}$ in \req{d2}.
}
\label{fig3}%
\end{figure}

As usual in the logarithmic series, one has to look at the
perturbation theory for the interaction vertices. The lowest order
diagram of interest is shown on Fig.~\ref{fig3} and all the
remaining terms are enumerated and discussed in Sec.~\ref{sec5c}.

According to the rules of the diagrammatic technique,
Fig.~\ref{fig201}, the term of  Fig.~\ref{fig3}, (let us denote it
by ${\cal J}_5$) can schematically be presented as \be
\begin{split}
{\cal J}_5^{ij}=&-\frac{4T}{\nu}\sum_{\w_n\neq 0}
\int \frac{d^d\k}{\left(2\pi\right)^d}
f(\k-\k_f)f(\k-\k_i)
\\
&\times \gamma_i(\widehat{\n_1\n_2})\gamma_j(\widehat{\n_1\n_2})
\\
&\times \Str(\hat{\tau}_3
\hat{Q}_1\hat{G}_{0}\left(\w_n,\k;\n_1\right)\hat{\tau}_3\hat{Q}_2
\hat{G}_{0}\left(\w_n,\k;\n_2\right)),
\end{split}
\label{d2} \ee where the supermatrices $\hat{Q}_{1,2}$ indicate
all the combinations of the supermatrices and direct products of
the supervectors standing on the left and on the right of the
Green functions, see Fig.~\ref{fig3}. The only important point is
that we will neglect their momentum dependence, which suffices our
aim for the logarithmic accuracy. The extra matrix $\hat{\tau}_3$
and the numerical coefficient in front are introduced for the
convenience.

Substituting \req{d1} into \req{d2}, and keeping only non-vanishing
terms, we find
\be
\begin{split}
&{\cal J}_5^{ij}=-
\frac{4T}{\nu}\sum_{\w_n\neq 0}
\int \frac{d^d\k}{\left(2\pi\right)^d}
f(\k-\k_f)f(\k-\k_i)
\\
&\times
\frac{\gamma_i(\widehat{\n_1\n_2})\gamma_j((\widehat{\n_1\n_2}))}
{\left(\omega^2_n+v_F^2(k^\parallel_1)^2\right)
\left(\omega^2_n+v_F^2(k^\parallel_2)^2\right)}
\\
&\times
\left(v_F^2k^\parallel_1 k^\parallel_2\Str
\hat{Q}_1\hat{\Sigma}_3\hat{Q}_2
\hat{\Sigma}_3
-\w^2_n
\Str
\hat{Q}_1\hat{\tau}_3\hat{\Lambda}_1\hat{\tau}_3\hat{Q}_2\hat{\Lambda}_1
\right);
\\
& k^\parallel_{1,2}\equiv \k\cdot \n_{1,2}.
\end{split}
\label{d3}
\ee

Direct examination of \req{d3} shows that the integral can exhibit
logarithmic divergence for $\n_1\to \n_2$ only if the two terms in
the last factor have the same sign. The last condition is
conveniently taken care of by representing each $Q$ matrix in the
form of Eq.\rref{d7b}, and using the relations \rref{d7c}. To
facilitate further manipulations, we introduce \be
\begin{split}
& \n=(\n_1+\n_2)/{2};\quad
\delta\n=\n_1-\n_2;\\
&k_\parallel=\k\cdot\n; \quad \k_\perp=\k-k_\parallel \n,
\end{split}
\label{d300} \ee and consider $|\delta\n|\ll 1$. After integration
over $k_\parallel$ in \req{d3}, one finds [using the decomposition
\rref{d7b}] \be
\begin{split}
&{\cal J}_5^{ij}={\gamma_i(|\delta\n|)\gamma_j(|\delta\n|)}
\Str\left[
\hat{Q}_1^{(3)}\hat{Q}_2^{(3)}-\hat{Q}_1^{(1)}\hat{Q}_2^{(1)}\right]
\\
&\times
\frac{4T}{ \nu v_F}\sum_{\w_n>0}^{[v_F/r_0]}
\int \frac{d^{d-1}\k_\perp }{\left(2\pi\right)^{d-1}}
\frac{4\omega_nf(\k_\perp-\k_f)f(\k_\perp-\k_i)}
{4\omega^2_n+v_F^2(\delta\n\k_\perp)^2},
\end{split}
\label{d301} \ee where the upper limit on the frequency summation
appears because we neglected $k_\parallel$ dependence of the
cut-off function $f$ and the typical momenta contributing to the
integral are of the order of $\w_n/v_F$.

Summation over Matsubara frequency leads immediately to a
logarithmic result independently on the dimensionality of the
system: \be
\begin{split}
&{\cal J}_5^{ij}={\gamma_i(|\delta\n|)\gamma_j(|\delta\n|)}
\Str\left[
\hat{Q}_1^{(3)}\hat{Q}_2^{(3)}
-\hat{Q}_1^{(1)}\hat{Q}_2^{(1)}\right]
\\
&\quad\quad\quad\times f^{(2)}(\k_f;\k_i)
\ln\left(\frac{v_F}{r_0\tilde{\epsilon}}\right)
,
\end{split}
\label{d6} \ee where the infrared cut-off of the logarithm is
determined by \be \tilde{\varepsilon}=\max\left\{  T,\left\vert
\delta{n}\right\vert v_{F}r_{0}^{-1}\right\}  \ll \frac{v_F}{r_0}.
\label{d6a} \ee
 The supertrace in \req{d6} can be re-expressed
using \req{d7g} as
\be
\begin{split}
&\Str\left[
\hat{Q}_1^{(3)}\hat{Q}_2^{(3)}
-\hat{Q}_1^{(1)}\hat{Q}_2^{(1)}\right]
\\
&=-\frac{1}{4}\sum_{i=1,3}
\sum_{k=1}^4\left(-1\right)^{\frac{i-1}{2}}\lambda_{ik}
\Str\left[
\hat{Q}_1\hat{\Pi}_k\hat{Q}_2\hat{\Pi}_k\right].
\end{split}
\label{d6000}
\ee

Although the logarithmic divergence \rref{d6} is present in any
dimension, the coefficient in front of the logarithm in \req{d6}
is determined by the dimensionality and  the details of the
ultraviolet cutoff: \be f^{(2)}(\k_f;\k_i)= \frac{2}{  \pi \nu
v_F} \int \frac{d^{d-1}\k_\perp }{\left(2\pi\right)^{d-1}}
f(\k_\perp-\k_f)f(\k_\perp-\k_i), \label{d60} \ee
where the cut-off function $f(\k)$ is given by \req{a7c}. For
$d=1$ there is no integration over the transverse momentum, $f(k)$
cuts the  logarithmic divergence only, so
\begin{subequations}
\label{d27} \be f^{(2)}_{d=1} =\mu_1=2 \label{d27a} \ee for
$\k_{f,i} \lesssim (1/r_0)$ and decreases rapidly for the larger
momenta. For $d=2,3$ we notice that logarithmic contributions
originate from the region $|k_\parallel |\ll |\k_\perp| \ll
1/r_0$, and this feature will persist in all the further terms of
the perturbation theory. Neglecting the parallel components in
\req{d60}, we find \be
\begin{split}
&f^{(2)}(\k_f;\k_i) =
\mu_d
\int \frac{d^{d-1}r}{(r_0)^{d-1}}e^{i(\k_i-\k_f)\r}
\left[\bar{f}_\perp\left(\frac{|\r|}{r_0}\right)\right]^2;
\\
&\mu_2  =4
\left(  p_{F}r_{0}\right)  ^{-1}; \quad \mu_3=4\pi\left(
  p_{F}r_{0}\right)  ^{-2};
\\
&\bar{f}_\perp\left(\frac{|\r|}{r_0}\right)=r_0^{d-1}
\int\frac{d^{d-1}\k_\perp}{(2\pi)^{d-1}}{e^{i\k_\perp\r}}f(\k),
\end{split}
\label{d27b} \ee where the coordinate integration is in the plane
$\r\cdot\n=0$, The significance of $\bar{f}_\perp$ is the
regularization of the fields that are otherwise singular functions
of the transverse coordinates.
\end{subequations}

Equation (\ref{d6}) demonstrates that the field theory under  study is
logarithmic in any dimensions. The corrections coming from the interaction
diverge in the limit $\left\{  T,\left|  \delta\n\right|  \right\}
\rightarrow 0$. Therefore, we can use a renormalization group scheme for
calculation of physical quantities. This can be done in the limit of small
$\Gamma$ considered here.

A remarkable feature of the corrections is that the main
contribution comes from configurations with either parallel or
antiparallel alignment of the vectors $\n$. To some extent, the
spin degrees of freedom of the electron system have a tendency to
forming a one dimensional structure and this happens in all
dimensions.

\subsection{Integration over fast variables}
\label{sec5c}

This subsection is devoted to summation of the perturbation series
in the leading logarithmic approximation.
It means that the expansion for a physical quantity $y$ is classified not in powers of
a coupling constant $y=\sum_na_n\gamma^n$ but as an expansion
of the type
$y=\sum_n \left[\gamma\ln(\dots)\right]^n  a_n(\gamma)$.
The  renormalization group (RG) corresponds to the
Taylor series expansion of each function $a_n(\gamma)$, whereas the
value of
the logarithmic factor itself can be large.

Following the conventional scheme, see e.g.,
Ref.~\onlinecite{wilson}. we subdivide the supervectors $\psi$
into a slow, $\bPsi\left(  X\right)  $,
and fast, $\bUpsilon\left(  X\right)  $, parts%
\begin{equation}
\bpsi\left(  X\right)  =\bPsi\left(  X\right)  +\bUpsilon\left(  X\right),
\label{d17}%
\end{equation}
and integrate over the fast variable $\bUpsilon\left(X\right)$ using the
perturbation theory in the effective interaction.

One should be careful defining the fast and slow variables because, as
we saw from \reqs{d3}--\rref{d6},
the logarithmic contributions originate from the configuration of the
fields  highly anisotropic in space (smooth along directions of the
momentum $\n$ and sharp in the transverse direction). Therefore, we
can not separate the fast and slow variables considering the moduli
of the momenta\footnote{It should be contrasted to the consideration
  of Ref.~\onlinecite{shankar} where RG language was merely used to
  reformulate known\cite{landau,landau1,agd} results. In the RG
  language, the Fermi liquid constants arise in $d>1$ as the
  dimensionally irrelevant
couplings and, thus, RG procedure for calculating their values is useless.}.

Fortunately, this problem can be avoided because we can define
the fast and slow variables with respect to frequencies only. As the main
contribution in the integral over $k_{\parallel}$, see
Eqs.~(\ref{d3}), \rref{d301}, comes from
$k_{\parallel}\sim\omega/v_{F}$, this type of the separation is sufficient. As
concerns the perpendicular components $k_{\perp}$, they do not participate in
the renormalization group treatment entering equations as a parameter (like
the other variable $u$).

After decomposition \rref{d17} the action acquires the form
\be
{\mathcal S}\left\{  \bPsi,\bUpsilon\right\}
={\mathcal S}_{0}\left\{  \bPsi\right\}
+ {\mathcal S}_{0}^\Upsilon\left\{  \bUpsilon\right\}
+{\mathcal S}_{2} \left\{ \psi\right\}
+ {\mathcal S}_{\rm int}\left\{ \bPsi, \bUpsilon\right\},
\label{d170}
\ee
where
\be
{\mathcal S}_{\rm int}\left\{ \bPsi, \bUpsilon\right\}
={\mathcal S}_{2}\left\{\bUpsilon\right\}+
{\mathcal S}_{3}\left\{\bPsi+\bUpsilon\right\}
+{\mathcal S}_{4}\left\{\bPsi+\bUpsilon\right\}.
\label{d171}%
\end{equation}
The free action for the fast fields has the form, cf. \req{c37},
\be
{\mathcal S}_{0}^{\Upsilon}\left\{ \Upsilon\right\}  =
-2i\nu\int\overline{\Upsilon}_{\gamma}
\left[
\hat{\mathcal L}_{0}
-i\left(\varkappa\w_c\right)\hat{\Lambda}
\right]
\Upsilon_{\gamma}  dX,
\label{d172}%
\ee where $\hat{\mathcal L}_{0}$ is given by Eq.~(\ref{c261}). The
second term in brackets leaves the contribution only from
frequencies \be \varkappa \w_c \lesssim |\w|. \label{d18} \ee
where $\w_c$ is the running cutoff-of the problem, so that $|\w|
\lesssim \w_c$ and $\varkappa < 1$. 

With
such a choice, one step of renormalization group transforms
the running cut-off as
\be
\w_c \to \varkappa \w_c.
\label{d1730}
\ee

Our goal is to obtain the correction to the action of the slow
variables $\Psi$ arising due to the interaction with the fast fluctuations:
\be
\delta {\mathcal S}_\Psi
=
-\ln\left\langle \exp\left[-{\mathcal S}_{\rm int}\left\{ \bPsi,
      \bUpsilon\right\}
\right]\right\rangle_\Upsilon
-{\mathcal S}_{\rm int}\left\{ \bPsi,0\right\}.
\label{d173}
\ee
Hereinafter, the averaging over the fast fields $\Upsilon$ is defined as
cf. \req{d101}
\be
\left\langle\dots \right\rangle_\Upsilon
=\int \dots\
\exp\left(  -{\mathcal S}_0\left\{  \bUpsilon\right\}  \right) D\bUpsilon.
\label{d174}
\ee

The integration over the fast field $\left\{  \bUpsilon\right\}$
is performed using the Wick theorem and, thus, all the
machinery of \ref{sec5a} is still applicable.
The only difference is that in the intermediate lines one
has to replace $\hat{G}_{0} \to \hat{G}_{\Upsilon} $, where
\be
\hat{G}_{\Upsilon}\left(\w,\k;\n\right)
=\frac{f_1\left(\frac{|\w|}{\w_c}\right)}{i\omega\hat{\Lambda}_{1}+v_{F}\k\n
\hat{\tau}_{3}\hat{\Sigma}_{3}-i \hat{\Lambda}\w_c\varkappa}
\label{d175}
\ee
which differs from \req{d175} by the regularization term restricting
the domain of the frequency integration from below,
$\hat{\Lambda}\w_c$, and from above
$f_1\left(\frac{|\w|}{\w_c}\right)$. Smooth function $f_1(x)$ has the
asymptotic behavior $f_1(x) \to 1,\ x\ll 1$ and $f(x\to\infty)\to 0$
\footnote{The particular functional form of
  the
cut-off will be not important
in at least first loop renormalization group calculation, though
it might be that the more accurate choice will be required for
the higher loop calculations [we did not investigate such loops in
details].
}.

Analogously to the bare action, its correction can be decomposed
\be
\delta {\mathcal S}_\Psi
= \delta {\mathcal S}_4+\delta {\mathcal S}_3+\delta {\mathcal S}_2+
\delta {\mathcal S}_0.
\label{d176}
\ee
We will  consider each of those contributions separately.

\subsubsection{Renormalization of the quartic term, $\delta{\mathcal S}_4\left\{\Psi\right\} .$}

\begin{figure}[ht]

\setlength{\unitlength}{2.3em}

\begin{picture}(10,16)
\put(0,14) 
   {
    \begin{picture}(10,1)
      \put(0.2,0)
      {\includegraphics[width=3\unitlength
        ]{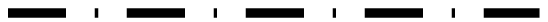}$\ =\ \frac{i}{4\nu}\hat{G}_\Upsilon\left(\w,\k,\n\right)$}
      \put(1.1, 0.4){$\w,k,n,u$}
    \end{picture}
  }

\put(0,11) 
   {
    \begin{picture}(10,3)
      \put(2.2,0)
      {\includegraphics[width=2\unitlength
        ]{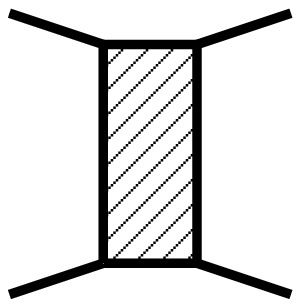}}
\put(4.3,0.8){$=-{\mathcal S}_4$}

    \end{picture}
  }

\put(0,8) 
   {
    \begin{picture}(10,3)
      \put(0.3,0)
      {\includegraphics[width=2\unitlength
        ]{fig4z1}}
\put(0.4,0.8){$
\delta$}
  \put(2.7,0.8){$=$}

\put(3.3, 0.7){$\displaystyle{\left[\frac{1}{4}\right]}$}
\put(4.1,-0.3)
      {\includegraphics[width=5\unitlength]{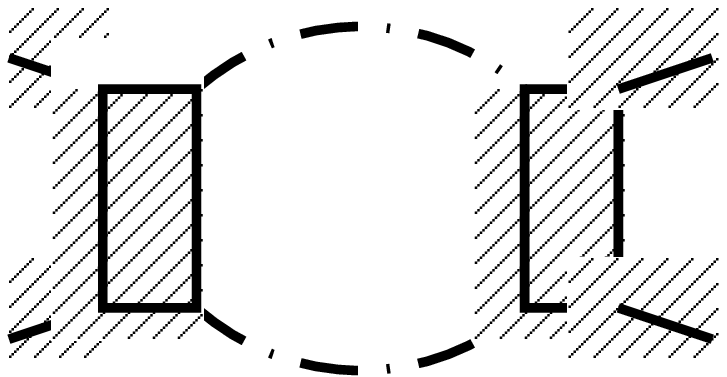}}
\put(8.9, 0.9){a)}

\put(0,-0.5){$\beta_2$}
\put(0,2){$\beta_1$}

\put(2,-0.5){$\gamma_2$}
\put(2.3,2){$\gamma_1$}

\put(3.7,-0.5){$\beta_2$}
\put(3.7,2){$\beta_1$}

\put(6.7,-0.8){$\delta_2$}
\put(6.7,2.3){$\delta_1$}

\put(9,-0.5){$\gamma_2$}
\put(9,2){$\gamma_1$}

    \end{picture}
  }

\put(1.4,5){b)}
\put(1,3){$+$}
\put(2.1,1.3)
      {\includegraphics[width=1.7\unitlength]{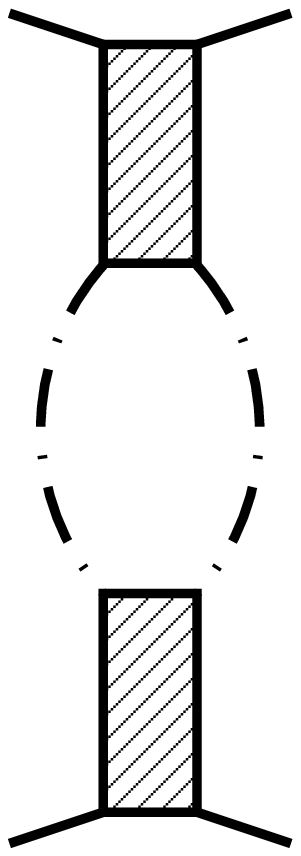}}

\put(5.2,5){c)}
\put(4.8,3){$+$}
\put(6.1,1.9)
      {\includegraphics[width=2.1\unitlength]{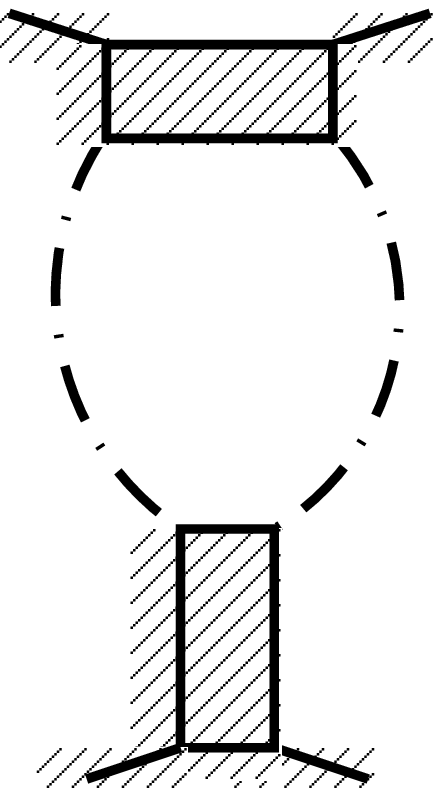}}

\end{picture}
\caption{First loop renormalization of the quartic interaction.
The notation is introduced in Fig.~\ref{fig201}, and filling means
the vertices with renormalized interaction constants $\gamma_i$.
The Green function for the fast fields, $\hat{G}_{\Upsilon}$,
 is introduced in \req{d175}.
Factor $[1/4]$ in diagram a) account for the symmetries of the diagram.
Diagram (b) vanishes because of the
 supersymmetry, see \req{d102}. Diagram c) does not produce logarithmic
contribution, see Eqs.~(\ref{d178}) -- (\ref{d179}).
}%
\label{fig4}%
\end{figure}

The first loop diagrams leading to the renormalization of the
quartic interaction are shown in Fig.~\ref{fig4}. Only the
diagram, Fig.~\ref{fig4}a), may produce a logarithm. Indeed, the
diagram Fig.~\ref{fig4}b) contains a closed loop and vanishes
because of supersymmetry, see \req{d102}. The diagram \ref{fig4}c)
does not produce a logarithm because of the locations of the poles
in the corresponding Green functions, as it will be more formally
discussed in the end of this subsubsection.

To obtain the analytic expression for the diagram~\ref{fig4}a),
we apply the rules of Fig.~\ref{fig201} and
notice that the result has the structure of
\req{d2}, see also Fig.~\ref{fig3}, with the replacement of
$\hat{G}_{0}\to \hat{G}_{\Upsilon}$
and
\be
\begin{split}
\left[\hat{Q}_{1,2}\right]_{\delta_1\delta_2}^i
&=\sqrt{\nu}\sum_{k=1}^4\lambda_{ik}\epsilon_{\alpha\beta_1\delta_1}
\epsilon_{\alpha\beta_2\delta_2}
\\
&\times
u\hat\Pi_k \Psi_{\beta_1}\otimes \overline{\Psi}_{\beta_2} \hat\Pi_ku
\hat\tau_3
\end{split}
\label{d177}
\ee
where the indices in the spin space are written explicitly,
and index $i=1,2,3,4$ labels  the coupling  constant in \req{d2}.

It is easy to see that the appearance of the cut-offs
in \req{d175} leads to the
replacement of $\ln\left(\frac{v_F}{r_0\tilde{\epsilon}}\right)
\to \ln(1/\varkappa)$ in \req{d6} without affecting the matrix
structure of the latter. Using \req{d6000} and applying \req{d7h}
twice, we find \be
\begin{split}
\delta{\cal S}_4 &=2\nu  \Xi_{\beta_1\gamma_1}^{\beta_2\gamma_2}
\sum_{i,j=1}^4\lambda_{ij} \int dX \left(  \overline{\Psi}_{\gamma_1
}\left(  X\right)  \hat{\tau}_{3}\hat{\Pi}_j\Psi_{\beta_1}\left(  X\right)  u\right)
\\
&\times
{\delta \hat{\Gamma}}_{i}\left(  u\overline{\Psi}_{\beta_{2}}\left(
X\right)\hat{\tau}_{3}\hat{\Pi}_j\Psi_{\gamma_{2}}\left(  X\right)  \right),
\end{split}
\label{d1000}
\ee
where
the action of the operator ${\delta \hat{\Gamma}}_{i}$ is
defined by \req{d801} with the kernels
\be
\begin{split}
&\delta{\Gamma}_{2}=\delta{\Gamma}_{4}=0;
\\
&\delta{\Gamma}_{1}\left(\theta;u,u_1;\r_\perp\right)
=-\mu_duu_1 \bar{f}_\perp(\r_\perp)\ln \varkappa
\left[{\Gamma}_{1}\left(\dots\right)\right]^2;
\\
&
\delta{\Gamma}_{3}\left(\theta;u,u_1;\r_\perp\right)
=\mu_duu_1\bar{f}_\perp(\r_\perp) \ln \varkappa
\left[{\Gamma}_{1}\left(\dots\right)\right]^2,
\end{split}
\label{d16c} \ee where we suppressed the arguments in the right
hand side implying that they are the same as in the left hand
side. Equation \rref{d16c} is valid for $|\theta| \lesssim
\w_cr_0/v_F$, otherwise the logarithmic renormalization vanish.
The function $\bar{f}_\perp(\r_\perp)$ is defined in \req{d27b}.

The tensor $\Xi_{\beta_1\gamma_1}^{\beta_2\gamma_2}$ is given by
\be
\begin{split}
\Xi_{\beta_1\gamma_1}^{\beta_2\gamma_2}&=
2\epsilon_{\alpha_1\beta_1\delta_1}
\epsilon_{\alpha_1\beta_2\delta_2}\epsilon_{\alpha_2\beta_1\delta_1}
\epsilon_{\alpha_2\beta_2\delta_2}
\\
&=2\delta_{\gamma_1\gamma_2}\delta_{\beta_1\beta_2}
+ 2\delta_{\gamma_1\beta_1}\delta_{\gamma_2\beta_2}
\\
&=\epsilon_{\alpha\beta_1\gamma_1}\epsilon_{\alpha\beta_2\gamma_2}
+\delta\Xi_{\beta_1\gamma_1}^{\beta_2\gamma_2};
\\
\delta\Xi_{\beta_1\gamma_1}^{\beta_2\gamma_2}
&=2\delta_{\gamma_1\beta_1}\delta_{\gamma_2\beta_2}
+ \delta_{\gamma_1\gamma_2}\delta_{\beta_1\beta_2}
+ \delta_{\gamma_1\beta_2}\delta_{\beta_1\gamma_2}.
\end{split}
\label{d179} \ee
As matrices \rref{d7a} are self-conjugate and
$\hat{\tau}_3$ is anticonjugate, one finds using \reqs{c24a}--
\rref{c24b}
\[
\left(  \overline{\Psi}_{\beta_1
}\left(  X\right)  \hat{\tau}_{3}\hat{\Pi}_j\Psi_{\gamma_1}\left(
  X\right)  \right)
= - \left(  \overline{\Psi}_{\gamma_1
}\left(  X\right)  \hat{\tau}_{3}\hat{\Pi}_j\Psi_{\beta_1}\left(
  X\right)  \right),
\]
and the contribution proportional to $\delta\Xi$ will vanish after
substitution in \req{d178}.

Therefore, the resulting action \rref{d1000} is nothing but
the original quartic interaction \rref{d8}  with the
couplings ${\hat{\Gamma}}_{i}$ renormalized according to
\req{d16}.
This is sufficient to write down the renormalization
group equation. We will do it in the next section after
we consider the transformation of the remaining terms in the action
under the RG step.

Closing our consideration of the quartic interaction, let us give
a formal proof that  the diagram Fig.~\ref{fig4}c does not give a
logarithmic contribution. Once again, we apply the rules of
Fig.~\ref{fig201} and notice that the result has the structure of
\req{d2} with [cf. \req{d177}] \be
\begin{split}
\left[\hat{Q}_{2}\right]_{\delta_1\delta_2}^i
&=\sqrt{\nu}u_{1}\sum_{k=1}^4\lambda_{ik}\epsilon_{\alpha\beta_1\delta_1}
\epsilon_{\alpha\beta_2\delta_2}
\\
&\times
\hat\Pi_k \Psi_{\beta_1}\otimes \overline{\Psi}_{\beta_2} \hat\Pi_k
\hat\tau_3;\\
 \left[\hat{Q}_{1}\right]_{\delta_1\delta_2}^i
&=\sqrt{\nu}u_{1}\sum_{k=1}^4\lambda_{ik}\epsilon_{\alpha\beta_1\delta_1}
\epsilon_{\alpha\beta_2\delta_2}
\\
&\times
\hat\Pi_k \left(\overline{\Psi}_{\beta_2}\hat\Pi_k
\tau_3\Psi_{\beta_1}\right),
\end{split}
\label{d178}
\ee
Using \req{d6000}, we find
\be
\begin{split}
&\Str\left[
\hat{Q}_1^{(3)}\hat{Q}_2^{(3)}
-\hat{Q}_1^{(1)}\hat{Q}_2^{(1)}\right]
\\
&
\propto\sum_{i=1,3}
\sum_{k=1}^4\left(-1\right)^{\frac{i-1}{2}}\lambda_{ik}
\Str\left[\Pi_l
\hat{\Pi}_k\hat{Q}_2\hat{\Pi}_k\right]
\\
&
=\sum_{i=1,3}
\sum_{k=1}^4\left(-1\right)^{\frac{i-1}{2}}\lambda_{ik}
\Str\left[\hat\Pi_l\hat{Q}_2\right]=0,
\end{split}
\label{d1790}
\ee
as the matrices $\hat{\Pi_k}$, see \req{d7a}, commute with each other,
$\left[\hat{\Pi}_k\right]^2=1$,
and second of the properties \rref{d7e} is used.

\subsubsection{
Renormalization of the cubic term, $\delta{\mathcal S}_{3}\left\{\Psi\right\} .$}

\begin{figure}[ht]

\setlength{\unitlength}{2.3em}

\begin{picture}(10,13.8)

\put(0,11) 
   {
    \begin{picture}(10,3)
      \put(2.2,0)
      {\includegraphics[width=2\unitlength
        ]{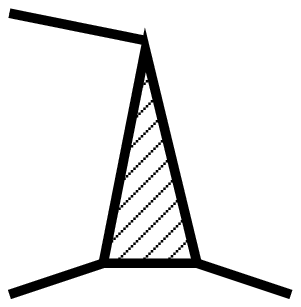}}
\put(4.3,0.8){$=-{\mathcal S}_3$}

    \end{picture}
  }

\put(0,8) 
   {
    \begin{picture}(10,3)
      \put(0.3,0)
      {\includegraphics[width=2\unitlength
        ]{fig5z1}}
\put(0.4,0.8){$
\delta$}
  \put(2.7,0.8){$=$}

\put(4.1,-0.3)
      {\includegraphics[width=5\unitlength]{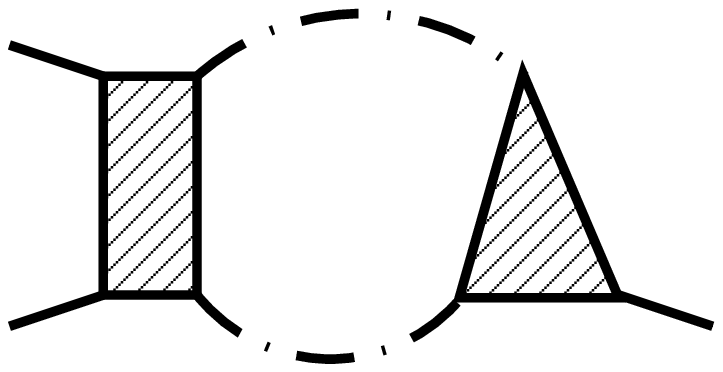}}
\put(8.3, 2.2){a)}

\put(0,-0.5){$\beta_2$}
\put(0,2){$\beta_1$}

\put(2,-0.5){$\gamma_2$}

\put(3.7,-0.5){$\beta_2$}
\put(3.7,2){$\beta_1$}

\put(6.7,0.1){$\delta_2$}
\put(5.4,-0.9){$\w+{\Omega},\q+\k$}

\put(6.7,1.6){$\delta_1$}
\put(6.1,2.4){$\w,\k$}

\put(9,-0.5){$\gamma_2$}
\put(8.8,0.3){$\Omega,\q$}

    \end{picture}
  }

\put(0.4,5){b)}
\put(0,3){$+$}
\put(1.1,1.3)
      {\includegraphics[width=1.7\unitlength]{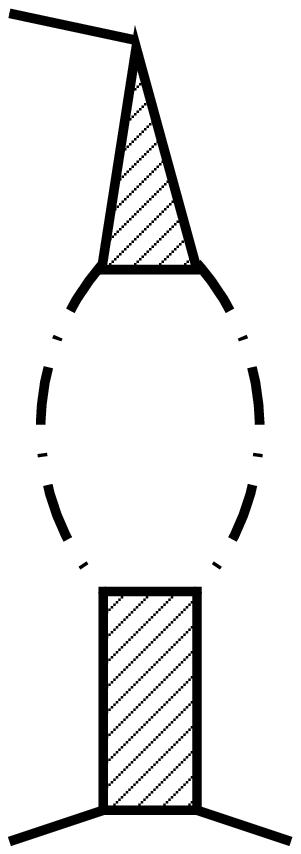}}

\put(3.9,5){c)}
\put(3.4,3){$+$}
\put(4.4,1.3)
      {\includegraphics[width=2.1\unitlength]{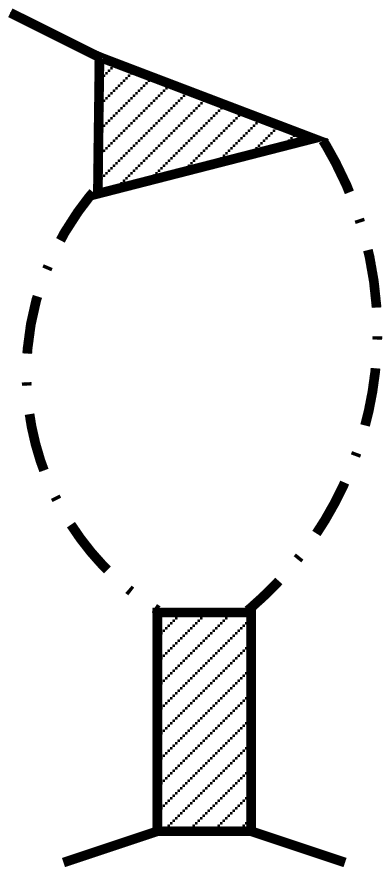}}

\put(7.7,5){d)}
\put(7.2,3){$+$}
\put(8.2,1.3)
      {\includegraphics[width=2.1\unitlength]{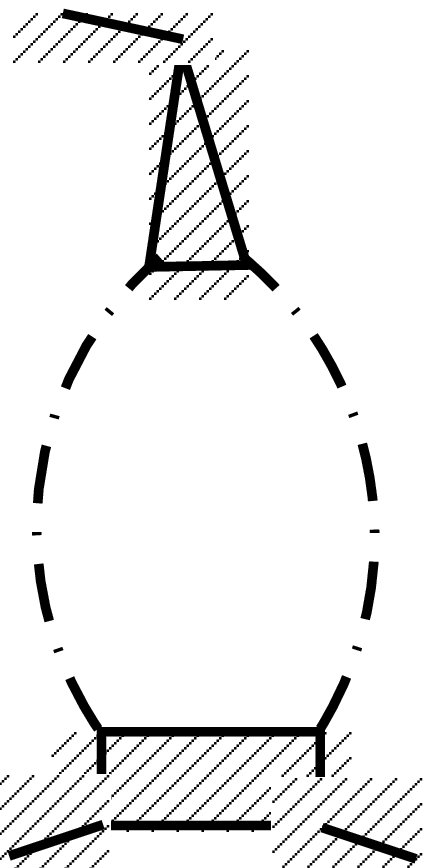}}

\end{picture}

\caption{First loop renormalization of the cubic interaction. The
notation is introduced in Figs.~\ref{fig201}, \ref{fig4} and the
filling means the vertices with renormalized interaction constants
$\beta_i$.
 Diagram (b) vanishes due to the
 supersymmetry, similarly to Fig.~\ref{fig4}b.
Diagrams c) d) do not contain logarithmic
contribution, similarly to Fig.~\ref{fig4}c.
}%
\label{fig5}%
\end{figure}

The first loop diagrams leading to the renormalization of the cubic
interaction are shown on Fig.~\ref{fig5}.
Similarly to what we saw when calculating $\delta {\mathcal S}_4$.
only the diagram Fig.~\ref{fig5}a)  produces a logarithm
and we turn to the calculation of this contribution now.

The only difference in this calculation from the one for
the quartic term is the presence of the differential operator
$\hat{\mathbb D}_-$ in the expression for the vertex.
Using \reqs{d13} and \rref{c410} we write
\be
\begin{split}
&\overline{\mathbb D}_{-}(\w,\k)
=\frac{u}{2}\overline{\cal F}_{0}
\left[\left(2\alpha-1\right)\hat{\cal L}_{0}
(\w,\k) - \hat{\cal L}_{0}^\dagger(\w,\k)\right]
\hat\tau_{-}\\
&\quad=
\frac{u}{2}\overline{\cal F}_{0}
\left[\left(2\alpha-1\right)\hat{\cal L}_{0}
(\w,\k) - \hat{\cal L}_{0}^\dagger(\w+\Omega,\q+\k)\right]\hat\tau_{-}
\\
&\quad-\frac{u}{2}\overline{\cal F}_{0}\hat{\cal
  L}_{0}^\dagger(-\Omega,-\k)
\hat\tau_{-}
.
\end{split}
\label{d350}
\ee
where frequencies and momenta are arranged as in Fig.~\ref{fig5}a).

The terms in the second line cancel the small denominator in one
of the Green function. Integration over $\w,\k$ does not produce
terms $\propto \ln\varkappa$ because \be
\begin{split}
&\int d\w d k_\parallel
\hat{A}\hat{\cal L}_{0}^\dagger
(\w, k_\parallel)\hat{G}_\Upsilon(\w, k_\parallel)
\hat{B}
\hat{G}_\Upsilon^\dagger(\w, k_\parallel)
\\
 &
\simeq\int d\w d k_\parallel \hat{A}\hat{B} \hat{G}_\Upsilon(\w,
k_\parallel) \simeq
-i\left(\kappa\w_c\ln\kappa\right)\hat{\Lambda},
\end{split}
\label{d3500}
\ee
 i.e, it is determined by the lower limit of the integration and must
be excluded from RG scheme.

The term in the last line is not affected by the integration, so
the result can be once again recast in the form of \req{d2}, and
the calculation proceeds similarly as it was done for the quartic
term. Instead of \req{d177}, we find \be
\begin{split}
\left[\hat{Q}_{1}\right]_{\delta_1\delta_2}^{i,\sigma}
&=\sqrt{2i\nu}\epsilon_{\delta_1\gamma_2\delta_2}\sum_{k=1}^4\lambda_{ik}
\hat{\Pi}_k\Psi_{\gamma_2}\otimes\overline{\mathbb D}_{\sigma}
\hat{\Pi}_k\hat\tau_3
\\
\left[\hat{Q}_{2}\right]_{\delta_1\delta_2}^i
&=\sqrt{\nu}\sum_{k=1}^4\lambda_{ik}\epsilon_{\alpha\beta_1\delta_1}
\epsilon_{\alpha\beta_2\delta_2}
\\
&\times
u\hat\Pi_k \Psi_{\beta_1}\otimes \overline{\Psi}_{\beta_2} \hat\Pi_k u
\hat\tau_3,\\
\end{split}
\label{d351}
\ee
where we introduced the notation
\be
\label{d352}
\overline{\mathbb D}_{\sigma} \equiv \overline{\mathbb D}_{\sigma}\left(
\alpha=1/2\right).
\ee

Repeating the same steps as when deriving  \req{d1000} and using
the identity
\begin{equation}
\left(\varepsilon_{\alpha\beta_1\delta_1}\varepsilon_{\alpha\delta_{2}\beta_{2}%
}
\right)
\varepsilon_{\delta_2\gamma_{2}{\delta_1}}=\varepsilon_{\beta_{1}\beta
_{2}\gamma}, \label{d36}%
\end{equation}
we obtain, [cf. \req{d13}]
\begin{subequations}
{\setlength{\arraycolsep}{0pt}
\bea
&&\delta {\mathcal S}_3=\delta {\mathcal S}_3^++\delta {\mathcal S}_3^-
\label{d361a}\\
&&\delta {\mathcal S}_3^+
=-2\nu\sqrt{2i}\varepsilon_{ \beta\gamma\delta}\sum_{i,j=1}^4
 \lambda_{ij}\sum_{\sigma=\pm} \int dX
\label{d361b}\\
&&\quad\times
\left(  \overline{\Psi}_{\beta}\left(  X\right)
 \hat{\tau}_{3}\hat{\Pi}_j
 \Psi_{\gamma}\left(  X\right)  u
\right)\delta
\hat{{\mathcal B}}_{i}^+
\left(\overline{\mathbb D}_{+}
\tau_{3}\hat{\Pi}_j\Psi_{\delta}\left(  X\right)
\right),
\nonumber\\
&&\delta {\mathcal S}_3^-
=4\nu\sqrt{2i}\varepsilon_{ \beta\gamma\delta}\sum_{i,j=1}^4
 \lambda_{ij}\sum_{\sigma=\pm} \int dX u
\label{d361c}
\\
&&\times\Str
\left[\hat{\Pi}_j
 \Psi_{\gamma}\left(  X\right)
\otimes
\delta
\hat{{\mathcal B}}_{i}^-
\overline{\mathbb D}_{-}
\hat{\tau}_{3}
\right]
\left[
\hat{\Pi}_j\Psi_{\delta}\left(  X\right)
\otimes
 \overline{\Psi}_{\beta}\left(X\right)\hat{\tau}_{3}
\right]
,
\nonumber
\eea
\label{d361}
where the sign difference between \reqs{d361b} and \rref{d361c}
appears because of the definition of the supertrace \rref{d7f0}.
}
\end{subequations}

The action of the operators $\delta\hat{{\mathcal B}}_{i}^\pm$
in \reqs{d361} are defined by \req{d12} with
the kernels given by [cf. \req{d16c}]
\be
\begin{split}
&\delta{\mathcal B}_2^{\pm}=\delta{\mathcal B}_4^\pm=0;
\\
& \delta{\mathcal B}_{1}^+
=-2\mu_duu_1\bar{f}_\perp(\r_\perp)\ln \varkappa\
{\Gamma}_{1}
{\mathcal B}_{1}^+
;
\\
&\delta{\mathcal B}_{1}^-
=-\mu_duu_1 \bar{f}_\perp(\r_\perp)\ln \varkappa\
{\Gamma}_{1}{\mathcal B}_{1}^-
\\
& \delta{\mathcal B}_{3}^+
=2\mu_duu_1\bar{f}_\perp(\r_\perp)\ln \varkappa\
{\Gamma}_{3}
{\mathcal B}_{3}^+
;
\\
&\delta{\mathcal B}_{3}^-
=\mu_duu_1 \bar{f}_\perp(\r_\perp)\ln \varkappa\
{\Gamma}_{3}{\mathcal B}_{3}^-
\end{split}
\label{d362} \ee for $|\theta| \lesssim \w_cr_0/v_F$, otherwise
the logarithmic renormalizations vanish. We did not write the
arguments of the kernels implying that they are the same as in the
left-hand side of \req{d16c}.

Equation \rref{d361c} is apparently not of the original form
\rref{d13} yet as the differentiation in \req{d361c} acts on two
fields on its right whereas the derivative of only one field is
present in \req{d13}. However, it can be transformed using
\[
\varepsilon_{ \beta\gamma\delta}
 \overline{\Psi}_\delta
 \hat{\tau}_{3}\hat{\Pi}
 \partial_z \Psi_{\gamma}
= \frac{\varepsilon_{ \beta\gamma\delta}}{2}   \partial_z
\left[\overline{\Psi}
 \hat{\tau}_{3}\hat{\Pi}_\delta
 \partial_z \Psi_{\gamma} \right]
\]
(as matrix $\hat\tau_3$ is anticonjugated), and $z$ denote either
$\tau$ or $\r$. We, thus, re-write \rref{d361c} as
\be
\begin{split}
&\delta {\mathcal S}_3^-
=4\nu\sqrt{2i}\varepsilon_{ \beta\gamma\delta}\sum_{i,j=1}^4
 \lambda_{ij}\sum_{\sigma=\pm} \int dX
\\
&\times
\Bigg[
\left(  \overline{\Psi}_{\beta}\left(  X\right)
 \hat{\tau}_{3}\hat{\Pi}_j
 \Psi_{\gamma}\left(  X\right)  u
\right)\delta
\hat{{\mathcal B}}_{i}^-
\left(\overline{\mathbb D}_{-}
\tau_{3}\hat{\Pi}_j\Psi_{\delta}\left(  X\right)
\right)
\\
&
+
\frac{1}{2}  \left(  \overline{\Psi}_{\beta}\left(  X\right)
 \hat{\tau}_{3}\hat{\Pi}_j
 \Psi_{\gamma}\left(  X\right)  u
\right)\delta
\hat{{\mathcal B}}_{i}^-
\left(\overleftarrow{\left[\overline{\mathbb D}_{-}\right]}
\tau_{3}\hat{\Pi}_j\Psi_{\delta}\left(  X\right)
\right)
\Bigg],
\end{split}
\label{d3620} \ee where the notation $\hat{\overleftarrow{\mathbb
D}}$ means that the differential operators included in ${\mathbb
D}$ act on the left: \be \label{d15a0} c
\overleftarrow{\partial}_z d \equiv \left({\partial}_z c\right) d
\ee for arbitrary functions $c,d$. After integration by parts in
the last term we obtain \be
\begin{split}
&\delta {\mathcal S}_3^-
=2\nu\sqrt{2i}\varepsilon_{ \beta\gamma\delta}\sum_{i,j=1}^4
 \lambda_{ij}\sum_{\sigma=\pm} \int dX
\\
&\times
\left(  \overline{\Psi}_{\beta}\left(  X\right)
 \hat{\tau}_{3}\hat{\Pi}_j
 \Psi_{\gamma}\left(  X\right)  u
\right)\delta
\hat{{\mathcal B}}_{i}^-
\left(\overline{\mathbb D}_{-}
\tau_{3}\hat{\Pi}_j\Psi_{\delta}\left(  X\right)
\right)
\end{split}
\label{d363} \ee that has the same form as \req{d13} for the
particular choice of the parameter $\alpha$ \be
\alpha=\frac{1}{2}. \label{d364} \ee In what follows we will use
only this value of $\alpha$. \footnote{Other choices of $\alpha$
would require additional shifts of the fields in order to
reproduce the cubic term in the RG. Therefore, the choice
\rref{d364} is the most convenient for the sake of the calculation
though the final physical answers can not depend on $\alpha$ at
all}.

Equations \rref{d361}, \rref{d363} shows that the cubic term is
reproduced under the RG step and \req{d362} determines the new
values of the coupling constants.

\subsubsection{
Renormalization of the quadratic interaction, $\delta{\mathcal
  S}_{2}\left\{\Psi\right\} $
and free action $\delta{\mathcal
  S}_{0}\left\{\Psi\right\}$ .}
\label{sec:renqua}

\begin{figure}
\setlength{\unitlength}{2.3em}
\begin{picture}(10,5)
 \put(0,0)
      {\includegraphics[width=8\unitlength
        ]{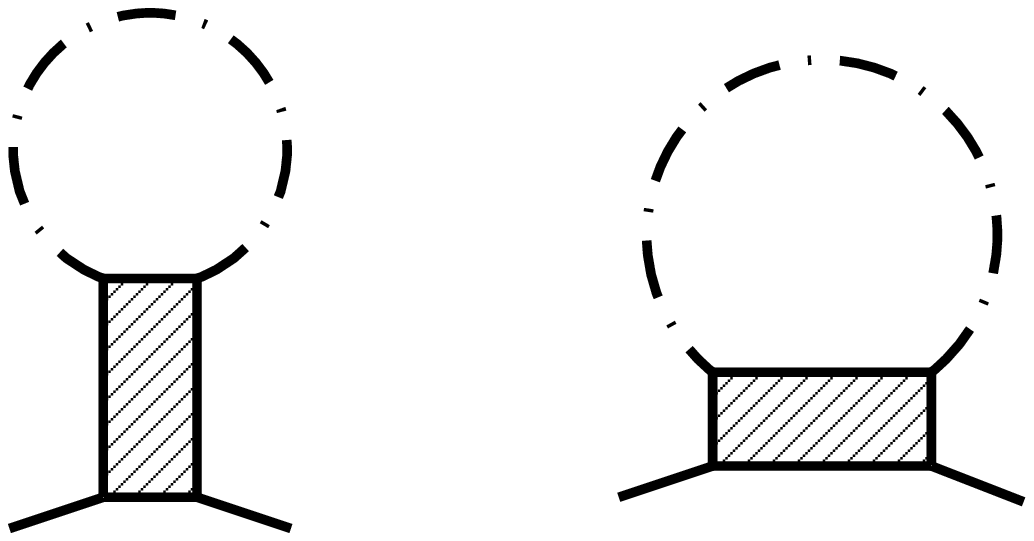}}
\put(0,4.5){a)} \put(5,4.5){b)}

\put(3.8,2.5){$=$} \put(8.8,2.5){$=0$.}
\end{picture}
\caption{First order corrections to the quadratic term. Diagram a)
vanishes due to the supersymmetry. Diagram b) vanishes after
$\w,\k$ integration, see \req{d3500}. } \label{fig60}
\end{figure}

\begin{figure}
\setlength{\unitlength}{2.3em}

\begin{picture}(10,16.8)

\put(0,14.8) 
   {
    \begin{picture}(10,5)
      \put(2,0)
      {\includegraphics[width=2.5\unitlength
        ]{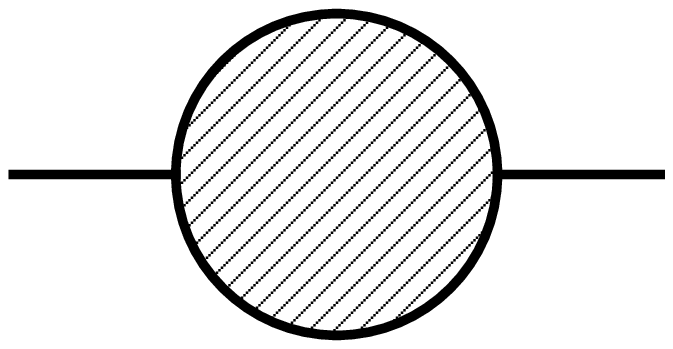}}
\put(5.6,0.5){$=-{\mathcal S}_2$}

    \end{picture}
  }

\put(0,11) 
   {
    \begin{picture}(10,3)
      \put(0.3,0)
      {\includegraphics[width=2.5\unitlength
        ]{fig6z}}
\put(-0.4,0.4){$
\delta$}
  \put(3.4,0.4){$=$}

\put(0.4,0.1){$\beta$}
\put(2.4,0.1){$\beta$}
    \end{picture}
}

\put(0.5,11) 
   {
    \begin{picture}(10,3)

\put(4.1,-0.3)
      {\includegraphics[width=5\unitlength]{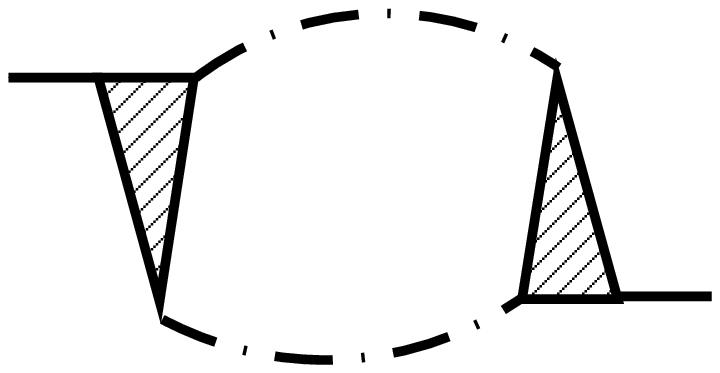}}
\put(8.8, 3.9){a)}

\put(3.8, 0.4){$\displaystyle{\left[\frac{1}{2}\right]}$}
\put(3.7,2){$\beta$}

\put(6.7,0.1){$\delta_2$}
\put(5.4,-0.9){$\w+{\Omega},\q+\k$}

\put(6.7,1.6){$\delta_1$}
\put(6.1,2.4){$\w,\k$}

\put(9,-0.7){$\gamma$}
\put(8.8,0.3){$\Omega,\q$}
   \end{picture}
}

\put(2.5,5){
\begin{picture}(3,4)
\put(0,4){b)}
\put(-0.7,2.2){$+\ \displaystyle{\left[\frac{1}{2}\right]}$}
\put(1.1,1.3)
      {\includegraphics[width=2.7\unitlength]{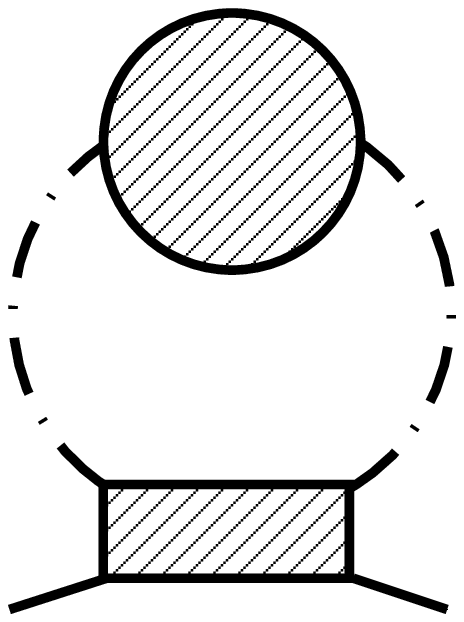}}
\put(0.7,1.4){$\beta$}
\put(4,1.4){$\beta$}

\put(1.6,2.4){$\delta$}
\put(3,2.4){$\delta$}
\put(4.1,3.3){$\w,\k$}
\end{picture}
}

\put(-4,2.2)
{
\begin{picture}(5,3)
\put(3.8,2.8){c)}
\put(3.7,1.8){$+$}
\put(4.4,1.3)
      {\includegraphics[width=4.1\unitlength]{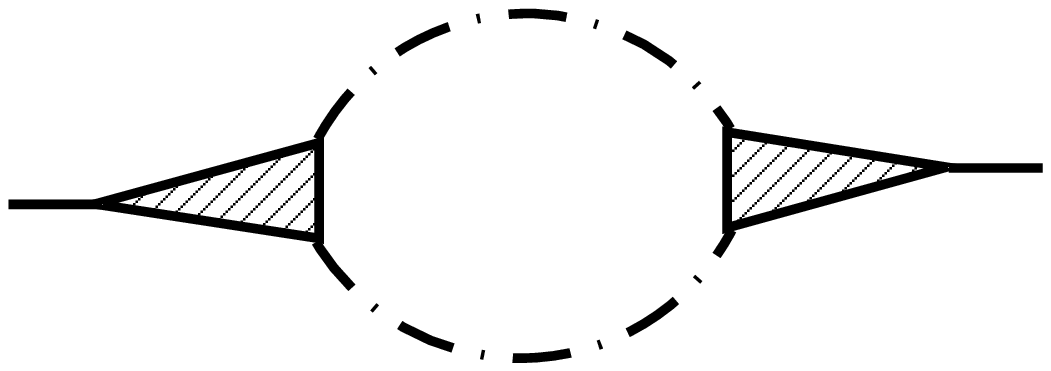}}
\end{picture}
}

\put(-4,-0.5)
{
\begin{picture}(5,3)
\put(3.8,2.8){d)}
\put(3.7,1.8){$+$}
\put(4.4,1.3)
      {\includegraphics[width=4.1\unitlength]{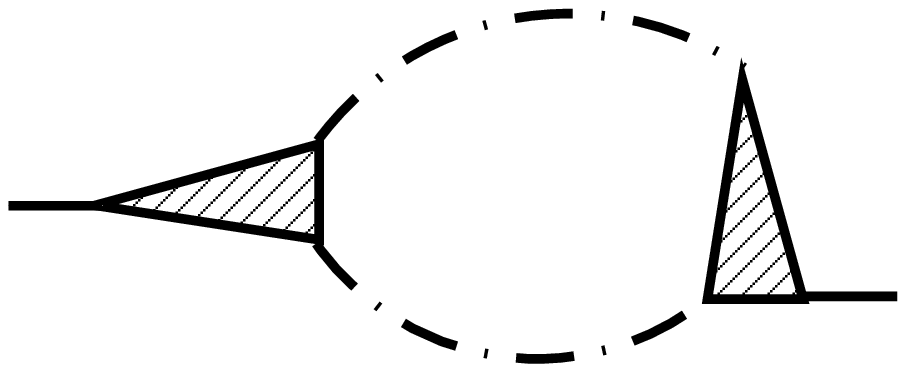}}
\end{picture}
}

\put(3,0)
{
\begin{picture}(3,4)
\put(3.9,5){e)}
\put(3.4,3){$+$}
\put(4.4,1.3)
      {\includegraphics[width=2.1\unitlength]{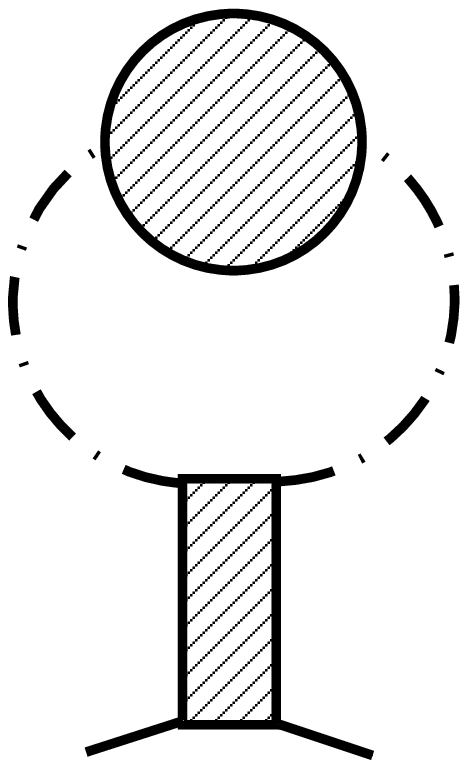}}
\end{picture}
}

\end{picture}

\caption{First loop corrections to the quadratic term ${\mathcal
S}_2$. The notation is introduced in Figs.~\ref{fig201},
\ref{fig4} and the filling means the vertices with renormalized
interaction constants $\Delta_i^{\sigma_1\sigma_2}$.
 Diagrams a,b) are logarithmic. Diagram c) vanishes due to the
 supersymmetry, similarly to Figs.~\ref{fig4}b, \ref{fig5}b.
Diagrams d) e) do not contain the logarithmic contribution,
similarly to Figs.~\ref{fig4}c, \ref{fig5}c,d. The factor $[1/2]$
on the panels a,b) appears because of the symmetry of the
diagrams, } \label{fig6}
\end{figure}

\begin{figure}
\setlength{\unitlength}{2.3em}
\begin{picture}(10,2.5)
\put(7.8,0.5){$=-\delta {\mathcal S}_0$}
\put(4.3,0.){$\n,u$}\put(1.7,0.){$\n,u$}\put(6.7,0.){$\n,u$}
\put(4.4,2.6){$\n_1$}
\put(1,0.9){$\displaystyle{\left[\frac{1}{2}\right]}$}
\put(2,0.3){
\includegraphics[width=4.9\unitlength]{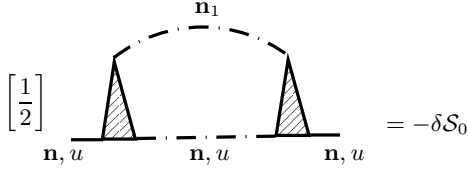}}
\end{picture}
\caption{First loop correction to the free action, ${\mathcal
S}_0$. This correction is logarithmic only for $d=1$. For $d>1$,
the logarithmic divergence vanish due to the integration over
$\widehat{\n\n_1}$.} \label{fig7}
\end{figure}

The one loop diagrams that may change the values of
$\delta{\mathcal
  S}_{0,2}\left\{\Psi\right\} $ are shown on Figs.~\ref{fig60}--\ref{fig7}.

One immediately notices that all the diagrams of the first order
in quartic interactions, Fig.~\ref{fig60}, vanish. Indeed, diagram
of Fig.~\ref{fig60}a) vanishes because of the supersymmetry.
Diagram of Fig.~\ref{fig60}b) vanishes because it involves
integration of one Green function only and cannot produce a
logarithmic divergence, see \req{d3500}.

Among the diagrams of the second order,
Figs.~\ref{fig6},\ref{fig7}, only Fig.~\ref{fig6} a,b) gives the
logarithmic contribution into the quadratic interaction. We
transform contributions involving $\overline{\mathbb
D}_{-}(\w,\k),\  {\mathbb D}_{-}(\w,\k)$ on diagrams
Fig.~\ref{fig6}a  according to \req{d350}, for $\alpha=1/2$. Then,
the contributions from the last line of \req{d350} are not
affected by the integration, so the result can be once again
recast in the form of \req{d2} with [cf. \reqs{d351}, \rref{d177}]
\be \left[\hat{Q}_{1,2}\right]_{\delta_1\delta_2}^{i,\sigma}
=\frac{1}{2}\sqrt{i\nu}\epsilon_{\delta_1\gamma_2\delta_2}\sum_{k=1}^4\lambda_{ik}
\hat{\Pi}_k\Psi_{\gamma_2}\otimes\overline{\mathbb D}_{\sigma}
\hat{\Pi}_k\hat\tau_3u. \label{d380} \ee

The first line of \req{d350} for the quadratic interaction does
not produce any logarithmic divergence for terms involving either
$\overline{\mathbb D}_{+}(\w,\k)\dots{\mathbb
  D}_{-}(\w,\k)$
or $\overline{\mathbb D}_{+}(\w,\k)\dots{\mathbb
  D}_{-}(\w,\k)$ due to the integral \req{d3500}.
The term involving
$\overline{\mathbb D}_{-}(\w+\Omega,\k+\q)\dots{\mathbb
  D}_{-}(\w,\k)$ have all the denominators cancelled and
the integral is ultraviolet divergent. This ultraviolet divergence
will be discussed later on in this subsection after we complete
the derivation of the logarithmic terms.

In the diagram Fig.~\ref{fig6}b only the terms involving
$\overline{\mathbb D}_{+}(\w,\k)\dots{\mathbb  D}_{+}(\w,\k)$ are
logarithmic. Diagrams involving $\overline{\mathbb
D}_{+}(\w,\k)\dots{\mathbb  D}_{-}(\w,\k)$ and $\overline{\mathbb
D}_{-}(\w,\k)\dots{\mathbb  D}_{+}(\w,\k)$ are of the type of
\req{d3500} and do not contribute to the logarithmic
renormalization group. Finally, the term $\overline{\mathbb
D}_{-}(\w,\k)\dots{\mathbb  D}_{-}(\w,\k)$ gives rise to the
ultraviolet divergence which will be discussed shortly, see
\req{d385}. The logarithmic part of  Fig.~\ref{fig6}b is of the
form \rref{d2} with \be
\begin{split}
\left[\hat{Q}_{1}\right]_{\delta_1\delta_2}^i
&=\frac{1}{2}\sqrt{\nu}\sum_{k=1}^4\lambda_{ik}\epsilon_{\alpha\beta_1\delta_1}
\epsilon_{\alpha\beta_2\delta_2}
\\
&\times
u\hat\Pi_k \Psi_{\beta_1}\otimes \overline{\Psi}_{\beta_2} \hat\Pi_k u
\hat\tau_3
\\
\left[\hat{Q}_{2}\right]_{\delta_1\delta_2}^{i}
&=\frac{1}{2}\sqrt{{\nu}}i\epsilon_{\delta_1\gamma_2\delta_2}\sum_{k=1}^4\lambda_{ik}
\hat{\Pi}_k{\mathbb D}_{+}
\otimes\overline{\mathbb D}_{+}
\hat{\Pi}_k\hat\tau_3.
\end{split}
\label{d381}
\ee

Collecting all the logarithmic contributions from
Fig.~\ref{fig6}a,b) with the help of
\reqs{d2}, \rref{d380}, \rref{d381} and the identity
\be
\epsilon_{\alpha\beta_1\gamma}\epsilon_{\alpha\beta_2\gamma}
=2\delta_{\beta_1\beta_2},
\label{d382}
\ee
we obtain
[cf.  \req{d15}]
\be
\begin{split}
\delta {\mathcal S}_{2}&\left[ \left\{ \psi\right\};\alpha\right]     =
- i\nu
\sum_{i,j=1}^4
\lambda_{ij}\sum_{\sigma_{1,2}=\pm}
\int dX
\\
&\times
\left(
\overline{\psi_{\delta}}\left(  X\right)\hat{\Pi}_j
 \tau_{3}
{\mathbb D}_{\sigma_1}
\right)\delta \hat{\mathbf \Delta}_i^{\sigma_1\sigma_2}
\left(
 \overline{\mathbb D}_{\sigma_2}\tau_{3}\hat{\Pi}_j\psi_{\delta}\left(  X\right)
\right).
\end{split}
\label{d383}
\ee

The action of the operators $\delta\hat{{\mathbf
\Delta}}_{i}^{\pm\pm}$
 for $i=2,3$ are given by
\req{d15a} with the kernels
[cf. \reqs{d16c}, \rref{d362}]
\be
\begin{split}
&\delta{\mathbf \Delta}^{\sigma_1\sigma_2}_2=0;\\
&\delta{\mathbf \Delta}_{3}^{++}
=2\mu_duu_1 \bar{f}_\perp(\r_\perp)\ln \varkappa
\left[
{\Gamma}_{3}
{\mathbf \Delta}_{3}^{++}
+ \left({\mathcal B}_{3}^+\right)^2
\right]
;
\\
&
\delta{\mathbf \Delta}_{3}^{+-}
=\delta{\mathbf \Delta}_{3}^{-+}=2\mu_duu_1 \bar{f}_\perp(\r_\perp)\ln
\varkappa\ {\mathcal B}_{3}^-{\mathcal B}_{3}^+.
\end{split}
\label{d384}
\ee
and
\be
\delta{\mathbf \Delta}_{3}^{--}
\stackrel{?}{=}2\mu_duu_1\bar{f}_\perp(\r_\perp) \ln \varkappa \left({\mathcal B}_{3}^-\right)^2.
\label{d3840}
\ee
 Equation \rref{d384} is valid for $|\theta|
\lesssim \w_cr_0/v_F$, otherwise, the logarithmic renormalizations
vanish. Once again, we did not write the arguments of the kernels
implying that they are the same as those in the left-hand side of
\req{d16c}. The reason why the correction $\delta{\mathbf
\Delta}_{3}^{--}$ is written separately from all the other
couplings will be explained momentarily.

To complete the calculation of the correction to the quadratic
interaction, we have to compute actually the  ultraviolet
divergent terms in \ref{fig6}a-b. We found \be \left[{\rm
Fig.}~\ref{fig6}a+ {\rm Fig.}~\ref{fig6}b\right]_{uv} \simeq
\sum_{i=2,3}{\cal O}(\w_c^2) \left\{ \left[\beta^{-}_{i}\right]^2
- \gamma_{i}\Delta^{--}_{i} \right\}. \label{d3850} \ee  The
ultraviolet divergences cancel each other for the initial
couplings \rref{d14}, \rref{d16}. In fact, the vanishing of such
divergences precludes the formation of the gap in the spectrum of
the excitations forbidden by the spin rotational symmetry, and
should be valid in any order of the perturbation theory. On the
other hand, the accuracy of our renormalization group procedure
does not allow us to the determine finite logartithmic terms from the
uncertainty ``$\infty - \infty$''. This makes the correction
\rref{d3840} meaningless. Fortunately, we can use the symmetry of
the system forbidding the formation of such ultraviolet
divergences to fix this coupling constant. Requiring the most
divergent part of the ultraviolet divergence to cancel at any
stage of the RG procedure, we find \be {\Gamma}_{3} {\mathbf
\Delta}_{3}^{--} = \left({\mathcal B}_{3}^-\right)^2 \label{d385}
\ee and use this equation for the further RG flow.

The last interesting diagram is shown in Fig.~\ref{fig7}. As it
conserves both the momentum direction $\n$ and the value of the
coordinates $\w,\k,u$, it is natural to classify it as the
correction to the free action, $\delta {\cal S}_0$. The pole
structure of the Green functions allows for the logarithmic
divergence at $\n_1 \to \n$. Calculation is performed using the
formula \be
\begin{split}
&{\mathbb K}^{\sigma_1\sigma_2}_k
\equiv
\left( \overline{\mathbb D}_{\sigma_1}(\w,\k,\n;u)\hat{\Pi}_k
\hat{\cal G}_0(\w,\k,\n)
{\mathbb
    D}_{\sigma_2}(\w,\k,\n;u)\right);
\\
&{\mathbb K}^{\sigma_1\sigma_2}_3= -{\mathbb
  K}^{\sigma_1\sigma_2}_1;
\quad  {\mathbb K}^{\sigma_1\sigma_2}_4= -{\mathbb
  K}^{\sigma_1\sigma_2}_2;\\
&{\mathbb K}^{++}_k={\mathbb K}^{--}=0;\\
&{\mathbb K}^{+-}_1(\w,\k,\n;u)={\mathbb K}^{-+}_1(\w,\k,\n;u)
=u \frac{\w^2-v_F^2(\n\cdot\k)^2}{\w^2+v_F^2(\n\cdot\k)^2};
\\
&{\mathbb K}^{+-}_2(\w,\k,\n;u)={\mathbb K}^{-+}_2(\w,\k,\n;u)
=u \frac{2iv_F\w(\n\cdot\k)}{\w^2+v_F^2(\n\cdot\k)^2},
\end{split}
\label{d386}
\ee
that can be checked directly using definitions
\rref{d1} and \rref{d13} for $\alpha=1/2$.
However, for $d=2,3$ integration over $\n_1$ cancels out this
logarithmic divergence.
For $d=1$, one finds
\[
\begin{split}
\delta{\mathcal S}_0
& \simeq
i\nu \ln\varkappa  \int\overline{\Psi}_{\gamma}\left( X\right)
 \left[
 \hat{\mathcal L}_{0}
 \hat{R}
 \right]
 \Psi_{\gamma}\left(  X\right)  dX,
 \\
 \hat{R}&\simeq {\mathcal B}_1^+{\mathcal B}_1^-\left(1-\hat{\Pi}_3\right) +
 {\mathcal B}_3^+{\mathcal B}_3^-\left(1+\hat{\Pi}_3\right).
\end{split}
\]
This correction can be eliminated by the rescaling of the fields
$\bPsi \to [1-(\ln\varkappa)\hat{R}]\bPsi$. This rescaling will
give the third order correction to the coupling constants in the
interaction part of the action, and thus has to be taken into
account only in the two loop RG equation. As we do not consider
such loop in the present paper, we will have to disregard $\delta
{\mathcal S}_0$ even for $d=1$.

Equations \rref{d1000}, \rref{d16}, \rref{d361} -- \rref{d363},
\rref{d383}-- \rref{d384} are the main results of this section.
They show that the functional form of the interaction part of the
action is reproduced after integration over the fast variables
and, moreover, describe the changes of the of the coupling
constants in \reqs{d8}, \rref{d13}, \rref{d15}, under the
renormalization. This will enable us to write proper
renormalization group equations in a standard way. These equations
and their solutions are presented in the next Section.

\section{Renormalization group equations and their solution.}
\label{sec6}

\subsection{General structure of RG equations.}

We have demonstrated in the previous  section that the functional
form of the interaction part of the action \rref{d8}, \rref{d13},
\rref{d15}, is reproduced after integration over the fast
variables, $\Upsilon$. On the other hand, each integration over
the fast variables corresponds to the transform \rref{d1730} of
the high-energy cut-off \be \ln \w_c \to \ln \varkappa + \ln \w_c.
\label{d17300} \ee This enables us to write a most general form of
the RG equations as \be
\begin{split}
\frac{d \hat{\Gamma}_i }{d\ln\w_c}
&=
 \hat{\mathfrak{B}}_{\Gamma_i}\left(
\hat{\Gamma}_j; \hat{\mathcal B}_j; \hat{\mathbf \Delta}_j
\right);
\\
\frac{d \hat{\mathcal B}_i }{d\ln\w_c}
&=
\hat{\mathfrak{B}}_{\mathcal B_i}\left(
\hat{\Gamma}_j; \hat{\mathcal B}_j; \hat{\mathbf \Delta}_j\right);
\\
\frac{d \hat{\mathbf \Delta}_i}{d\ln\w_c}
&=
\hat{\mathfrak{B}}_{{\mathbf \Delta_j}}\left(
\hat{\Gamma}_j; \hat{\mathcal B}_j; \hat{\mathbf \Delta}_j\right).
\end{split}
\label{e0} \ee The renormalization group flow starts from
$\w_c\simeq v_F/r_0$ with the initial conditions \rref{d81},
\rref{d14}, \rref{d16} and it should stop at $\w_c \simeq {\rm
max}\left(T,|\theta|v_F/r_0\right)$. One sees immediately a
significant difference between the standard RG scheme and the
problem in hand. In our case, the entire coupling operators may be
renormalized, its renormalization is a functional of all the other
operators, etc. Thus, we are dealing with the {\em functional
renormalization group}.

Surprisingly, in the one loop approximation, the functional RG
equations can be obtained explicitly and solved in a closed form.

\subsection{One loop RG equations.}

The one loop equations for the kernels $\Gamma_i,\ {\mathcal
B}_i,\ {\mathbf \Delta}_i$ in Eqs.~\rref{d801}, \rref{d12},
\rref{d15a} are obtained by dividing both sides of \reqs{d16c},
\rref{d362},
 \rref{d384}, and \rref{d385}
by $\ln\varkappa$ and taking the limit $\ln\varkappa
=d\ln\w_c \to 0$.
As the result, we find
\begin{subequations}
\label{e01e02}
\be
\begin{split}
&\frac{d{\Gamma}_{1}\left(\theta;u,u_1;\r_\perp\right)}
{d\ln\w_c}
=-\mu_duu_1 \bar{f}_\perp(\r_\perp)
\left[{\Gamma}_{1}\left(\theta;u,u_1;\r_\perp\right)\right]^2;
\\
&
\frac{d{\mathcal B}_{1}^+\left(\dots\right)}
{d\ln\w_c}
=-2\mu_duu_1 \bar{f}_\perp(\r_\perp)
{\Gamma}_{1}\left(\dots\right)
{\mathcal B}_{1}^+\left(\dots\right);
\\
&
\frac{d{\mathcal B}_{1}^-\left(\dots\right)}
{d\ln\w_c}
=-\mu_duu_1 \bar{f}_\perp(\r_\perp)
{\Gamma}_{1}\left(\dots\right)
{\mathcal B}_{1}^-\left(\dots\right);
\\
&\frac{d {\Gamma}_{4}}
{d\ln\w_c}
=\frac{d {\mathcal B}_{4}^{\sigma}}
{d\ln\w_c} = 0.
\end{split}
\label{e01}
\ee
for the set of couplings not having counterpart in
the quadratic part of the action.
Here $\left(\dots\right)$ is the short hand notation
for the omitted arguments $\left(\theta;u,u_1;\r_\perp\right)$.

For the couplings affecting the quadratic part of the action, we
find \be
\begin{split}
&\frac{d{\Gamma}_{3}\left(\theta;u,u_1;\r_\perp\right)}
{d\ln\w_c}
=\mu_duu_1 \bar{f}_\perp(\r_\perp)
\left[{\Gamma}_{3}\left(\theta;u,u_1;\r_\perp\right)\right]^2;
\\
\\
&
\frac{d{\mathcal B}_{3}^+\left(\dots\right)}
{d\ln\w_c}
=2\mu_duu_1 \bar{f}_\perp(\r_\perp)
{\Gamma}_{3}\left(\dots\right)
{\mathcal B}_{3}^+\left(\dots\right);
\\
&
\frac{d{\mathcal B}_{3}^-\left(\dots\right)}
{d\ln\w_c}
=\mu_duu_1 \bar{f}_\perp(\r_\perp)
{\Gamma}_{3}\left(\dots\right)
{\mathcal B}_{3}^-\left(\dots\right);
\\
\\
&
\frac{d{\mathbf \Delta}_{3}^{++}\left(\dots\right)}
{d\ln\w_c}
=2\mu_duu_1 \bar{f}_\perp(\r_\perp)
\\
&\quad\quad\times
\left\{
{\Gamma}_{3}\left(\dots\right)
{\mathbf \Delta}_{3}^{++}\left(\dots\right)
+ \left[{\mathcal B}_{3}^+\left(\dots\right)\right]^2
\right\};
\\
&\frac{d{\mathbf \Delta}_{3}^{-+}\left(\dots\right)}
{d\ln\w_c} =\frac{d{\mathbf \Delta}_{3}^{+-}\left(\dots\right)}
{d\ln\w_c}
\\
&
\quad\quad=2\mu_duu_1 \bar{f}_\perp(\r_\perp)
{\mathcal B}_{3}^+\left(\dots\right){\mathcal B}_{3}^-\left(\dots\right);
\\
\\
&
{\Gamma}_{3}\left(\dots\right)
{\mathbf \Delta}_{3}^{--}\left(\dots\right)
=
\left[{\mathcal B}_{3}^-\left(\dots\right)\right]^2;
\\
\\
&\frac{d {\Gamma}_{2}}
{d\ln\w_c}
=\frac{d {\mathcal B}_{2}^{\sigma}}
{d\ln\w_c} =\frac{d {\mathbf \Delta}_{2}^{\sigma_1\sigma_2}}
{d\ln\w_c}= 0.
\end{split}
\label{e02}
\ee
\end{subequations}

The form of \reqs{e01e02} suggests immediately the following
scaling form for the coupling kernels \be
\begin{split}
&{\Gamma}_{i}\left(\theta;u,u_1;\r_\perp\right)
=\gamma_i\left[\xi\left(\theta;u,u_1;\r_\perp\right);
\gamma_i^0(\theta)\right]
\\
&{\mathcal B}_{i}^\sigma\left(\theta;u,u_1;\r_\perp\right)
=\beta_i^\sigma\left[\xi\left(\theta;u,u_1;\r_\perp\right)
;\gamma_i^0(\theta)
\right]
\\
&{\mathbf \Delta}_{i}^{\sigma_1\sigma_2}\left(\theta;u,u_1;\r_\perp\right)
=\Delta_i^{\sigma_1\sigma_2}\left[\xi\left(\theta;u,u_1;\r_\perp\right);
\gamma_i^0(\theta)
\right]
\\
&\xi\left(\theta;u,u_1;\r_\perp\right)= uu_1\mu_{d}\bar{f}_\perp(\r_\perp)\
\ln\left[{\rm min}\left(\frac{1}{\theta},
\frac{v_F}{r_0T}
\right)\right]
\end{split}
\label{e03} \ee where $\gamma_1^0=\gamma_2^0=\gamma_f$,
$\gamma_3^0=\gamma_4^0=\gamma_b$, see \reqs{c380}, \rref{d81}.
Comparing Eq. \rref{e03} with \reqs{e01e02}, we obtain
\begin{subequations}
\label{e04e05}
\be
\begin{split}
&\frac{d{\gamma}_{1}\left(\xi\right)}
{d\xi}
=
\left[{\gamma}_{1}(\xi)\right]^2;
\\
&
\frac{d{\beta}_{1}^+\left(\xi\right)}
{d\xi}
=2
{\gamma}_{1}\left(\xi\right)
{\beta}_{1}^+\left(\xi\right);
\\
&
\frac{d{\beta}_{1}^-\left(\xi\right)}
{d\xi}
=
{\gamma}_{1}\left(\xi\right)
{\beta}_{1}^-\left(\xi\right);
\\
&\frac{d {\gamma}_{4}}
{d\xi}
=\frac{d {\beta}_{4}^{\sigma}}
{d\xi} = 0,
\end{split}
\label{e04}
\ee
and
\be
\begin{split}
&\frac{d{\gamma}_{3}\left(\xi\right)}
{d\xi}
=-
\left[{\gamma}_{3}(\xi)\right]^2;
\\
&
\frac{d{\beta}_{3}^+\left(\xi\right)}
{d\xi}
=-2
{\gamma}_{3}\left(\xi\right)
{\beta}_{3}^+\left(\xi\right);
\\
&
\frac{d{\beta}_{3}^-\left(\xi\right)}
{d\xi}
=-
{\gamma}_{3}\left(\xi\right)
{\beta}_{3}^-\left(\xi\right);
\\
&
\frac{d\Delta_{3}^{++}\left(  \xi\right)  }{d\xi}=-2\Delta_{3}^{++}\left(
\xi\right)  \gamma_{3}\left(  \xi\right)  -2\left[  \beta_{3}^{+}
\left(\xi\right)
\right]
^{2}
\\
&
\frac{d\Delta_{3}^{-+}\left(  \xi\right)  }{d\xi}=
\frac{d\Delta_{3}^{+-}\left(  \xi\right)  }{d\xi}
=  -2  \beta_{3}^{-}
\left(\xi\right)
\beta_{3}^{+}
\left(\xi\right)
\\
&
\Delta_{3}^{--}\left(  \xi\right) \gamma_3(\xi)=
\left[  \beta_{3}^{-}
\left(\xi\right)
\right]
^{2}
\\
&\frac{d {\gamma}_{2}}
{d\xi}
=\frac{d {\beta}_{2}^{\sigma}}{d\xi}
=
\frac{d {\Delta}_{2}^{\sigma_1\sigma_2}}
{d\xi} = 0,
\end{split}
\label{e05}
\ee
Equations \rref{e04} -- \rref{e05} have to be solved with the
initial conditions [cf. \reqs{d81}, \rref{d14}, \rref{d16}]
\be
\gamma_i(\xi=0)=\beta_i^\pm(\xi=0)=\Delta_i^{\pm}(\xi=0)=\gamma_i^0,
\label{e06}
\ee
where $\gamma_1^0=\gamma_2^0=\gamma_f$, $\gamma_3^0=\gamma_4^0=\gamma_b$.
\end{subequations}

There is a good intuitive reason to separate the equations for the
coupling constants for $i=1,4$ from those for $i=2,3$. The latter
group contains the quadratic interaction breaking the
supersymmetry. Moreover, this would be the only group if we did
not introduce the hermitization procedure of Sec.~\ref{sec4a}.
Those are the modes that will directly contribute to the
observable quantities, see next Section, and we will call this
sector ``physical''.

The coupling constants  with $i=1,4$ are related to the fields
that appear as a result of the Hermitization procedure. Their
quadratic parts remain supersymmetric and that is why this sector
by itself does not contribute to any observables. This is the
reason why we will call this sector ``non-physical''.

In the one loop approximation, Eq. \rref{e04e05} these two
sectors do not talk to each other and we will consider them
separately.

\subsection{Solution of RG equations in the ``physical'' sector.}

Equations \rref{e05} is the system of the first order non-linear
equations with the triangular structure (there is no feedback of
constants $\Delta,\ \beta$ into evolution of the four particle
vertex $\gamma$). As such, it can be easily solved  with the
initial conditions, Eqs. \rref{e06}: \be
\begin{split}
&\gamma_{2}\left(  \xi\right) =\beta_{2}^{\pm}\left(  \xi\right)=
\Delta_{2}^{\pm\pm}\left(  \xi\right)
=\gamma_{f}(\theta);\\
\\
&\gamma_{3}\left(  \xi\right)  =\frac{1}{\xi_{b}^{\ast}+\xi};
\\
\\
&\beta_{3}^{+}\left(  \xi\right)  =\frac{\xi_{b}^{\ast}}{\left(  \xi_{b}^{\ast
}+\xi\right)  ^{2}};
\\
&\beta_{3}^{-}\left(  \xi\right)  =\frac{1}{\xi_{b}^{\ast}+\xi};
\\
\\
&\Delta_{3}^{++}\left(  \xi\right)  =\frac{2\xi_{b}^{\ast2}}{\left(  \xi
_{b}^{\ast}+\xi\right)  ^{3}}-\frac{\xi_{b}^{\ast}}{\left(  \xi_{b}^{\ast}%
+\xi\right)  ^{2}};
\\
& \Delta_{3}^{--}\left(  \xi\right)  =\frac{1}{\xi_{b}^{\ast}+\xi};
\\
&\Delta_{3}^{+-}\left(  \xi\right)  =\Delta_{3}^{-+}\left(  \xi\right)
=\frac{\xi_{b}^{\ast}}{\left(  \xi_{b}^{\ast}+\xi\right)  ^{2}},
\end{split}
\label{e07} \ee where we introduced the notation \be
\xi_{b}^{\ast}(\theta) \equiv \frac{1}{\gamma_b(\theta)} > 0.
\label{e08} \ee and the backscattering amplitude $\gamma_b^0$ is
defined in \req{c380}.

From \req{e07} we see that the forward scattering amplitude is not
renormalized in contrast to the backscattering ones. As we
consider the repulsive case, the amplitudes
$\gamma_{3},\beta_3,\delta_3$ tend to zero in the limit
$T,\left|\theta\right|  \to 0$ and this is a ``zero charge''
situation. This behavior should, in principle, be seen using the
conventional diagrammatic analysis. However, as the spin
excitations considered here correspond to two particle electron
Green functions, one should consider four particle electron Green
functions in order to identify the behavior described by Eq.
(\ref{e07}). Such interactions in four particles Green functions
have not been studied previously and the result of \req{e07} is a
major and decisive step in this direction. The ``zero charge''
behavior does not indicate any drastic changes of the ground state
of the system but is interesting on its own because it is
definitely not present in the orthodox Fermi liquid picture.

As the result corresponds to the zero-charge flow, the one loop
renormalization group and \req{e07} would solve the problem
completely. The renormalized amplitudes $\Delta$ can be used for
calculating the thermodynamic properties of the system, as it will
be done in the next section. On this route, however, a potential
reef may rise and we turn to the statement of this problem now.

\subsection{Solution of RG in ``non-physical'' sector and possible instability.}

Solving \req{e04} with the initial conditions
\rref{e06}
 \be
 \begin{split}
 &\gamma_{1}\left(  \xi\right) =\beta_{1}^{\pm}\left(  \xi\right)
 =\gamma_{f};\\we obtain
 \\
 &\gamma_{1}\left(  \xi\right)  =\frac{1}{\xi_{f}^{\ast}-\xi};
 \\
 \\
 &\beta_{1}^{+}\left(  \xi\right)  =\frac{\xi_{f}^{\ast}}
{\left(  \xi_{f}^{\ast }-\xi\right)  ^{2}};
 \\
 &\beta_1^{-}\left(  \xi\right)  =\frac{1}{\xi_{f}^{\ast}-\xi},
 \end{split}
 \label{e09}
 \ee
where we introduced the notation \be \xi_{f}^{\ast}(\theta) \equiv
\frac{1}{\gamma_f(\theta)} > 0, \label{e010} \ee and forward the
scattering amplitude $\gamma_f^0$ is defined in \req{c380}.

The behavior of the amplitude $\gamma_{1}\left(  \xi\right)$
from \req{e09} comes as a real surprise because it demonstrates the existence of
a logarithmic pole. This pole should be reached at $\xi=\xi_{f}$ and may
signal on an instability of the ground state because, at first glance, the
scenario looks similar to the one leading to the BCS theory of
superconductivity. Does the logarithmic pole in \req{e010} lead to a phase
transition? We do not try to answer this question in this paper but the
situation looks more complicated than usual because, within the RG scheme, the
 scattering amplitude $\gamma_{1}$ does not enter thermodynamic
quantities like e.g. the specific heat, see the next section, and
it does not couple to the physical sector at least on the level of
one loop renormalization group.

We can envision two different scenarios that may follow from the existence of
the pole in $\gamma_{1}$ in \req{e010}:
\begin{enumerate}
\item

 The amplitude $\gamma_{1}\left(  \xi\right)  $ does not enter any physical
quantity and, therefore, the pole does
not mean anything. In this case, we would be able to use all the equations for
the backscattering amplitudes down to $\xi=0,$ which would allow us to go in
temperature down to $T=0$. This would mean that there are non-analytical
corrections to the Fermi liquid but otherwise the Landau theory of Fermi
liquid is a correct low-temperature limit.

\item
 The logarithmic pole means that at $\xi=\xi_{f}^{\ast}$ a
reconstruction of the ground state (either in a form of phase
transition or sharp crossover) occurs at a critical temperature
$T_{c}$, which would manifest itself in a formation of the gap in
the non-physical sector, that would affect physical degrees of
freedom. In this case the ground state would change and one would
expect formation of an order parameter. Then, the present RG
treatment of the effective action would be applicable for
$T>T_{c}$ only. The formation of the gap means a breakdown of the
Fermi liquid description for the spin excitations. As concerns the
charge degrees of freedom, we have seen that they decouple from
the spin ones on the very early stage, see Sec.~\ref{sec3}, and
the Landau kinetic equation is applicable for them for arbitrary
small temperatures.
\end{enumerate}

It is still to be seen which of these two scenarios corresponds to
the original action and we relegate this question to the further
study\cite{elsewhere}. We emphasize, however, that the answer may
differ for $d=1$ and $d>1$, because there is a large set of
diagrams logarithmic in $1D$ and non-logarithmic for $d=2.3$, see
{\em e.g.} Fig.~\ref{fig7}.

\section{Specific heat}
\label{sec7}

We have performed the renormalization group calculations for the
case when the vectors $\n$ and $\n^{\prime}$ of two spin
excitations were close to each other (parallel or antiparallel
motion). Only in this limit one obtains large logarithms that
determined the renormalization of the vertices. A crucial question
is whether  or not this narrow region of the phase space can bring
an important contribution to thermodynamics or other physical
quantities. This is  not quite trivial question because the system
was not  assumed to be one- or quasi-one-dimensional, and one
could imagine that all the effect of the singularities in the
vertices would be washed out after the summation over the whole
phase space.

In fact, this almost parallel motion of the spin excitations does
not contribute much into the thermodynamic potential
$\Omega_s\left(  T\right)$ itself. Fortunately, this is not a very
interesting quantity and what one would like to know are
derivatives of the thermodynamic potential with respect to
temperature or other sources. In the present paper, we restrict
ourselves with the specific heat \be
C=-T\frac{\partial^2\Omega}{\partial T^2}. \label{f00} \ee Our
goal is to identify the non-analytic contributions to the specific
heat using the properties of the effective theory established in
the previous sections.

As we will show, our low-energy field theory is applicable for
the calculation of
\begin{equation}
\delta\Omega_s\left(  T\right)  =\Omega_s\left(  T\right)  -\Omega_s\left(
T=0\right),  \label{f3}%
\end{equation}
and we will focus in this section on the calculation of the latter
quantity.

We will present the main approximations and manipulations suitable
for any dimensions in Sec.~\ref{sec7a}.
We will collect the final results for two- and three- dimensional
systems in Sec.~\ref{sec7d} and discuss their relation to
the contribution of the Cooper channel in \ref{sec7c}.
For one dimensional systems approximations of
Secs.~\ref{sec7a},
will turn out not to be sufficient and we will have to take
additional terms into account specific for one-dimensional systems,
see Sec.~\ref{sec7c}.

\subsection{General formulae.}
\label{sec7a}

 Using the diagrammatic method of the calculations
 one can always cut one of the Green functions and
 express the  thermodynamic potential
 $\Omega\left(  T\right)  $ in terms of a sum
 over bosonic Matsubara frequencies $\omega_{n}=2\pi T n$ transmitted
through this particular Green function:
\begin{equation}
\Omega_s\left(  T\right)  =T\sum_{\omega_{n}}{\mathbb R}\left(
\omega_{n}\right)
\label{f3a}%
\end{equation}
where ${\mathbb R}\left(\omega_{n}\right)  $ is a function of the
  frequency to be calculated later.

The sums of the type \rref{f3a} are very often divergent at high
frequencies if one uses expressions available from a low energy
effective theory. This problem can be avoided calculating the
quantity $\delta\Omega\left(  T\right)$  from \req{f3}. Using the
Poisson formula, we  represent $\delta \Omega\left(  T\right)  $
in the form
\begin{equation}
\delta\Omega\left(  T\right)  =\sum_{l\neq 0}\int
{\mathbb R}\left(  \omega\right)  \exp\left(
  -\frac{il\omega}{T}\right)
\frac{d\omega}{2\pi
},
 \label{f4}%
\end{equation}
which improves the convergence significantly. The essential frequencies
in \req{f4}) are of the order of $T$ and are smaller then those frequencies that
form logarithms in the vertices. That is why the renormalized vertices
calculated in the previous section become useful.

To proceed with actual calculation,
we notice that if we kept in the action ${\cal S}\left\{  \psi\right\}$
only the supersymmetric part
${\cal S}_{0}\left\{  \psi\right\} + {\cal S}_{4}\left\{
  \psi\right\}$, see \reqs{c37}, \rref{c38}, \rref{d8}
only, we would obtain unity for the partition
function and, thus, no contribution to  the thermodynamic potential $\Omega$.
The interaction terms
$ {\cal S}_{2}\left\{  \psi\right\}$ and  $ {\cal S}_{3}\left\{  \psi\right\}$,
see \reqs{c40}, \rref{d13}, \rref{c42}, \rref{d15} violate
the supersymmetry and, as a result, one obtains finite contribution
to $\Omega_s$ only when expanding in such terms.

\begin{figure}[t]
\setlength{\unitlength}{2.3em}
\begin{picture}(10,11)
\put(-0.6,8.6){$\displaystyle{\frac{\delta\Omega_1}{T}=-\left[\frac{1}{2}\right]}$}
\put(2.1,7.4){\includegraphics[width=2.5\unitlength]{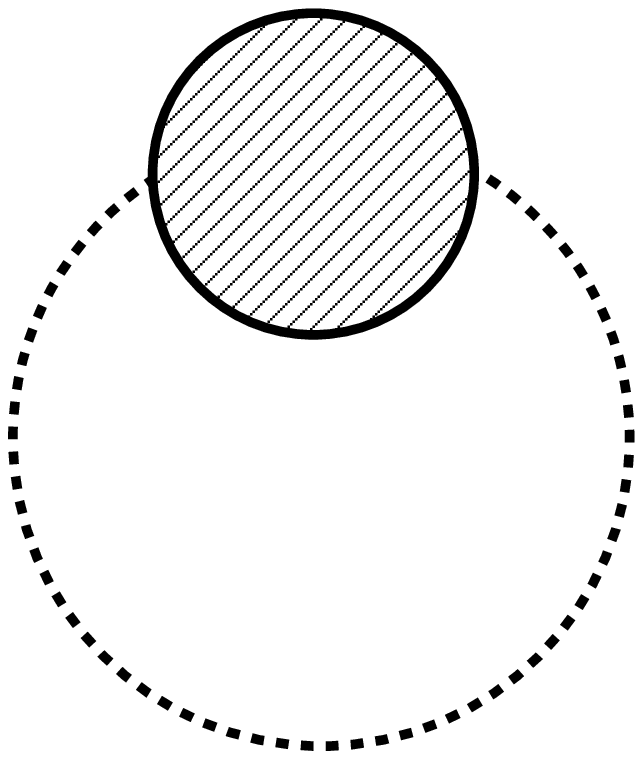}}

\put(7.9,7){\includegraphics[width=2.4\unitlength]{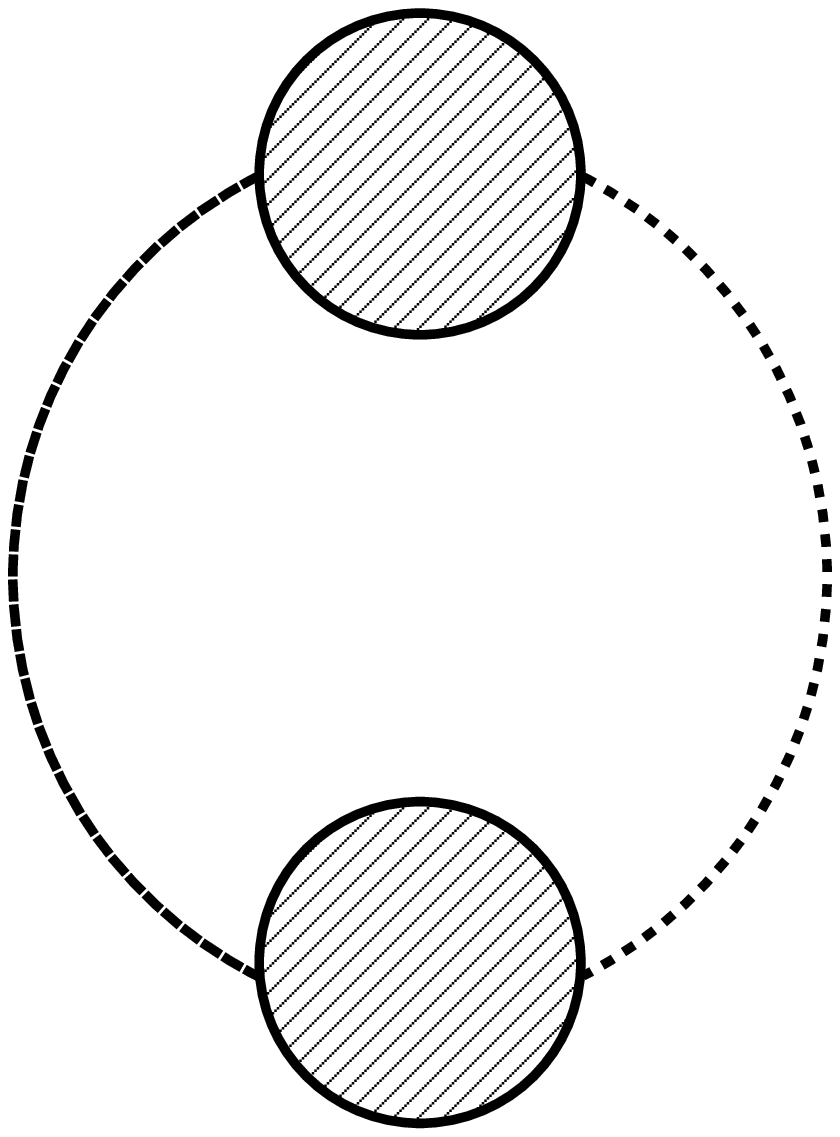}}
\put(4.8,8.6){$\displaystyle{;\ \, \frac{\delta\Omega_2}{T}=-\left[\frac{1}{4}\right]}$}

\put(1.5,10){a)} \put(6.5,10){b)}

\put(0,4.7){\includegraphics[width=3.9\unitlength]{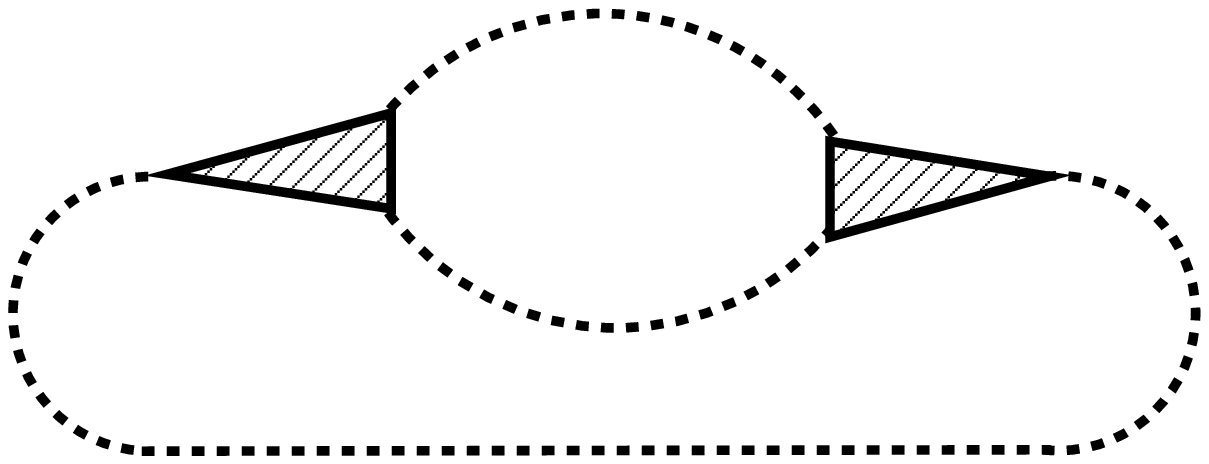}}
\put(5,4.3){\includegraphics[width=3.7\unitlength]{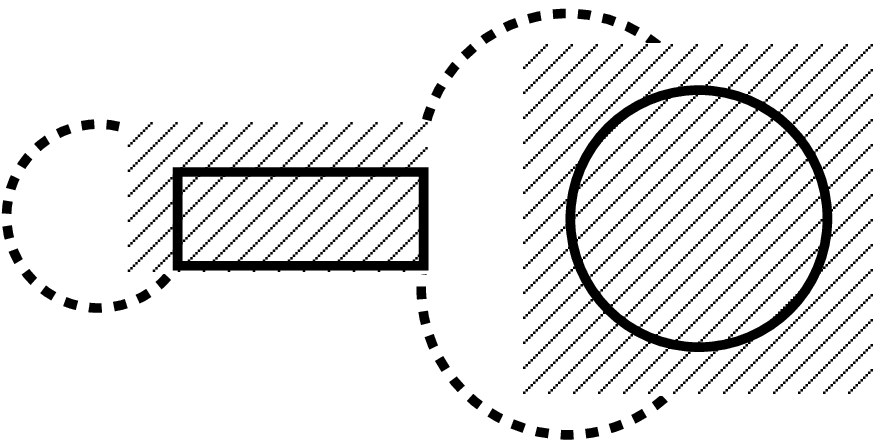}}

\put(1,6.5){c)} \put(6.5,6.5){d)}

\put(4.3,5){$=$}\put(9.3,5){$=0$}

\put(0.0,0.7){\includegraphics[width=3.4\unitlength]{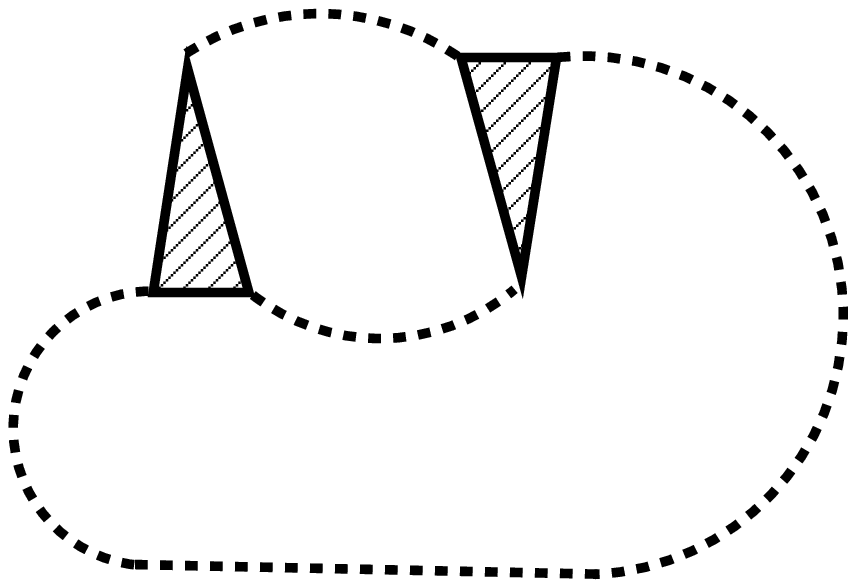}}
\put(4.8,0.7){\includegraphics[width=4.2\unitlength]{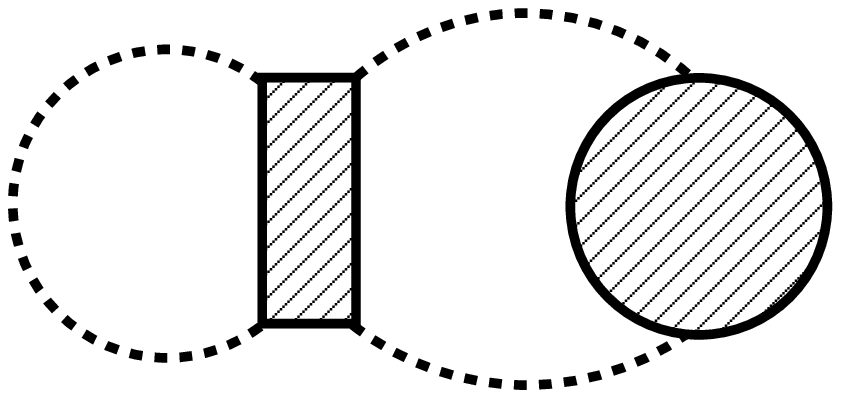}}
\put(4.,1.3){$+$}\put(9.3,1.3){$=0$}
\put(0.,2.6){e)} \put(9.1,2.6){f)}
\end{picture}
\caption{ Lowest order corrections to the temperature dependent
part of the thermodynamic potential $\delta \Omega(T)$. The
notations are explained in Sec.~\ref{sec5}, see Figs.~\ref{fig201}
-- \ref{fig6}. The coefficients in the brackets correspond to the
number of symmetries of the diagram. The dotted line for the Green
function means that summation over the frequency of this line is
performed with \req{f4} replacing the summation \rref{f3a}.
Diagrams c,d) vanish due to the supersymmetry. Diagrams e,f) are
separately of higher order in the forward scattering amplitude
and, moreover, cancel each other. }

\label{fig8}
\end{figure}

As all the high-frequency $\w \gtrsim T$ contributions are already
included into the renormalized value of the vertices, the
thermodynamic potential $\Omega\left(  T\right)  $ can be expanded
in terms of the renormalized action $\mathrm{{S}}_{2}\left\{
\Psi\right\}$, \req{d15}, and the lowest non-vanishing orders take
the form
\begin{equation}
\begin{split}
&\Omega_s\left(  T\right)  =\Omega_{1}\left(  T\right)  +\Omega_{2}\left(
T\right);
\\
&\Omega_{1}\left(  T\right)  =T\left\langle {\mathcal S}_{2}\left\{
\psi\right\}  \right\rangle _{0}\text{; \ \ \ \ }
\\
&
\Omega_{2}\left(  T\right)
=-\frac{T}{2}\left\langle
\left[{\mathcal S}_{2}\left\{
\psi\right\}\right]
 ^{2}\right\rangle _{0}, \label{f6}%
\end{split}
\end{equation}
and $\left\langle ...\right\rangle _{0}$ was defined in \req{d101}. The
corresponding diagrams are depicted in Fig.~\ref{fig8}a,b).

It will turn out that correction $\Omega_{1}$ is analytic function
of temperature whereas the most interesting term, $\Omega_{2}$ is
non-analytic. For the pedagogical reasons, we will consider an
analytic correction first and then use the gained knowledge for
more involved analysis of the non-analytic contribution.

\subsubsection{Analytic contribution.}

Using the rules of the diagrammatic technique of Sec.~\ref{sec5}
 we can obtain analytic expression for $\Omega_{1}\left(  T\right)$
(for unit volume) in terms of the function ${\mathbb K}$
introduced in \req{d386}: \be
\begin{split}
&\Omega_1=\frac{3 T}{4}  \sum_{\w_n}\sum_{i=2,3}\sum_{j=1}^4\lambda_{ij}
\int_0^1du\int d\n \int\frac{d^d\k}{(2\pi)^d}
\\
&\qquad\times
\sum_{\sigma_1\sigma_2=\pm}{\mathbb K}^{\sigma_1\sigma_2}_1(\w,\k,\n;u)
\Delta^{\sigma_1\sigma_2}_i(\theta=0;u,u;\k);\\
&\k_\perp=\k-\n(\k\cdot\n),\quad \r_\perp=\r-\n(\r\cdot\n)\\\\
&\Delta^{\sigma_1\sigma_2}_i(\theta;u,u;\k_\perp)
=\int\!\! d^d\r e^{-i\k\r}
\Delta^{\sigma_1\sigma_2}_i(\theta;u,u;\r_\perp)\bar{f}(|\r|),
\end{split}
\label{f70} \ee where the cutoff function $\bar{f}\left(
\r\right)$ is defined by \req{a20a}, all the other entries were
introduced in Secs.~\ref{sec4}, \ref{sec5}, and the convention
\rref{c110} for the angular integration is used. Replacing the
summation over the Matsubara frequency with the integrations
according to \reqs{f3a}--\rref{f4}, and using $\lambda_{ij}$ from
\req{d70e}, we obtain \be
\begin{split}
  \delta\Omega_{1}\left(  T\right)  &=  3
\int_{0}^{1}udu\int d\n
\sum_{l=1}^{\infty}\int\frac{d\omega}{2\pi}
\exp\left(\frac{-il\omega}{T}\right)
\\
&  \times
\int\frac{f\left(  \k\right) d\k}{(2\pi)^d}
\left(  \frac{i\omega+v_{F}\k\n}
{i\omega-v_{F}\k\n}\right)
\Delta
_{2}^{+-}\left(  \theta=
0\right)
,
\end{split}
\label{f7}%
\ee and the value of $\Delta_{2}^{+-}$ is given by \req{e07}. The
integral in \req{f7} contains integration over all directions of
the unit vector $\mathbf{n}$ with the normalization of Eq.
(\ref{a11a}). As we obtained in the previous section,
$\Delta_{2}^{+-}$ is not re-normalized by interaction and it is
given by its bare value $\gamma_f\equiv \gamma_f(\theta)=0$.
 Nevertheless, the remaining integral in Eq.~(\ref{f7})
gives a temperature dependent contribution and let us show how one
can calculate this integral.

The integration over $u$ is trivial and gives $1/2.$ The
integration over $\mathbf{k}$ can be performed separately for the
component $k_{\parallel}$ parallel to $\mathbf{n}$ and
$\mathbf{k}_{\perp}$ perpendicular to $\mathbf{n.}$ Essential
$k_{\parallel}$ are of the order of $T/v_{F}$, whereas $k_{\perp}$
are of the order of the maximum momentum $k_{c}\simeq r_{0}^{-1}$.
Therefore, we can neglect $k_{\parallel}$ in the cutoff function.
Integration over  $k_{\parallel}$ is, then, immediately performed
with the result

\be
\begin{split}
\delta\Omega_{1}\left(  T\right)   &  =\lim_{\eta\rightarrow+0}\sum
_{l=1}^{\infty}\int\frac{d\omega}{2\pi}\left\vert \omega\right\vert
\exp\left(  -i\frac{l\omega}{T}-\eta\left\vert \omega\right\vert \right)
\\
&  \times\frac{3\gamma_{f}}{2v_{F}}\int f_{0}\left(  k_{\perp}r_{0}\right)
\frac{d^{d-1}\mathbf{k}_{\perp}}{\left(  2\pi\right)  ^{d-1}} \label{f8a}%
\end{split}
\ee
The small parameter $\eta$ in the exponential is added to provide the
convergence of the integral over $\omega.$

Expression in the first line   of Eq.~(\ref{f8a}) takes after the
$\omega$-integration the following form:
\begin{equation}
\lim_{\eta\rightarrow0}\sum_{l=1}^{\infty}\left[
\frac{1}{\left(  \eta
+\frac{il}{T}\right)^2}
+
\frac{1}{\left(  \eta
+\frac{-il}{T}\right)^2}
\right]  =-\frac{\left(
\pi T\right)  ^{2}}{3}. \label{f9}%
\end{equation}
Substituting Eqs. (\ref{f8a})--(\ref{f9}) into Eq. (\ref{f00}) we obtain the
corresponding contribution $\delta C_{1}\left(  T\right)  $ to the specific
heat
\begin{equation}
\begin{split}
\frac{\delta C_{1}}{T} &=\frac{\pi \left(3\gamma_{f}\right)}{6v_{F}\lambda_{0}%
^{d-1}}, \quad
\lambda_{0}^{1-d}=\int f_{0}\left(  k_{\perp}r_{0}\right)  \frac
{d^{d-1}\mathbf{k}_{\perp}}{\left(  2\pi\right)  ^{d-1}}.
\end{split}
\label{f10}%
\end{equation}
The parameter $\lambda_{0}\simeq 1/r_0$ becomes of the order of
$k_{F}^{-1}$ on the limit of the applicability of the theory.

We see from \req{f10} that the correction $\delta C_{1}\left(
  T\right)$
does not change the linear dependence of the specific heat on
temperature.

Let us discuss the significance of this result and its relation to
the free mode consideration of Sec.~\ref{sec3c}. First of all,
direct comparison shows that \req{f7} is equivalent to the first
term in the perturbative expansion of temperature dependent parts
of \reqs{b31prime}, \rref{b30prime}. Therefore, the singlet
contribution \rref{b30prime} also has to be taken into account in
\req{f10}, which leads to the replacement
$\left(3\gamma_{f}\right) \to
\left(3\gamma_{f}-\gamma_f^\rho\right)$, where
$\gamma_f^\rho\equiv \gamma^\rho(\theta=0)$.

Equation \rref{f10} gives the contribution which does not depend
on the cutoff in one dimension. In this case it describes the
renormalization of the velocities of the spin and charge modes in
Luttinger liquid regime. In contrast, in higher dimensions, $d>1$,
the coefficient does depend on the cut-off. However, all this
contribution coming from the small distances can be ascribed to
the renormalization of the effective mass and included into the
partition function of non-interacting quasiparticles, [$Z_0$ from
\req{b13}].

The effective theory of the interacting collective modes being the
effective low energy theory does not describe such ultraviolet
corrections and that is why we cannot identify the numerical
coefficient from our theory. However, the effective low energy
theory\footnote{Without any drawbacks one can formulate the
effective low-energy theory for the Fermi liquid function such
that $3\gamma_{f}=\gamma_f^\rho$. The ultraviolet correction to
the specific heat does not appear in this case at all.}
 does describe the non-analytic
corrections as the latter are associated with the spatial
scales $\simeq v_F/T \gg 1/k_F$. We turn to the derivation
of such non-analytic corrections now.

\subsubsection{Non-analytic contribution.}

The analytic expression for the diagram Fig.~\ref{fig8}b reads
[see \req{f70} for the notations] \be
\begin{split}
&\Omega_2=-\frac{3 T}{4}  \sum_{\w_n}\sum_{i=2,3}\sum_{j=1}^4\lambda_{ij}
\iint_0^1du_1du_2\iint d\n_1d\n_2
\\
&\qquad\times
\int\frac{d^d\k}{(2\pi)^d}\sum_{\sigma_1,\dots,\sigma_4=\pm}
\\
&\qquad\times\Big\{
{\mathbb K}^{\sigma_1\sigma_2}_j(\w,\k,\n_1;u_1)
\Delta^{\sigma_2\sigma_3}_i(\widehat{\n_1\n_2};u_1,u_2;\k_\perp)
\\
&\qquad\times
\ \ {\mathbb K}^{\sigma_3\sigma_4}_j(\w,\k,\n_2;u_2)
\Delta^{\sigma_4\sigma_1}_i(\widehat{\n_2\n_1};u_2,u_1;\k_\perp)\Big\}
;\\
\end{split}
\label{f11} \ee where the convention \rref{c110} for the angular
integration is implied, and in definition of $\k_\perp$, the
direction  $\n$ means $\n=\left(\n_1+\n_2\right)/2$. Using
\reqs{d386} and \rref{d70e}, keeping only terms with the pole
location such that the result does not vanish after integration
over $k_\parallel=\k\cdot\n$, and using replacements
\reqs{f3a}--\rref{f4}, we find \be
\begin{split}
&  \delta\Omega_{2}\left(  T\right)  =-6\lim_{\eta\rightarrow+0}\
\sum_{l=1}^{\infty}
\int\frac{d\w}{(2\pi)}
\exp\left(  -i\frac{l\omega}{T}-\eta\left\vert
\omega\right\vert \right) \\
&  \times
\int\!\! d\n_1d\n_2
\int \frac{
d^{d}\k}{\left(  2\pi\right)  ^{d}}
Y\left(\widehat{\n_1\n_2};\k_\perp;k_\parallel\right){\mathscr P}
_{d}\left(  \omega,\k;\n_1,\n_2\right),
\end{split}
\label{f12}
\ee
and the convention \rref{c110} for the angular integration is implied.
The function $Y\left(\theta;\k_\perp\right)  $ defined
as
\be
\begin{split}
 & Y\left(\theta;\k_\perp,k_\parallel\right)  =
\iint_{0}^{1}u_1u_2\,du_1du_2
\\
&\qquad\times
\Big\{\left[{\Delta}_{3}^{+-}\left(\theta;u_1,u_2;\k_\perp,
k_\parallel\right)  \right]^{2}\\
&\qquad +
{\Delta}_{3}^{++}\left(\theta;u_1,u_2;\k_\perp,k_\parallel\right)
{\Delta}_{3}^{--}\left(\theta;u_1,u_2;\k_\perp,k_\parallel\right)
\Big\}
\end{split}
\label{f13}
\ee
will be the most important entry in the final expression for the specific
heat. In \req{f13}, we wrote explicitly the transverse
and the longitudinal momenta of $\k$, see \req{f7}, as their
role will be different.

The formfactor
\be
{\mathscr P}_{d}\left(  \omega,{\k};\n_1,\n_2\right)
=\frac{\left(  i\omega+v_{F}\k\n_2\right)  \left(
i\omega-v_{F}\k\n_1\right)  }{\left(  i\omega-v_{F}\k\n_2\
\right)  \left(  i\omega+v_{F}\k\n_1\right)  }
\label{f14}%
\ee
 depends on the
dimensionality of the system and it describes basically
the free propagation of the two spin excitations in almost opposite
directions.

If one used the bare values of $\Delta$, see \req{d16}, we would
obtain the second term in the expansion of Eq. \rref{b31prime} in
powers of $\gamma$. The main advantage of \req{f14} is that it
accounts for the logarithmic renormalization of the quadratic
interaction obtained in Secs.~\ref{sec5},\ref{sec6}, see
\req{e08}.

The non-analytic contributions originate from the small region of
the phase space $|\n_1-\n_2| \ll 1$. It enables us to introduce
[cf. Sec.~\ref{sec5b}] \be
\begin{split}
& \n=(\n_1+\n_2)/{2};\quad
\delta\n=\n_1-\n_2;\\
&k_\parallel=\k\cdot\n; \quad \k_\perp=\k-k_\parallel \n,
\end{split}
\label{f140}
\ee
and integrate over $k_\parallel$ in \req{f13}.

To facilitate the integration we introduce the function
\be
\begin{split}
&{\mathcal P}_{d}\left(\delta\n;{k}_\perp\right)\equiv
\lim_{\eta\rightarrow+0}\
\sum_{l=1}^{\infty}
\int\frac{d\w dk_\parallel}{\left(2\pi\right)^2}
\exp\left(  -i\frac{l\omega}{T}-\eta\left\vert
\omega\right\vert \right)
\\
&
 \times  \frac{1}{2}
\sum_{\pm}
{\mathscr P}_{d}\left(  \omega,\pm{\k}_\perp+
k_\parallel\n;\n+\frac{\delta\n}{2},\n-\frac{\delta\n}{2}\right).
\end{split}
\label{f141}
\ee
Using
\[
\lim_{\eta\rightarrow+0}\
\sum_{l=1}^{\infty}
\int{d\w}
\exp\left(  -i\frac{l\omega}{T}-\eta\left\vert
\omega\right\vert \right)
\left(\int dk_\parallel {f^2(\k)}\right)=0,
\]
we obtain
\begin{subequations}
\label{f142ab} \be {\mathcal P}_{d=1}=0, \label{f142a} \ee which
means that for one dimensional systems there is no contribution to
the specific heat of the second order in the backscattering
amplitude in accord with known results\cite{lukyanov}.
This does not mean, however that there are no logarithmic
contributions to the specific heat in $1d$ at all. Corrections of
the third order in the effective amplitude will be recovered in
Sec.~\ref{sec7c}.

For the higher dimensions we obtain
\be
\begin{split}
{\mathcal P}_{d=2,3}=
\lim_{\eta\rightarrow+0}\
\sum_{l=1}^{\infty}
\int\frac{d\w}{\left(2\pi\right)}
\exp\left(  -i\frac{l\omega}{T}-\eta\left\vert
\omega\right\vert \right)
\\
\times
\left(  \left\vert \omega\right\vert -\frac{4\left\vert
\omega\right\vert ^{3}}{4\omega^{2}+\left( \delta\k_{\perp}\n\right)
^{2}v_{F}^{2}}\right).
\end{split}
\label{f142b} \ee The first term in \req{f142b} is similar to
\req{f8a} and produces only analytic contribution to the specific
heat. We will disregard this term and focus only on the second
contribution. This is this contribution that quantifies the effect
of the small region of the phase space where the interaction
renormalization is strong. It is worth noting that the
characteristic $k_\parallel$ contributing into the result where of
the order of $T/v_F$ and, therefore, the separation of the
integration on the transverse and longitudinal parts is well
justified.
\end{subequations}

Substituting \req{f142b} in \req{f13} and performing the Fourier
transform in the transverse direction we find for $d>1$ \be
\begin{split}
& \delta\Omega_{2}\left(  T\right)  =-\frac{12
}{v_{F}^{2}}
\iint\frac{ d\n d\delta\n }{|\delta\n |}\int d x
{\bar{Y}\left(  |\delta\n|; x\right)}
\\
&
\times
\lim_{\eta\to +0}
\sum_{l=1}^{\infty
}\int\frac{\w^2d\omega}{2\pi}
\exp\left(  -\frac{2\left\vert \omega x\right\vert }
{\left\vert \delta{\n}\right\vert v_{F}
}-\frac{i\omega l}{T}-\eta\left\vert \omega\right\vert \right),
\end{split}
 \label{f19}
\ee
where the integration over $\n$ and $\delta\n$ are performed using conventions
\rref{c110} and \rref{a11a} respectively, and
\be
 \bar{Y}\left(\theta;|\r_\perp|\right)
=\int \frac{d^{d-1}\k_\perp}{(2\pi)^{d-1}}
e^{i\k_\perp\r_\perp}Y\left(\theta;\k_\perp;\k_\parallel=0\right).
\label{f190}
\ee

Integration over $\w$ and $\n$ in \req{f19} can be immediately
performed with the result
\be
\begin{split}
 \delta\Omega_{2}\left(  T\right)  & =-\frac{6T^3}{\pi^2 v_{F}^{2}}
\sum_{l=1}^{\infty} \int
 \frac{d^{d-1}\delta\n}
{2^{d-2} |\delta\n |}
\int dx
\\
&
\times
{\bar{Y}\left(  |\delta\n|;\ x\right)}
{\rm Re}\left(\frac{1}{il+
\frac{
2 T \vert \left\vert x\right\vert
}{
\left\vert \delta{\n}\right\vert v_{F}
}}\right)^3,
\end{split}
 \label{f191}
\ee
where integration over $\delta\n$
has to be understood as integration over usual $d-1$ dimensional
vector with ${|\delta n|\lesssim 1}$.

The remaining integrations in \req{f19} are slightly different for
$d=2,3$ and we will describe them separately.
For $d=2$, $\delta\n$ is a one-dimensional variable.
After obvious rescaling, \req{f191} gives
\be
\begin{split}
 \delta\Omega_{2}^{d=2}\left(  T\right)  & =
-\frac{12 T^3}{\pi^2 v_{F}^{2}}\sum_{l=1}^{\infty}
\frac{1}{l^3}
\int dx
\\
&
\times
{\rm Re}\int\limits_{0}^{\frac{l v_F}{2 |x| T}}
\frac{ {d\phi}\ \phi^2
\ \bar{Y}
\left(
\frac{2\phi |x| T}{l v_F};
\ x
\right)
}
{\left(i\phi + 1\right)^3}
.
\end{split}
 \label{f192}
\ee
The integral in \req{f192} convergent at $\phi\simeq 1$.
Because $x \simeq r_0$, and $T \ll v_F/r_0$, the upper
limit in the integral can be put to infinity. On the other hand,
 according to \req{e03}, $Y(\theta)$ does not depend on
$\theta$ at $\theta < T r_0/v_F$, therefore, we can put the first
argument in $\bar{Y}$ equal to zero. The remaining integral takes
the form
\[
\int_0^\infty
\frac{ {d\phi} \phi^2}
{\left(i\phi
+ 1\right)^3}=-\frac{\pi}{2},
\]
 The sum over $l$ is trivial, and we obtain
\be
\delta\Omega_{2}^{d=2}\left(  T\right) =
\frac{6\zeta\left(  3\right)  T^{3}}{\pi
v_{F}^{2}}Y\left(\theta=0\right),  \label{f24}%
\ee where $\zeta(3)\approx 1.202\dots$ is the Riemann
$\zeta$-function and
\be Y\left(\theta\right)\equiv
Y\left(\theta;\k=0\right), \label{f241} \ee see \req{f13}. The
value $\theta=0$ in the function $Y\left(  \theta\right)  $
entering Eq. (\ref{f24}) corresponds to the exactly backward
scattering.

Let us consider, now the  three-dimensional case. The result turns
to be logarithmic with the main contribution from $Tx/v_F\lesssim
|\delta_n|\gg 1$. Expanding the last factor in Eq. \ref{f191}and
keeping only the first non-vanishing contribution we find \be
\begin{split}
 \delta\Omega_{2}^{d=3}\left(  T\right)  & =\frac{36 T^4}{\pi^2 v_{F}^{3}}
\sum_{l=1}^{\infty}
\frac{1}{l^4}
\!\int
 \frac{d^{2}\delta \n}
{ |\delta\n |^2}
\!\!\int\limits_0^{\frac{v_F|\delta_n|}{T}}\!\!  x dx
{\bar{Y}\left(  |\delta\n|;\ x\right)},
\end{split}
 \label{f242}
\ee As the integrand decays rapidly at $x\simeq r_0$, the upper
limit in the last integral can be put to infinity for
$\theta=|\delta\n| \gtrsim Tr_0/v_F$. Using
\[
2\pi \int\limits_0^{\infty}\!\!  x dx
{\bar{Y}\left(\theta;\ x\right)} = Y(\theta),
\]
see \req{f241},
we obtain
\be
\delta\Omega_{2}^{d=3}\left(  T\right)  =\frac{2\pi^{2}T^{4}}{5v_{F}^{3}}%
\int\limits_{\frac{Tr_0}{v_F}}^{1}
\frac{d\theta\ Y\left(  \theta\right)}{\theta}.
\label{f30}%
\ee

Equations \rref{f24} and \rref{f30}
are the main results of this subsection. They show that the thermodynamic
potential
for the both two and three dimensional systems can be expressed in
terms
of the function
$Y\left(  \theta\right)  ,$ that has to be determined from \reqs{f13},
and \rref{e07}.
Carrying on this program and obtaining the final
expressions for the specific heat will be
subject of the next subsection.
Here, we just emphasize the difference between
\reqs{f24} and  \rref{f30}. Namely, in $2d$ case one has to take the function
$Y$ at strictly $\theta=0,$ whereas in $3d$
it involves integrals over all angles. At the same time,
even in $3D$, the main contribution comes with
the logarithmic accuracy from small angles $\theta\lesssim 1\,,$ that
describe again the scattering close to backward one.

\subsection{Final results for two- and three- dimensional systems.}
\label{sec7d}

From now on we will concentrate only on the
non-analytic contribution $\delta\Omega_2^{d=2,3}$
and that is why we will omit the subscript.
Before writing the final results in a general form,
it is instructive to perform the calculation
replacing the quadratic interaction constants $\Delta^{\pm\pm}$
in \reqs{f13} and \rref{f241} with  their bare values \rref{d16}.
After integration over $u_{1,2}$ in \req{f13}
one obtains
\begin{equation}
Y^{\left(  0\right)  }\left(  \theta\right)
=\frac{1}{2}\left[\gamma_{b}(\theta)\right]^2
f^{2}\left(  0\right)  =\frac{1}{2}\gamma_{b}^{2}(\theta), \label{f32}%
\end{equation}
where $\gamma_{b}(\theta)$ is defined in \req{c380},
and $f\left(  \mathbf{k}\right)  $ is  given by Eq. (\ref{a7c}).
Substitution of \req{f32} into \reqs{f24} and \rref{f30}
yields
\begin{subequations}
\begin{equation}
\delta\Omega^{d=2}_{\left(t,  {\rm bare}\right)  }\left(  T\right)  =
 3\left[\gamma_{b}\right]^2
\frac{\zeta\left(
3\right) T^{3}}{\pi v_{F}^{2}};
\quad
\gamma_{b}\equiv\gamma_{b}(\theta=0),
\label{f33}%
\end{equation}%

where we restored the explicit subscript ``t'' for the triplet
contribution. If  $Y^{(0)}\left(  \theta\right)$ is a smooth
function for $\theta\to 0$, the three-dimensional thermodynamic
potential reads with the logarithmic accuracy
\begin{equation}
\delta\Omega^{d=3}_{\left( t,{\rm bare}\right)  }\left(  T\right)  =
3\left[\gamma_{b}\right]^2
\frac{\pi^{2}%
T^{4}}{15v_{F}^{3}}\ln\frac{\varepsilon_{F}}{T}. \label{f34}%
\end{equation}
\label{f3334}
\end{subequations}

Similarly to \req{f10}, formula \rref{f3334} is
 equivalent to the second term
in the perturbative expansion of temperature dependent parts of
\reqs{b31prime}, \rref{b30prime}. Therefore, the singlet
contribution \rref{b30prime} has also to be taken into account in
\req{f3334}, which leads to the replacement
$\left(3\gamma_{b}^2\right) \to
\left(3\gamma_{b}^2+\left[\gamma_b^\rho\right]^2\right)$, where
$\gamma_b^\rho\equiv \gamma^\rho(\theta=\pi)$. Unlike the analytic
contribution, \reqs{f3334} do not contain the ultraviolet cut-off
and their contribution can not be ascribed to the renormalization
of the effective mass. These are clear effects of the contribution
of the bosonic collective modes of the Fermi liquid into the
thermodynamics of the system.

Using \reqs{f3334} in \req{f00} and introducing the density of
particles
\begin{equation}
N_{d=2}=\frac{p_{F}^{2}}{2\pi},\text{ }N_{d=3}=\frac{p_{F}^{3}}{3\pi^{2}}
\label{f35}%
\end{equation}
we write the correction to the specific heat per particle $\delta c
=\delta C/N$ as
\begin{subequations}
\begin{equation}
\delta c_{d=2}^{\left(  {\rm bare}\right)  }=-\frac{3\zeta\left(  3\right)  }{\pi
}\left(  \frac{T}{\varepsilon_{F}}\right)  ^{2}\left(
3\gamma_{b}^2+\left[\gamma_b^\rho\right]^2
\right);  \label{f36}%
\end{equation}%
\begin{equation}
\delta c_{d=3}^{\left( {\rm bare}\right)  }\!=-\frac{3\pi^{4}}{10}\left(  \frac
{T}{\varepsilon_{F}}\right)  ^{3}\!\ln\left(  \frac{\varepsilon_{F}}{T}\right)
  \label{f37}%
\left(
3\gamma_{b}^2+\left[\gamma_b^\rho\right]^2
+{\cal O}\left(\gamma^3\right)
\right).
\end{equation}
\label{f3637}
\end{subequations}
[The last term in \req{f3637} is
the contribution of the third order in coupling constant
which was obtained in Ref.~\onlinecite{chubukov3}
and we refer the reader to this reference for
the explicit form of the coefficients in this term and
will not write explicitly in the subsequent considerations.]
It is worth reminding that $p_F$ is not renormalized by
interaction, whereas $\epsilon_F\equiv v_Fp_F/2$
is significantly affected\cite{landau}. Actually,
$v_F$ has the meaning only as a quantity describing the
slope in a leading linear in temperature quasiparticle contribution to
the specific heat.

Equations \rref{f3637}  agree with the corresponding expressions
obtained previously in a number of works using conventional diagrammatic
expansions (see {\em e.g.}
Refs.~\onlinecite{chubukov3,chubukov4}, for the latest
 developments consisting in accurate evaluation
of the angular and $q$ integrals in expressions similar to \reqs{b30},
\rref{b31} and obtaining correct analytic expressions for the first time)
\footnote{
The definitions of the coupling constants
$\gamma$, \req{b310}, differs by the factor $2$ from the one of
Ref.~\onlinecite{chubukov3,chubukov4} because we used the density of states
$\nu$ per one
spin orientation but not the total density of states.}.

Using the conventional diagrammatic technique one can hardly go
beyond the first orders, which would be definitely enough for the
singlet channel. At the same time, using the present bosonization
scheme we have found for the first time the logarithmic
contributions discussed in the previous Sections and have derived
and solved proper renormalization group equations. Now we can
include the logarithmic contributions into the formulas for the
specific heat. The only thing that remains to be done is to
calculate the function $Y\left(\theta\right)  $ from \reqs{f241},
\rref{f13}.

Using the explicit expressions \rref{e07} for the couplings
$\Delta^{\pm\pm}_3$, and the formula
\[
\begin{split}
\iint_0^1du_1du_2
&\left(u_1u_2\right)^2
\Bigg[\frac{\left(1+x_1 u_1u_2\right)^2+\left(1+x_2 u_1u_2\right)^2}
{\left(1+x_1 u_1u_2\right)^3\left(1+x_2
      u_1u_2\right)^3}
\\
&\quad
-\frac{(x_1+x_2)u_1u_2}{2\left(1+x_1
      u_1u_2\right)^2\left(1+x_2
      u_1u_2\right)^2}
\Bigg]
\\
&
=\frac{1}{2\left(1+x_1 \right)\left(1+x_2 \right)},
\end{split}
\]
we reduce \req{f13}  to
the form
\be
\begin{split}
 & Y\left(  \theta\right)   =\frac{\gamma_{b}^{2}(\theta)}{2}
\left[\int\frac{d^{d-1}\r_{\perp}}{r_0^{(d-1)}}
 \frac{
 \bar {f}_{\perp}\left(\frac{|\r_{\perp}|}{r_0}\right)
 }{  1+
  \bar {f}_{\perp}\left(\frac{|\r_{\perp}|}{r_0}\right)
 {\mathscr X}(\theta)}\right]^2
,
\end{split}
\label{f37c}
\ee
where function $\bar{f}_{\perp}\left(  r_{\perp}%
/r_{0}\right)  $ is given by \req{d27b}.
The variable ${\mathscr X}$ is defined as
\begin{equation}
\begin{split}
&{\mathscr X}(\theta)
=-\mu_{d}\gamma_{b}(\theta)\ln\left[  \max\left\{  \theta,T/\varepsilon
_{0}\right\}  \right];
\\
&{\mathscr X}(T)\equiv{\mathscr X}(\theta=0)
=\mu_{d}\gamma_{b}\ln\left(\frac{\varepsilon
_{0}}{T}  \right),
\end{split}
 \label{f37b}
\end{equation}
with the parameter $\mu_{d}$ given by \req{d27},
and the cut-off energy defined as $\varepsilon_0=\frac{v_F}{r_0}\simeq \varepsilon_F$.

Equation (\ref{f37c}) gives the most general form of the function
$Y\left( \theta\right)  $ for any function $f$. The asymptotics of
the function $Y\left(  \theta\right)  $ in the limit $X\ll1$ can
easily be written as
\begin{equation}
Y\left(  \theta\right)  =\gamma_{b}^{2}
(\theta)\left(  \frac{1}{2}-{\mathscr X}(\theta)
r_{0}^{d-1}\int f^{2}\left(  \mathbf{k}_{\perp}\right)  \frac{d^{d-1}%
\mathbf{k}_{\perp}}{\left(  2\pi\right)  ^{d-1}}\right),  \label{f37d}%
\end{equation}
where the cutoff function $f\left(  \mathbf{k}\right)  $ is given
Eq. (\ref{a7c}).

The first term in Eq.~(\ref{f37d}) is what has been used when
deriving \reqs{f3637}. The second term leads to an additional
logarithm in the $T^{d}$-dependence of the specific heat in both
two and three dimensional systems. Notice, that the latter term
depends on the form of the ultraviolet cut-off $f(\k)$.  As the
function $f\left(  \mathbf{k}\right)  $ has been introduced in our
theory phenomenologically, the complete results depend on its
form, which is a deficiency of our low energy bosonization
approach, where the bare coupling constants were introduced
independently of the ultraviolet cutoff. At the same time, we will
see that it gives at least a good qualitative description of the
interesting temperature behavior, and will address the issue of
the dependence of the coupling constants vs. the cutoff function
in the next subsection.

A simple expression for the specific heat for arbitrary ${\mathscr X}$
can be obtained if
we choose the function $\bar{f}_{\perp}\left(  r_{\perp}/r_{0}\right)  $,
\req{d27a} as
\begin{equation}
\bar{f}_{\perp}\left(  r_{\perp}/r_{0}\right)  =\exp\left(  -r_{\perp}%
/r_{0}\right)  \label{f38}%
\end{equation}

The choice \rref{f38} in the coordinate space corresponds
to
\be
f_{d=2}(\k)=\frac{1}{1+\k^2\r_0^2};
\quad f_{d=3}(\k)=\frac{1}{\left(1+\k^2\r_0^2\right)^{3/2}}
\label{f38a}
\ee
in the momentum space.

Substituting Eq. (\ref{f38}) into Eq. (\ref{f37c}), and performing
the remaining integration, we obtain
\begin{equation}
\begin{split}
&Y_{d=2}\left(  \theta\right)  =
\frac{\gamma_{b}^{2}\left\{\ln\left[1+{\mathscr
        X}(\theta)\right]\right\}^2}
{2\left[{\mathscr X}(\theta) \right]^2};
\\
&Y_{d=3}\left(  \theta\right)  =
\frac{\gamma_{b}^{2}\left\{{\mathrm Li}_2\left[-{\mathscr
        X}(\theta)\right]^2\right\}^2}
{2\left[{\mathscr X}(\theta) \right]^2},
\label{f39}%
\end{split}
\end{equation}
where ${\mathrm Li}_2(x)=\sum_{k=1}^\infty x^k/k^2$ is the
polylogarithm function.
 Using Eq.~(\ref{f39}) we can write the asymptotics of the function $Y\left(
\theta\right)  $ for $X\gg1$ as%
\begin{equation}
Y_{d=2}\left(  \theta\right)
\simeq\frac{\gamma_{b}^{2}\ln^2 {\mathscr X}}{2{\mathscr
    X}^{2}};
\quad Y_{d=3}\left(  \theta\right)
\simeq\frac{\gamma_{b}^{2}\ln^4 {\mathscr X}}{8{\mathscr
    X}^{2}}.
 \label{f40}%
\end{equation}
The asymptotic behavior ${\mathscr X}^{-2}$ in \req{f40} is not
very sensitive to the form of the function $f\left(
\mathbf{k}\right)$, although the power of $\ln {\mathscr X}$ is
not universal. On the other hand, more accurate treatment of those
double logarithmic temperature dependencies would be an
overstepping of the accuracy of the one loop approximation anyway.

The final formula for the specific heat $\delta c$ per particle can be written
with the logarithmic accuracy as
\begin{subequations}
\label{f4142}
\begin{equation}
\delta c_{d=2}=-\frac{3\zeta\left(  3\right)  T^2}
{\pi\varepsilon_{F}^2}
\left\{
\left[ \gamma_{b}^{\rho}\right]^2+\frac{3\gamma_b^2
\left\{\ln\left[1+{\mathscr
        X}(T)\right]\right\}^2}
{2\left[{\mathscr X}(T) \right]^2}
\right\};  \label{f41}%
\end{equation}%
\begin{equation}
\begin{split}
\delta c_{d=3}=&-
\frac{3\pi^{4}}{10}\left(  \frac{T}{\varepsilon_{F}}\right)
^{3}
\\
&\times
\left\{
\left[
    \gamma_{b}^{\rho}\right]^2
\ln\frac{\varepsilon_F}{T}
+\frac{3\gamma_b}{2\mu_3}
\int_{0}^{{\mathscr X}(T)}
\frac{dz}{z^2}\left[{\mathrm Li}_2\left(-z
       \right)\right]^2
\right\};
\end{split}
\label{f42}
\end{equation}
where the variable ${\mathscr X}(T)$ is defined in \req{f37b}.
Equations \rref{f4142}
refine \reqs{f3637} by including all the leading logarithmic
corrections originating from the interaction of the spin modes.
\end{subequations}

Equations \rref{f37b}, \rref{f37d}, \rref{f39},
\rref{f40}) give the final results for
different cases and demonstrate non-trivial logarithmic dependencies on
temperature. This behavior is more complicated than what one usually expects
for the Fermi liquid picture. The unusual behavior is due to the interaction
between the spin excitations. As concerns the charge excitations, they
contribute in a more simple way, and their contribution is completely
expressible in terms of the Fermi-liquid interaction function.

\subsection{On the role of the Cooper channel and
choice of ultraviolet cut-off function.} 
\label{sec7c0}

Strictly speaking, all the results we present here can be
justified for $r_{0}p_{F}\gg 1,$ where $r_{0}$ has been introduced
as the shortest length of our low energy theory. However, we hope
that they remain relevant for the initial model of the Fermi gas
with a repulsion. The scale $r_{0}$ in this case has to be found
from the explicit calculation of the logarithmic corrections in
original model of the interacting fermions rather than in the
reduced model \rref{a7}. Such calculation will not be done in the
present paper, however, we will try to outline the steps which
should be performed, without claiming too much rigor.

\begin{figure}[ht]
\setlength{\unitlength}{2.3em}
\begin{picture}(10,9)

\put(-0.3,0){
\begin{picture}(10,8)

\put(2,4){\includegraphics[width=3.5\unitlength]{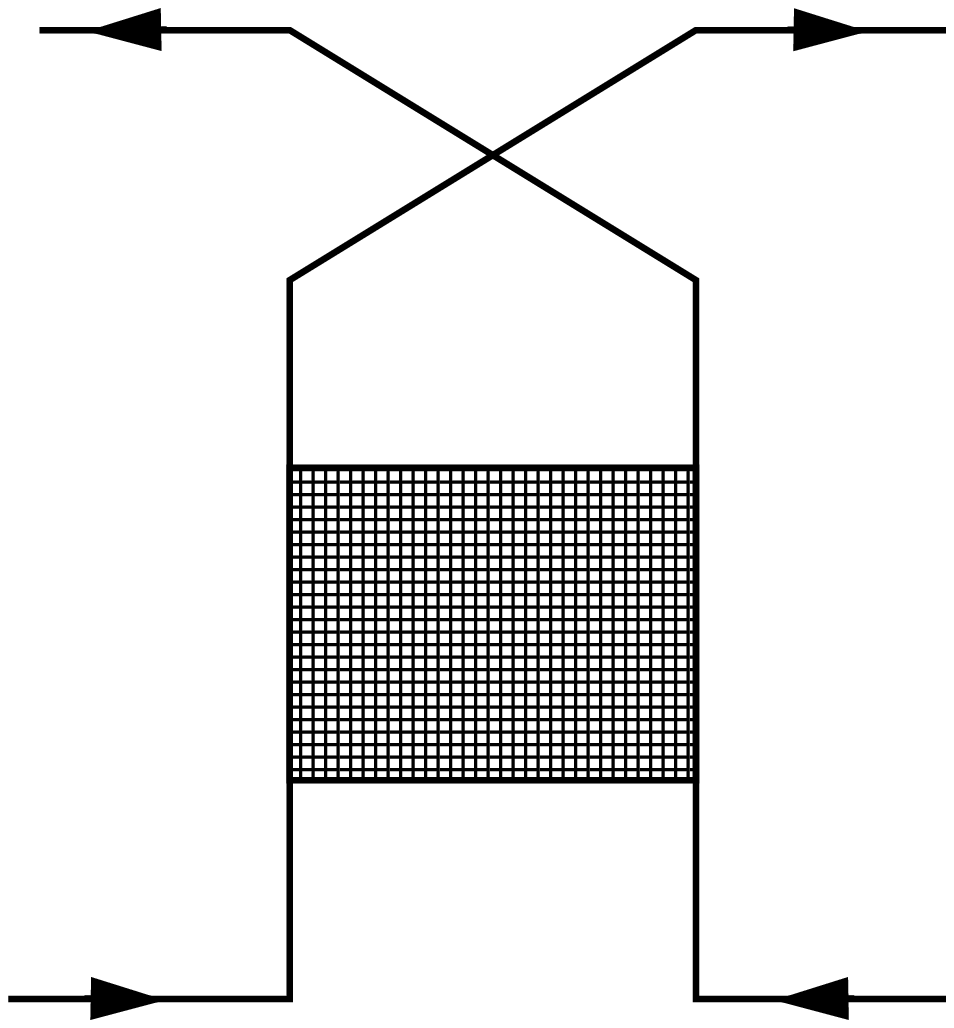}}
\put(5,5.5){$
\displaystyle{\gamma_c\left(\pi-\frac{|\k_\perp|}{p_F};
\p_1+\p_2;\epsilon_1+\epsilon_2\right)}$}
\put(2.,7.1){$\sigma_1$} \put(5.3,7.1){$\sigma_2$}
\put(2.,4.5){$\sigma_2$} \put(5.3,4.5){$\sigma_1$}

\put(0,8.1){$\p_1+\frac{\k}{2},\epsilon_1+\Omega$}
\put(5.3,8.1){$\p_2-\frac{\k}{2},\epsilon_2-\Omega$}

\put(1.4,3.5){$\p_1-\frac{\k}{2},\epsilon_1$}
\put(5.3,3.5){$\p_2+\frac{\k}{2},\epsilon_2$}
\end{picture}
}

\put(0,0.7){\setlength{\unitlength}{2.3em}
\begin{picture}(10,8)

\put(0,0){\includegraphics[width=9\unitlength]{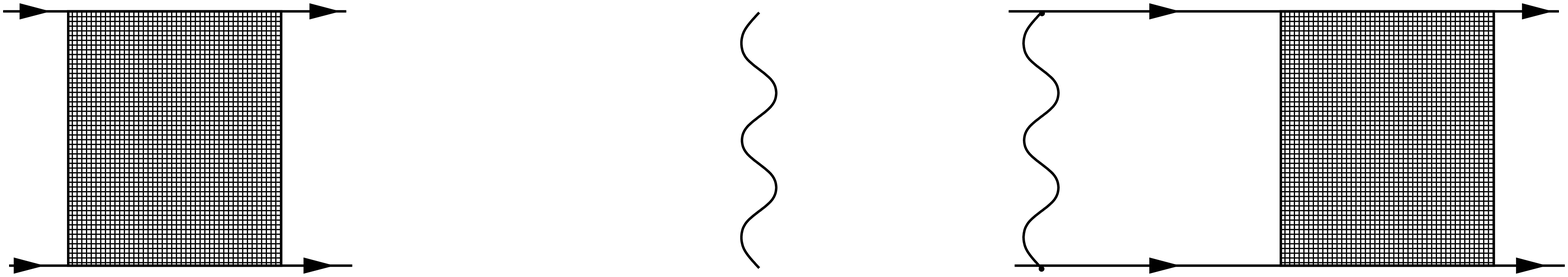}}
\put(3,0.6){$=$}
\put(5,0.6){$+$}

\put(6.6,1.7){$\p_i$} \put(5.5,1.7){$\p_1$} \put(9.1,1.7){$\p_2$}
\put(6.4,-0.3){$-\p_i$} \put(5.1,-0.3){$-\p_1$} \put(8.9,-0.3){$-\p_2$}
\end{picture}
}

\end{picture}
\caption{Leading logarithmic renormalization of the vertex
$\gamma_b$. The integration over the intermediate   momenta $P_i$
has to exclude the region $|\p_i - \p_{1,2}|\lesssim 1/r_0$ as the
latter has already been included in the effective energy theory
for the spin excitations, see also Figs.~\ref{fig2}, \ref{fig20}.}
\label{figcooper}
\end{figure}

It is well known\cite{agd} that the Fermi liquid functions
experience the strong logarithmic renormalization for scattering
directions close to backwards. (Such a logarithmically divergent
term for the Fermi liquid function has been obtained for the first
time in Ref.~\onlinecite{AKh}.) This is because the Cooper
channel, see Fig.~\ref{figcooper}, at such angles has a strong
mixing with the electron-hole channel. If the logarithmic
renormalization were present only in the Cooper channel the result
would read \be
\begin{split}
&\hat{\gamma_c}
\left({\mathbf Q},\omega\right)
=\frac{\hat{\gamma}_0}{1+
\hat{\gamma_0}
{\mathscr L}};
\\
&{\mathscr L}= \ln\left[\frac{\epsilon_F}{
{\rm max}\left(|\omega|,v_F|{\mathbf Q}|\right)
}\right],
\end{split}
\label{f139} \ee 
i.e, one had a zero-charge situation provided
operator $\hat{\gamma_0}$ is positive definite. The needed
function $\hat{\gamma_c}(\theta;Q,\omega)$ are defined as kernels
of the operator $\hat{\gamma_c}$.
Notice, however, that the result dependends on all of the
angular harmonics of the bare interaction $\hat{\gamma_0}$.

There is a region in the phase space, however, where the Cooper
channel and the triplet channel cannot be distinguished from each
other, see also Figs.~\ref{fig2}, \ref{fig20}, and this is the
region that was studied in previous sections.
Even though the structure of the results \rref{f4142}, indicate that,
indeed, the result can be factorized to the logarithmic
renormailization of the backscattering amplitudes
as it was argued in Ref.~\onlinecite{chubukov4}, however,
it does not mean that they coincide with the renormalization
by the Cooper channel only.

Closing this section, we  notice that the main contribution in the
Kohn-Luttinger\cite{kl} scenario of the superconducting
instability also originates from the region of the phase space
studied in our paper. As in this region the Cooper channel
intervenes the particle-hole channel, the simple use of the second
order screened interaction in the Cooper channel\cite{kl} does not
appear to be justified.

\subsection{Peculiarities for one dimensional systems.}
\label{sec7c}

The purpose of this subsection is to find the leading singular
correction to the specific heat the in one-dimensional case and
compare the result with the one obtained in
Ref.~\onlinecite{lukyanov} for the spin chain. The low-energy
properties of the latter model are the same as for the spin
dynamics for the interacting electrons.

\begin{figure}[h]
\setlength{\unitlength}{2.3em}
\begin{picture}(10,21.8)
\put(0,10.5){
\begin{picture}(10,12)
\put(-0.6,8.6){$\displaystyle{\frac{\delta\Omega_3}{T}=-}$}
\put(1.8,6.2){\includegraphics[width=4.7\unitlength]{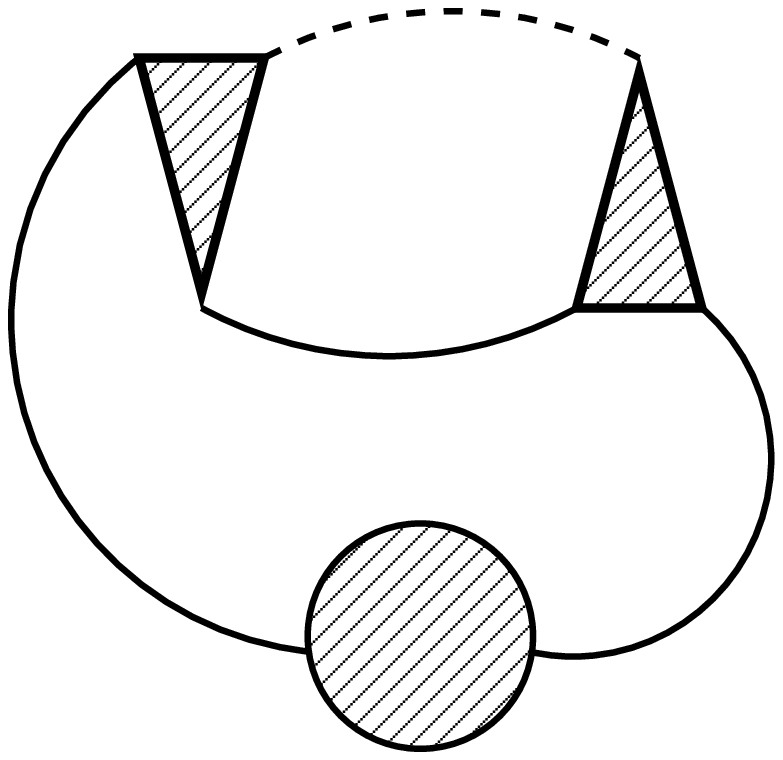}}

\put(6.7,8.0){$\w_1+\w_2$,}
\put(6.7,7.6){$k_1+k_2$}

\put(4.1,10.9){$\w_2,k_2$}
\put(4.1,8.2){$\w_1,k_1$}

\put(1.2,10){a)}
\end{picture}
}

\put(0,5){
\begin{picture}(10,12)
\put(1.5,4.8){c)}

\put(1.9,1.4){\includegraphics[width=6.6\unitlength]{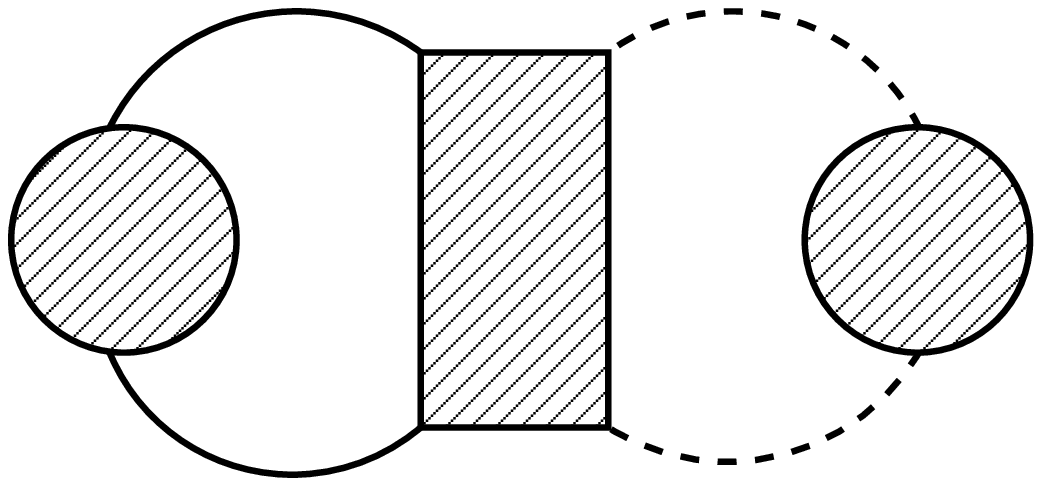}}
\put(2.8,4.6){$\w_1,k_1$}

\put(6.1,4.6){$\w_2,k_2$}

\put(0.4,2.7){$\displaystyle{-\left[\frac{1}{2}\right]}$}
\end{picture}
}

\put(0,4.8){
\begin{picture}(10,12)
\put(0.,8.6){$\displaystyle{-\left[\frac{1}{2}\right]}$}
\put(1.8,6.2){\includegraphics[width=4.7\unitlength]{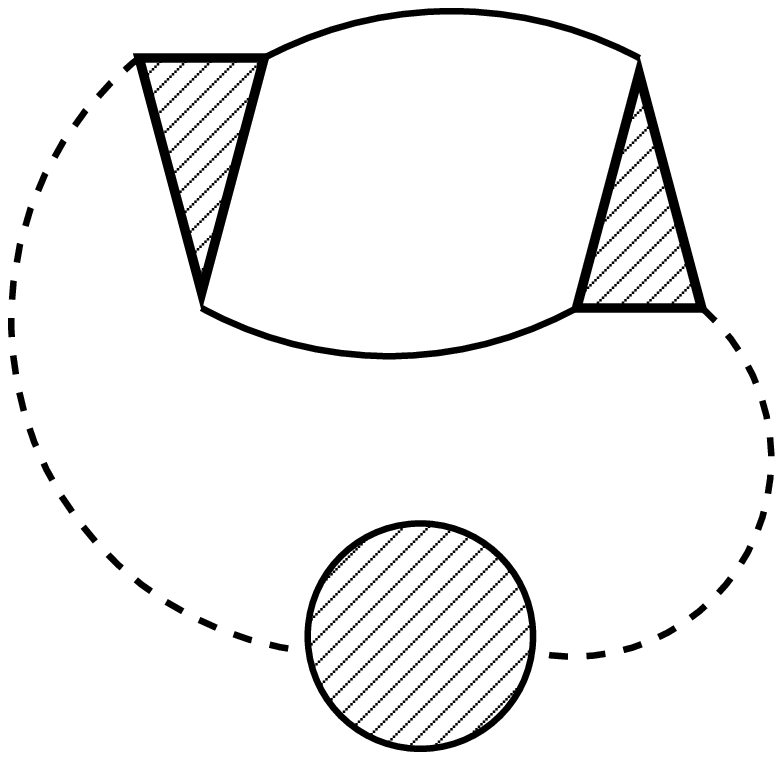}}

\put(6.7,8.0){$\w_2,k_2+k_1$}

\put(4.1,10.9){$\w_1,k_2$}
\put(3.2,8.2){$\w_2-\w_1,k_1$}

\put(1.2,10){b)}
\end{picture}
}

\put(-2,0){
\begin{picture}(10,12)
\put(1.5,4.8){d)}

\put(1.9,0.4){\includegraphics[width=2.1\unitlength]{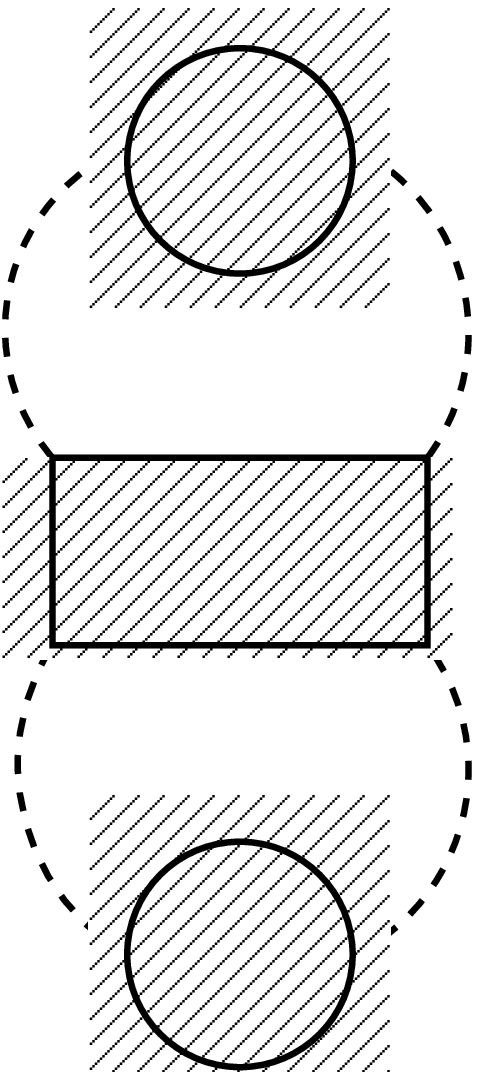}}

\put(4.2,2.7){$\displaystyle{=0}$}
\end{picture}
}

\put(3,0){
\begin{picture}(10,12)
\put(1.5,4.8){e)}

\put(2.3,0.4){\includegraphics[width=4.1\unitlength]{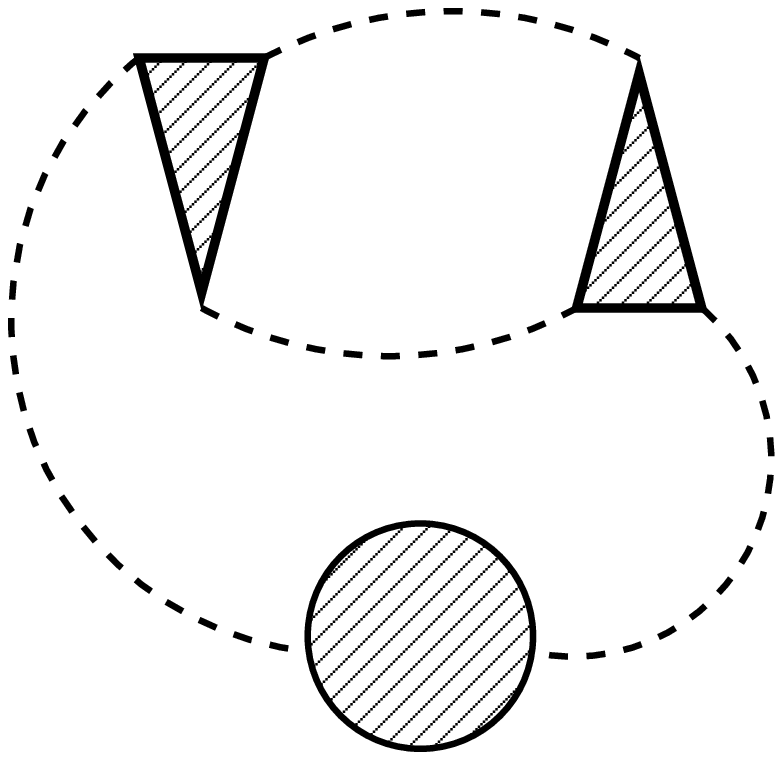}}

\put(4.5,4.6){$\w_1$}\put(4.5,2.8){$\w_2$}
\put(0.9,2.7){$\displaystyle{-\left[\frac{1}{2}\right]}$}
\end{picture}
}

\end{picture}
\caption{Diagrams of the third order, $\delta \Omega_3(T)\propto
\gamma_b^3$, giving the leading logarithmic contribution the in
one-dimensional case. The notation are introduced in
Sec.~\ref{sec5}, see Figs.~\ref{fig201} -- \ref{fig6}. The
coefficients in brackets corresponds to the number of symmetries
of the diagram. The dotted line for the Green function means that
the summation over the frequency of this line is performed with
\req{f4} replacing the summation \rref{f3a}. The solid line
implies the simple integration over frequency as at $T=0$. Diagram
d) does not contribute to the specific heat.}

\label{fig9}
\end{figure}

We represent the temperature dependent part of the desired correction as
[cf. \req{f4}]
\be
\begin{split}
\delta\Omega_3\left(  T\right)  =&\sum_{l\neq 0}
\int\frac{d\omega_2}{2\pi
}\exp\left(
  -\frac{il\omega_2}{T}\right)
\\
&\times
\left[{\mathbb R}_a
\left(  \omega_2\right)+
 {\mathbb R}_b
\left(  \omega_2\right)
+ {\mathbb R}_c
\left(  \omega_2\right)
\right]
\\
&\hspace*{-1.5cm}
+ \sum_{\substack{l_1,l_2\neq 0\\l_1\neq l_2}}
\iint\frac{d\w_1 d\omega_2}{\left(2\pi\right)^2}\exp\left(\frac{il_1\omega_1}{T}
  +\frac{il_2\omega_2}{T}\right)
{\mathbb R}_e
\left( \w_1, \omega_2\right),
\end{split}
\label{f601}
\ee
where the subscripts $a,b,c,e$ indicate
the analytic expression for the corresponding diagrams on
 Fig.~\ref{fig9}.

Summation over the matrix indices is performed using the
formula analogous to \req{d386} with the result
\be
\begin{split}
{\mathbb R}_{a,b}
&=-\frac{6}{\nu}
\sum_{\sigma_{1,2}=\pm} Y^{\sigma_1\sigma_2}
{\mathcal R}^{\sigma_1\sigma_2}_{a,b,e}(\w_2);
\\
{\mathcal R}^{\sigma_1\sigma_2}_{a}(\w_2)
&=2
\lim_{\eta_1\to 0}
\lim_{\eta_2 \to \eta_1}
\int\frac{d\w_1}{2\pi}\int\frac{dk_1dk_2}{(2\pi)^2}
\\
&
\times
\cos\left(k_1\eta_1+k_2\eta_2\right)
{\mathscr R}^{\sigma_1\sigma_2}(\w_1,\w_2;k_{1,2}),
\\
{\mathcal R}^{\sigma_1\sigma_2}_{b}(\w_2)
&=\lim_{\eta_1\to 0}
\lim_{\eta_2 \to \eta_1}\int\frac{d\w_1}{2\pi}\int\frac{dk_1dk_2}{(2\pi)^2}
\\
&
\times\cos\left(k_1\eta_1+k_2\eta_2\right)
{\mathscr R}^{\sigma_1\sigma_2}(\w_2-\w_1,\w_1;k_{1,2});
\\
{\mathcal R}^{\sigma_1\sigma_2}_{e}(\w_2)
&=\lim_{\eta_1\to 0}
\lim_{\eta_2 \to \eta_1}\int\frac{dk_1dk_2}{(2\pi)^2}
\\
&
\times\cos\left(k_1\eta_1+k_2\eta_2\right)
{\mathscr R}^{\sigma_1\sigma_2}(\w_2-\w_1,\w_1;k_{1,2});
\end{split}
\label{f6020}
\ee
where $Y^{\sigma_1\sigma_2}$ depends on the interaction constants only:
\be
{Y}^{\sigma_1\sigma_2}=
\iint_{0}^1\left(u_1u_2\right)^2du_1du_2
\left[
\Delta_3^{\sigma_1\sigma_2}\beta_3^{-\sigma_1}
\beta_3^{-\sigma_2}
\right]
.
\label{f6021}
\ee
The  most interesting factors
${\mathscr R}^{\sigma_1\sigma_2}(\w_1,\w_2;k_{1,2})$
are given by
\begin{subequations}
\label{f602}
{\setlength{\arraycolsep}{0pt}
\bea
&&
{\mathscr R}^{++}
=\frac{\left(i\w_2+k_2\right)\left(i\w_1-k_1\right)}{{\mathcal
    D}(\w_{1,2};k_{1,2})};
\\
&&
{\mathscr R}^{+-}
=\frac{\left(i\w_2+k_2\right)
\left[i\left(\w_1+\w_2\right)-\left(k_2+k_1\right)\right]
}
{{\mathcal
    D}(\w_{1,2};k_{1,2})};\\
&&
{\mathscr R}^{-+}
=\frac{\left[i\left(\w_1+\w_2\right)+\left(k_2+k_1\right)\right]
\left(i\w_1-k_1\right)
}
{{\mathcal
    D}(\w_{1,2};k_{1,2})};\\
&&
{\mathscr R}^{--}
=\frac{\left[i\left(\w_1+\w_2\right)+\left(k_2+k_1\right)\right]
\left[i\left(\w_1+\w_2\right)-\left(k_2+k_1\right)\right]
}
{{\mathcal
    D}(\w_{1,2};k_{1,2})};
\nonumber\\
\\
&&
\begin{split}
{\mathcal D}(\w_{1,2};k_{1,2})&=\left(i\w_1+v_Fk_1\right)\left(i\w_2-v_Fk_2\right)
\\
&\times
\left[i(\w_1+\w_2)-v_F(k_1+k_2)\right]
\\
&
\times\left[i(\w_1+\w_2)+v_F(k_1+k_2)\right].
\end{split}
\nonumber
\eea
}
\end{subequations}
Finally,
\be
\begin{split}
{\mathbb R}_c
&=\frac{12}{\nu}
\iint_{0}^1du_1du_2
\left(u_1u_2\right)^2
\gamma_3
\Delta_3^{-+}\Delta_3^{+-}\lim_{\eta_1\to 0}
\lim_{\eta_2 \to \eta_1}
\\
&
\times
\iint \frac{dk_1dk_2}{\left(2\pi\right)^2}
\int \frac{d\w_1}{2\pi}
\frac{\cos\left(k_1\eta_1-k_2\eta_2\right)}
{\left(i\w_1-v_Fk_1\right)\left(i\w_2+v_Fk_2\right)},
\end{split}
\label{f602c} \ee and all the other contributions ${\mathbb R}_c$
either vanish or produce contributions independent on $\w_2$.

As usual for  one-dimensional system with  the linearized
spectrum, the integrals \rref{f6020}, with the integrands
\rref{f602}, and \req{f602c} have the anomalous character: each
term could be eliminated by the shifts of the momentum if such
arbitrary shifts were allowed. That is why the chosen order of
limits is very crucial for the complete definition of the action.
In terms of the original model, it corresponds to the
regularization of the singular terms
$(\bphi^\dagger_L(x)\bphi_L(x))\bphi^\dagger_R(x)$, and
$(\bphi^\dagger_L(x)\bphi_L(x))(\bphi^\dagger_R(x)\bphi_R(x))$,
etc. in notation of \req{c1070}, by shifting the coordinate of the
left (right) movers by $\eta_1(-\eta_2)$. Such shifts eliminate
all the divergent terms.

Calculating the integrals \rref{f6020} we find
\begin{subequations}
\label{f604}
\bea
&&{\mathcal R}^{++}_a={\mathcal R}^{++}_b=-\frac{|\w_2|}{2 \pi v_F^2};
\\
&&{\mathcal R}^{--}_a={\mathcal R}^{--}_b=-\frac{|\w_2|}{4 \pi v_F^2};
\\
&&
{\mathcal R}^{--}_e= -\frac{{\rm sgn}\w_1{\rm sgn}\w_2}{4v_F^2};
\quad {\mathcal R}^{++}_e=0;
\\
&& {\mathcal R}^{+-}_{a,b,e}
+{\mathcal R}^{-+}_{a,b.e} =
 {\mathcal R}^{++}_{a,b,e}
+{\mathcal R}^{--}_{a,b,e},
\eea
and
\be
\begin{split}
{\mathbb R}_c
&=\left(\frac{|\w_2|}{4\pi v_F^2}\right)\frac{12}{\nu}
\iint_{0}^1du_1du_2
\left(u_1u_2\right)^2
\gamma_3
\Delta_3^{-+}\Delta_3^{+-}.
\end{split}
\label{f604c}
\ee
\end{subequations}

Substituting \reqs{f604} into \req{f601}, using \req{f9}, the
explicit form of the coupling constants, Eq. \rref{e07},
$\mu_1=2$, $\nu=1/(\pi v_F)$, and the formula
\[
\begin{split}
\iint_0^1du_1du_2
\left(u_1u_2\right)^2
&
\left[\frac{4}{\left(1+x u_1u_2\right)^5}-\frac{1}{\left(1+x
      u_1u_2\right)^4}\right]
\\
&
=\frac{1}{3\left(1+x \right)^3},
\end{split}
\]
we find the leading logarithmical contribution to the
thermodynamic potential \be \delta \Omega_3(T) =-\frac{\pi T^2}{16
v_F}
\left[\frac{2\gamma_b}{1+2\gamma_b\ln\frac{\epsilon_F}{T}}\right]^3.
\label{f605} \ee This correction agrees with the one previously
obtained one for the spin chains (see Eq.~(3.17) of
Ref.~\onlinecite{lukyanov}) and we conclude that the
supersymmetric low energy theory developed in this paper
reproduces  all of the known physical results despite the fact
that the intermediate degrees of freedom apparently differ from
those for the conventional bosonization.

\begin{figure}[h]
\setlength{\unitlength}{2.3em}
\begin{picture}(10,21.8)
\put(0,10.5){
\begin{picture}(10,12)
\put(0,8.6){$\displaystyle{\frac{\delta\Omega_3^{bf}}{T}=-\left[\frac{1}{2}\right]}$}
\put(2.8,5.1){\includegraphics[width=4.2\unitlength]{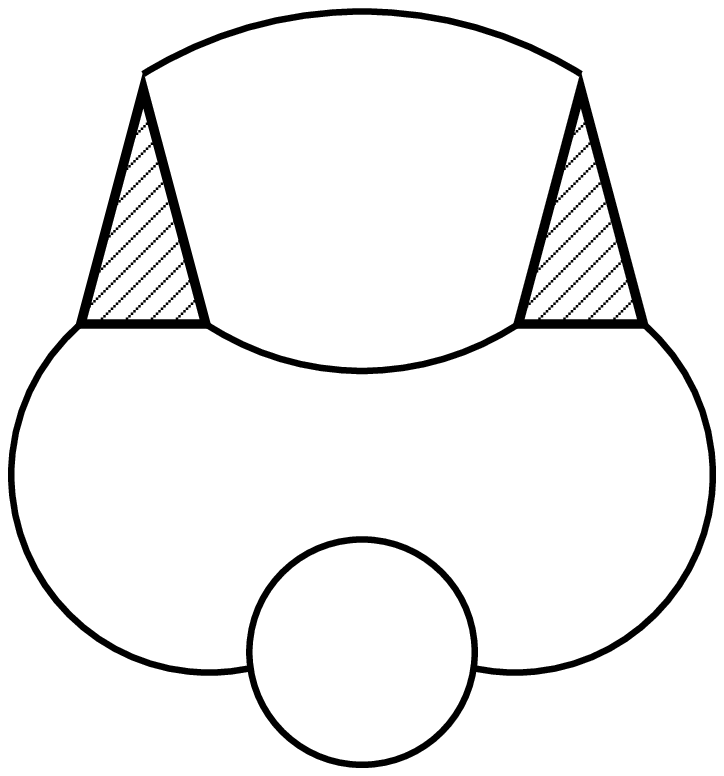}}

\put(1.2,10){a)}
\end{picture}
}

\put(0,5){
\begin{picture}(10,12)
\put(1.5,4.8){c)}

\put(1.9,1.4){\includegraphics[width=6.6\unitlength]{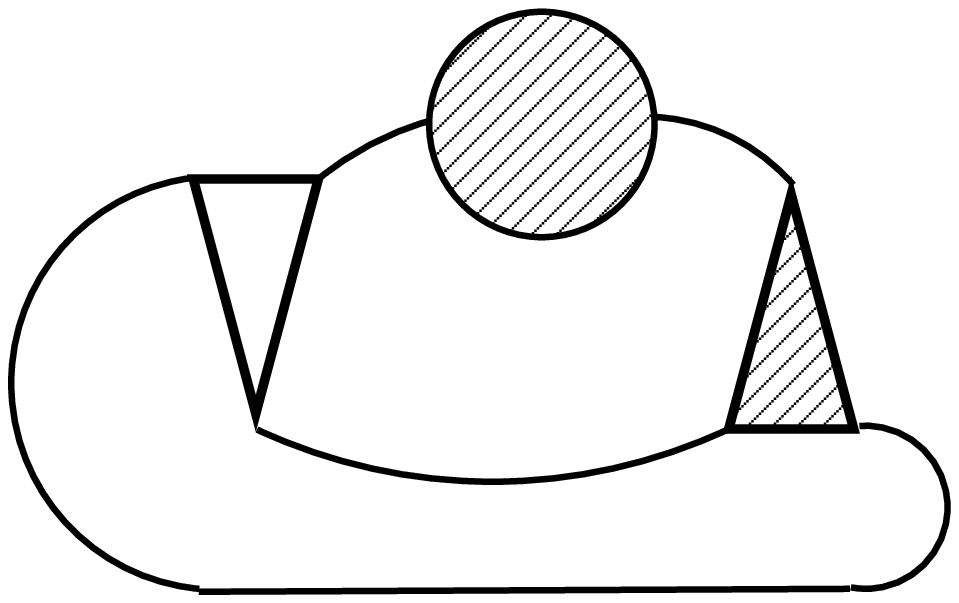}}

\put(0.4,2.7){$\displaystyle{-\left[\frac{1}{2}\right]}$}
\end{picture}
}

\put(0,4.8){
\begin{picture}(10,12)
\put(0.,8.6){$\displaystyle{-\left[\frac{1}{2}\right]}$}
\put(1.8,6.2){\includegraphics[width=4.1\unitlength]{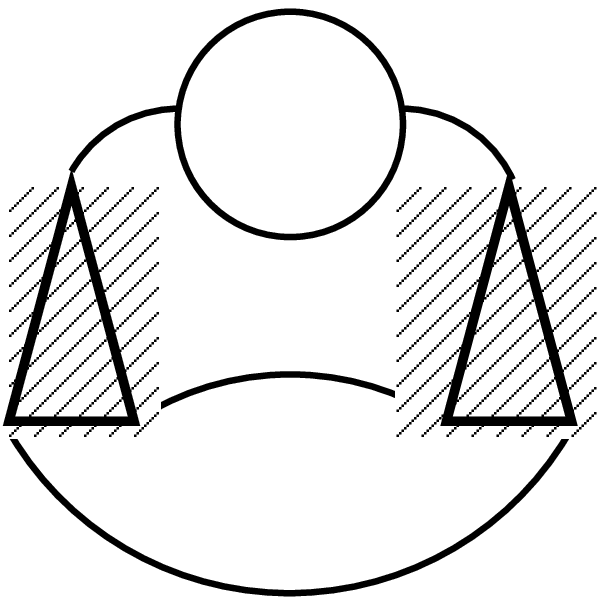}}

\put(1.2,10){b)}
\end{picture}
}

\put(0,0){
\begin{picture}(10,12)
\put(1.5,4.8){d)}

\put(2.5,1.4){\includegraphics[width=4.1\unitlength]{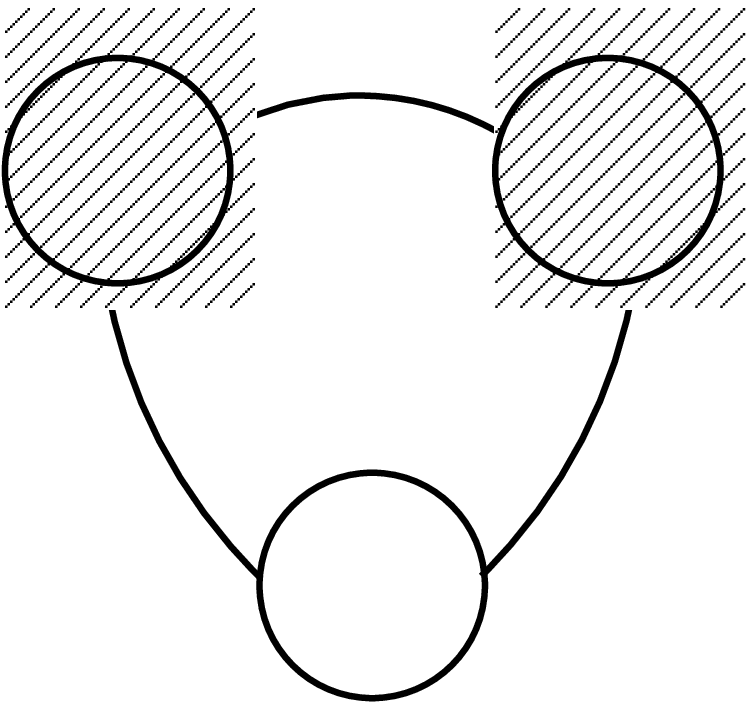}}
\put(1,2.8){$\displaystyle{-\left[\frac{1}{2}\right]}$}

\end{picture}
}

\end{picture}
\vspace*{-1cm}

\caption{Diagrams of the third order, $\delta
\Omega_3^{bf}(T)\propto \gamma_b^2\gamma_f$. Those diagrams do not
contribute to the specific heat. Non-filled vertices correspond to
the physical forward scattering amplitudes which are not
renormalized. Diagram b) vanishes due to the supersymmetry.
Diagram a) vanishes separately. }

\label{fig10}
\end{figure}

One can also prove by the explicit calculation
that there is no contribution to the specific heat
proportional to  $\gamma_f\gamma_b^2$. Corresponding
diagrams are shown on Fig.~\ref{fig10}.

\section{Discussion}
\label{sec8}

We have considered thermodynamics of an electron gas with a repulsion in
arbitrary dimensions. This model belongs to the class of systems that should
manifest a Fermi liquid behavior, which implies that the temperature behavior
of thermodynamic quantities is similar to the one of an ideal Fermi gas.

In order to investigate low lying excitations like spin or charge ones we
developed a new method of bosonization that allows us to replace the initial
electron model by a model for low lying excitations. Our approach is based on
a method of quasiclassical Green functions and differs from earlier high
dimensional bosonization schemes
\cite{haldane,houghton1,houghton,neto,kopietz,kopietzb,khveshchenko,khveshchenko1,castellani,metzner}%
. In contrast to the latter approaches we can consider not only charge
excitations but also spin ones. This advantage is crucial because the spin
excitations are much more interesting than the charge ones. The importance of
the spin excitations is seen from Eqs. (\ref{b10}, \ref{b11}). In contrast to
the charge excitations, the spin ones interact with each other via the
fluctuational magnetic field $\mathbf{h}$.

Studying the low lying spin excitations we have discovered non-trivial
logarithmic contributions to thermodynamic quantities originating from their
interaction and succeeded in summing them using a renormalization group
scheme. The logarithmic contributions come from momenta of the two excitations
parallel or antiparallel to each other (forward
[which doesnot contribute to the physical quantities studied in this paper]
 and backward scattering). To
some extent, the system manifests one dimensional properties even if we work
in two or three dimensions.

In principle, we could solve Eq. (\ref{b11}) using a perturbation
theory in the effective field $\mathbf{h}$. Then, substituting
such a solution into Eq. (\ref{b12}) we would be able to calculate
the thermodynamic quantities. However, this is not a completely
safe scheme. The problem is that the linear operator acting on the
variable $\mathbf{S}_{\mathbf{n}}$ is Eq. (\ref{b11}) is not
Hermitian because it contains linear derivatives in time and
coordinates. It is well established\cite{hatano,hermit} that
linear derivatives in the operators can lead to a new physics
because, the non-hermitian operators may have complex eigenvalues.
This problem can be avoided by the process of the hermitization
\cite{hermit} and everything can be reformulated in terms of a
field theory containing supervectors. This is why we developed a
supersymmetric scheme and used it for the calculations. This gives
also advantages because we can demonstrate the renormalizability
of the theory, which is difficult using the perturbation theory.

The method we developed is applicable also for one dimensional
systems and it exactly reproduced the known result for the
logarithmic correction to the specific heat. It gave us a great
deal of confidence in the correctness of our procedure, However,
our main interest is the higher dimensional systems and we do not
intend to compete with the very well developed methods in $1d$.

Using the method of the renormalization group we calculated all relevant
vertices of the theory, which allowed us to calculate the thermodynamic
potential and the specific heat, Eqs. (\ref{f41}, \ref{f42}). In the lowest
order one neglects the interaction between the spin excitations and obtains
Eqs. (\ref{f36}, \ref{f37}) that have been obtained previously by conventional
diagrammatic expansions (see the latest works\cite{chubukov3,chubukov4} and
references therein). These corrections are already non-analytic in $T^{2},$
which was the main motivation for their previous study. We see from the
results obtained here that the problem is even more interesting and the
temperature behavior of the thermodynamic quantities in really non-trivial.

We derived the results in the weak coupling limit and the approximations we
used are justified. Although we cannot apply the results in the strong
coupling limit, it is difficult to imagine that the non-trivial temperature
dependence would not be relevant in that region. This can lead to complicated
effects near quantum critical points.

The application of the RG scheme we developed has lead us to a very unusual
result, namely, the amplitude $\gamma_{1}$ describing the forward scattering
of the spin excitations has a logarithmic pole, Eq. (\ref{e09}), and can
diverge below a critical temperature. We emphasize that the appearance of this
diverging vertex is a consequence of the hermitization procedure we used for
the derivation of the field theory and its existence cannot be noticed using
conventional diagrammatic expansions.

Very often divergencies of scattering amplitudes lead to a phase transition
as, e.g., in theory of superconductivity. At the same time, the logarithmic
pole in the Kondo problem does not lead to any phase transition. The situation
now is even more tricky because the forward scattering amplitude does not
enter the thermodynamic quantities in the perturbation theory and the RG
scheme at all. Therefore, even if it diverges, this does not necessarily mean
a phase transition because it may drop out from physical quantities.
Clarifying this situation is the most challenging continuation of the
present study.

\section{ Acknowledgments}

We enjoyed interesting discussions with A. Chubukov, G. Schwiete
and R. Teodorescu. We are grateful to
F. Essler and A.A. Nersesyan for bringing Ref.~\onlinecite{lukyanov}
to our attention.  We are grateful to D. Basko, G. Catelani, A, Chubukov, 
L. Glazman, D. Maslov, and A. Millis for reading the manuscript and
valuable remarks.
K.B. Efetov is thankful to Columbia Physics Department
for the kind hospitality
and to grant
\textit{Transregio 12 \textquotedblleft Symmetries and Universality in
Mesoscopic Systems" } of the German Research for financial support.

\end{document}